Universitat Autònoma de Barcelona

# Modelling of Field-Effect Transistors based on 2D Materials targeting High-Frequency Applications

**Francisco Pasadas Cantos**

*A dissertation submitted in fulfilment of the requirements to obtain the Doctor degree in Electronic and Telecommunication Engineering*

*Advisor:*

**Dr. David Jiménez Jiménez**

Departament d'Enginyeria Electrònica

Escola d'Enginyeria

Barcelona, April 28, 2017

# UNIVERSITAT AUTÒNOMA DE BARCELONA

The undersigned, **Dr. David Jiménez Jiménez**, as Ph.D. Advisor and Professor of the Departament d'Enginyeria Electrònica (Escola d'Enginyeria) de la Universitat Autònoma de Barcelona

## CERTIFY:

That the thesis entitled ***Modelling of field-effect transistors based on 2D materials targeting high-frequency applications*** has been performed by the Ph.D. Candidate **Francisco Pasadas Cantos** under his supervision, in fulfilment of the requirements for the Doctor degree in Electronic and Telecommunication Engineering.

And hereby to acknowledge the above, sign the present.

Barcelona, 28th April, 2017

| | |
|---|---|
| Dr. David Jiménez Jiménez | Francisco Pasadas Cantos |
| Ph.D. Advisor | Ph.D. Candidate |

*To my Parents, Sister, Grandma and my Better Half*

"*Learning never exhausts the mind*"

**Leonardo da Vinci** (1452 – 1519)

# Acknowledgements

Although this thesis is written in English, some of the acknowledgements are written in Spanish for the sake of reinforcing my thanks.

En primer lugar, me gustaría dedicar mis primeras palabras a mi director de tesis David Jiménez. Agradecer la infinita dedicación prestada, eterna disponibilidad, contagiosa motivación e incansable apoyo. Sin su profesionalidad, capacidad y experiencia científica nunca podríamos haber llevado a cabo este trabajo. Un referente en lo profesional y en lo personal. Eternamente agradecido por brindarme la oportunidad de llevar a cabo esta investigación bajo su orientación.

También me gustaría agradecer a los miembros del Departament d'Enginyeria Electrònica su acogida, solidaridad y gran compañerismo. Un entorno idóneo para crecer personal y profesionalmente.

Deseo también expresar todo mi agradecimiento a mi familia, por su apoyo incondicional a lo largo de este camino. Dar las gracias a la persona que me acompaña desde hace media vida y hace que todo tenga sentido, gracias Blanca. Muy agradecido a mis amigos y compañeros de trabajo que, de un modo u otro, me han apoyado y ayudado todo este tiempo.

I must express my gratitude to Renato Negra for giving me the opportunity to have a stay in the High Frequency Electronics group from RWTH Aachen University. Thanks to HFE people who made enjoyable my stay in Aachen.

I would also like to give my gratitude to the funding institutions that made this work possible. It has been supported by the following projects:

- European Union Seventh Framework Programme: “Graphene-based revolutions in ICT and beyond – Graphene Flagship”, under grant agreement nº604391.
- European Union’s Horizon 2020 research and innovation programme: “Graphene-based disruptive technologies – GrapheneCore1”, under grant agreement nº696656.
- Ministerio de Economía y Competitividad: “Dispositivos electrónicos de baja dimensionalidad para aplicaciones de radiofrecuencia y digitales: simulación y desarrollo software”, under grant agreement TEC2012-31330.
- Ministerio de Economía y Competitividad: “Transporte de electrones y fonones en nanodispositivos para aplicaciones de bajo y cero consumo”, under grant agreement TEC2015-67462-C2-1-R.

# Modelling of field-effect transistors based on 2D materials targeting high-frequency applications

FRANCISCO PASADAS CANTOS

*Advisor:*

DAVID JIMÉNEZ JIMÉNEZ

*Departament d'Enginyeria Electrònica*

*Escola d'Enginyeria*

*Universitat Autònoma de Barcelona*

COVER: Front and back – artist's impression of a graphene sheet

Universitat Autònoma de Barcelona

# Abstract

New technologies are necessary for the unprecedented expansion of connectivity and communications in the modern technological society. The specific needs of wireless communication systems in 5G and beyond, as well as devices for the future deployment of Internet of Things has caused that the International Technology Roadmap for Semiconductors, which is the strategic planning document of the semiconductor industry, considered since 2011, graphene and related materials (GRMs) as promising candidates for the future of electronics. Graphene, a one-atom-thick of carbon, is a promising material for high-frequency applications due to its intrinsic superior carrier mobility and very high saturation velocity. These exceptional carrier transport properties suggest that GRM-based field-effect transistors could potentially outperform other technologies.

This thesis presents a body of work on the modelling, performance prediction and simulation of GRM-based field-effect transistors and circuits. The main goal of this work is to provide models and tools to ease the following issues: (i) gaining technological control of single layer and bilayer graphene devices and, more generally, devices based on 2D materials, (ii) assessment of radio-frequency (RF) performance and microwave stability, (iii) benchmarking against other existing technologies, (iv) providing guidance for device and circuit design, (v) simulation of circuits formed by GRM-based transistors.

In doing so, a key contribution of this thesis is the development of a small-signal model suited to 2D material based field-effect transistors (2D-FETs) that guarantees charge conservation. It is also provided a parameter extraction methodology that includes both the contact and access resistances, which are of upmost importance when dealing with low-dimensional FETs. Taking it as a basis, an investigation of the GFET RF performance scalability is performed, together with an analysis of the device stability. The presented small-signal model is potentially useful for fast prototyping, which is of relevance when dealing with the first stages of any new technology.

To complete the modelling task, an intrinsic physics-based large-signal compact model of graphene field-effect transistors (GFETs) has been developed, ready to be used in conventional electronic design automation tools. This is a necessary step towards the design of complex monolithic millimetre-wave integrated circuits (MMICs). Most of the demonstrated circuits based on GRMs so far are not integrated circuits (ICs), so requiring external circuitries for operation. At mm-wave frequencies, broadband circuits can practically only be realized in IC technology. The compact model presented in this thesis is the starting point towards the design of complex MMICs based on graphene. It has been benchmarked against high-performance and ambipolar electronics' circuits such as high-frequency voltage amplifiers, high-performance frequency doublers, radio-frequency subharmonic mixers and multiplier phase detectors.

The final part of the thesis is devoted to the bilayer graphene based FET. Bilayer graphene is a promising material for RF transistors because its energy bandgap might result in a better current saturation than the single layer graphene. Because the great deal of interest in this technology, especially for flexible applications, gaining control of it requires the formulation of appropriate models. A numerical large-signal model of bilayer graphene field-effect transistors has been realized, which allows: (i) understanding the electronic properties of bilayer graphene, in particular the tunable bandgap, (ii) evaluating the impact of the bandgap opening on the RF performance, (iii) benchmarking against other existing technologies, and (iv) providing guidance for device design. The model has been verified against measurement data reported.

# Contents

# List of Figures

# List of Tables

# Abbreviations

| | |
|---|---|
| **1D** | One-Dimensional |
| **2D** | Two-Dimensional |
| **2D-DOS** | Two-Dimensional Density Of States |
| **2D-FET** | Two-Dimensional material based Field-Effect Transistor |
| **2DM** | Two-Dimensional Material |
| **AC** | Alternating Current |
| **BLG** | Bilayer Graphene |
| **BLGFET** | Bilayer Graphene Field-Effect Transistor |
| **CB** | Conduction Band |
| **CMOS** | Complementary MOS |
| **CNP** | Charge Neutrality Point |
| **CVD** | Chemical Vapour Deposition |
| **DC** | Direct Current |
| **DD** | Drift-Diffusion |
| **DP** | Dirac Point |
| **DUT** | Device Under Test |
| **EDA** | Electronic Design Automation |
| **FET** | Field-Effect Transistor |
| $f_{max}$ | Maximum oscillation frequency |
| **FoM** | Figure of Merit |
| $f_{Tx}$ | Cut-off frequency |
| **GFET** | Single layer Graphene Field-Effect Transistor |
| **GNR** | Graphene Nanoribbon |
| **GRMs** | Graphene and Related Materials |
| **hBN** | Hexagonal Boron Nitride |
| **HEMT** | High-Electron-Mobility Transistor |
| **HF** | High-Frequency |
| **IC** | Integrated Circuit |
| **ITRS** | International Technology Roadmap for Semiconductors |
| **LPE** | Liquid-Phase Exfoliation |
| **MAG** | Maximum Available Gain |
| **MFP** | Mean Free Path |
| **MMIC** | Monolithic Millimeter-wave Integrated Circuit |
| **MOS** | Metal-Oxide Semiconductor |
| **MOSFET** | Metal-Oxide Semiconductor Field-Effect Transistor |
| **MSG** | Maximum Stable Gain |
| **OC** | Output Characteristic |
| **RF** | Radio-Frequency |
| **SCE** | Short-Channel Effect |
| **SFL** | Shift of the Fermi Level |
| **SLG** | Single Layer Graphene |
| **TC** | Transfer Characteristic |
| **TRL** | Technology Readiness Level |
| **VB** | Valence Band |
| **VLSI** | Very Large Scale Integration |
| **VNA** | Vector Network Analyser |
| **ZBDC** | Zero Bottom electric Displacement field Condition |

# Prologue

With the transistor invention, occurred in the late 1947, the word “end” was written on the vacuum tube era which was going to be replaced by the emerging semiconductor electronics that offered new scenarios, new possibilities and new challenges.

At that time the semiconductor field was not so known by the scientific community and the Bell Labs team, headed by W. Shockley and S. Morgan decided to work on the two simplest semiconductors: silicon and germanium. That was a right decision because two years later they reached the goal to realize the first transistor, but no one knew what was the theory that was mastering such a magic device. A clearly answer came out when S. Shockley wrote down the theory that the world was looking for and when he implemented the first germanium n-p-n junction transistor in 1950 [1]. In the same year Shockley’s theory became a book titled *Electrons and Holes in semiconductor with applications to transistor electronics* [2].

On the wave of enthusiasm during ‘50s, most scientists and engineers, attracted by the potential and the possibility to control those powerful materials, spent a lot of time in order to understand how semiconductors work and how humans can influence and manage such materials. Moreover, the Nobel Prize in physics awarded to W. Shockley, J. Bardeen and W. Brattain “*for their researches on semiconductors and their discovery of the transistor effect*” in 1956 amplified the scientific community interest in the semiconductors field. In 1952 the first unipolar transistor, whose concept was previously patented in 1926 by Lilienfeld [3], was realized by I. Ross and G.

Dacey [4] and in 1960 the first MOSFET was developed at Bell Labs by M. M. Atalla's group [5]. Once the device was created the semiconductor companies, born during '50s, were looking for a process in order to start mass-production. In this regard, Texas Instruments invented in 1957 the mesa transistor that allowed J. Kilby to build the first integrated circuit. The second step was done in Fairchild semiconductors when J. Hoerni developed the planar process for transistors and [6] R. Noyce made an integrated circuit using that technology in 1959. Shortly later, in a 1963 conference paper, C. T. Sah and Frank Wanlass from the Fairchild R&D Laboratory showed that logic circuits combining p-channel and n-channel MOS transistors in a complementary circuit configuration (CMOS) delivered close to zero power in standby mode. Thereupon the way of any electronic/semiconductor company was outlined, thanks to the CMOS concept and the planar process that allowed their connection [7].

From 1960 up to early 2000s, following Dennard's scaling rules [8], the semiconductor industry was able to shrink the MOSFET device following the exponential pace predicted by Moore [9] in 1965. Interestingly when the 130 nm node was reached, the classical Dennard scaling came to an end and a new era, termed "More Moore", started. In that new era, the CMOS physical principle is not changed, but new technological aids are introduced in order to further downscale the transistor. In a nutshell, the 130 nm node was the last CMOS generation that allowed better performance just thanks to its smaller dimensions. Therefore, since the 90 nm generation an important change happened: the channel lattice was strained to get better performance. This was a first signal indicating the possibility that new materials could replace silicon as the active element in the future. Later on, a number of material related enhancements have been demonstrated and included in the manufacturing process. Notably the high κ metal gate process, introduced in the 45 nm node, was key in controlling the transistor gate leakage current.

Nowadays the International Technology Roadmap for Semiconductors predicts the implementation of high-mobility CMOS channel materials in the near term [10]. IBM has recently started the program *7 nm and beyond* which is looking for new materials and circuit architecture designs compatible with the CMOS process, where GRMs belong to their catalogue. As an historical note, IBM has played an important role on graphene-based high-frequency electronics development and the world's first single layer graphene based integrated receiver front end for wireless communication was demonstrated by IBM people in the late 2013 [11].

After the Nobel Prize 2010 in Physics awarded to A. Geim and K. Novoselov "*for groundbreaking experiments regarding the two-dimensional material graphene*", (experiments carried out from 2004 [12], [13]), the research on graphene electronics has grown drastically and in 2013 the European Commission announced a 1 billion euro investment in graphene research and development that will be spread in the next 10 years. This big project, dubbed "*The Graphene Flagship*" [14], wants to start the commercialization of graphene-based electronics during the 2020s and it has proposed a roadmap for graphene [15].

Out of all the fields where GRMs could offer big opportunities, this thesis focuses in the analogue/radio-frequency (RF) applications. In particular, the work presented here, deals with the modelling of 2D material based field-effect transistors (2D-FETs), which are seen as potential candidates to further develop RF applications on both rigid and flexible substrates.

Regarding the organization of this thesis, Chapter 1 gives an introductory discussion about the RF technology state-of-the-art together with a short note on different types of models that can be considered for transistors. Chapter 2 introduces a small-signal model suited to 2D-FETs, which has been further applied to investigate the device RF performance and stability. After a brief note on the electronic properties of both single layer graphene (SLG) and bilayer graphene (BLG), a drain current, charge and capacitance models have been developed for SLG- and BLG-based field-effect transistors (FETs). This exercise has been done in Chapters 3 and 4, respectively. Importantly, the SLG-based FET model is within the category of compact models, meaning that it can be used in conventional electronic design automation software. The proposed models have been benchmarked against experimental prototype transistors. BLG has a special feature consisting in a tunable bandgap that might result in a better current saturation than the single layer counterpart. An analysis of the impact on the RF performance has been also carried out in Chapter 4. Finally, the conclusions and future prospects have been drawn in Chapter 5.

Chapter 1

# Introduction

The emergence of graphene to the world of solid-state electronics in 2004 [12] triggered lots of expectations and dreams for a future revolution in micro-electronics.

The field-effect transistor (FET) is the backbone of the semiconductor electronics. It represents the basic building block of the systems of modern information and communication technology and progress in this important field critically depends on rapid improvements of FET performance. An efficient option to achieve this goal is the introduction of novel channel materials into FET technology. In this regard, two-dimensional materials (2DMs) have drawn considerable attention of scientists and device engineers. Since a steadily increasing number of groups worldwide works intensively on 2DM-based FETs (2D-FETs), the chipmakers have paid attention to the progress in the field. That interest is reflected in the International Technology Roadmap for Semiconductors (ITRS) since 2011, the strategic planning document of the semiconductor industry, which has considered graphene and related materials (GRMs) as candidates for future electronics [10].

Since the emergence of graphene, over a surprisingly short period of time, entire classes of new 2DMs have been discovered. After the enthusiastic early days of graphene research it became clear, however, that graphene would not be able to replace silicon in mainstream electronics at least in the

near- and mid-term future [16], since it does not possess a bandgap, which is mandatorily needed for proper operation [17]. Instead, the main hope for graphene-based electronic devices lies on applications in analogue high-frequency (HF) devices. For these applications, the situation is completely different compared to the competition with silicon complementary metal-oxide-semiconductor (CMOS) technology.

HF transistors only get very fast for short gate and channel length and with a channel material having high mobility and high saturation velocity. Keeping this in mind, graphene has the potential to be the perfect channel material for radio-frequency (RF) transistors: graphene has the highest carrier mobility and highest saturation velocity of any semiconductor material so far [18]–[20]. In addition, any 2DM is the incarnation of an ultra-thin-body material and hence predestined for realizing ultra-short-channel devices.

Furthermore, from a manufacturing point of view, there seems to be no stopper for the success of graphene in electronic applications. Graphene is a planar and therefore well compatible with the planar processing technology used for semiconductors. In addition, the synthesis of graphene has been demonstrated on a square-meter scale [21], exceeding the size of silicon Si and III/V semiconductor wafers. One key advantage of graphene might be the flexibility in terms of substrate choice, as graphene can be transferred to nearly any handling substrate ranging from standard Si wafers to PET-foil for flexible electronics. Nevertheless, the homogeneity and reproducibility of large-scale graphene growth and, especially, the transfer process, are still demanding issues, which must be solved in order to meet the semiconductor industry’s high requests on device yield.

What is more, regarding the applications targeted by GRMs, to meet the fast-growing demand for telecommunication services, developing high-data-rate communication links in the range of multi-gigabit per second is necessary. The high-speed data links can be implemented using either wireless or fibre optic technologies. Wireless technology, particularly in urban areas, has several advantages over fibre optics such as portability, universal deployment, short installation time and cost effectiveness. However, to achieve data rates comparable to that of the fibre optics, there is a need to develop wireless systems with a very large bandwidth (~10 GHz). This may be achieved by operating at millimetre-wave (mm-wave) frequencies (30-300 GHz) [22]. In this regard, graphene is a promising material for the development of mm-wave electronics due to its excellent electron transport properties [23]. There has been a rapid progress in the development of graphene field effect transistors (GFETs) in short time. Many GFET-based circuits including frequency multipliers [24]–[27], mixers [27]–[31], amplifiers [32]–[35] and power detectors [36]–[39] have been presented. Most of the demonstrated circuits so far are not integrated circuits (ICs), so requiring external circuitries for operation. ICs allow for HF operation and complex circuits but at the cost of laborious fabrication process. At mm-wave frequencies, broadband circuits can practically only be realized in IC technology. Up to now, there are only few demonstrations of graphene-based ICs performing complex wireless communication functions such as signal modulation and demodulation (encoding/decoding information into/from a carrier signal) [11], [40]–[42].

So, the growing interest in GRM-based monolithic millimetre-wave integrated circuits (MMICs) results in a demand of GFET modelling in particular and 2D-FET modelling in general, which is needed to fill the gap between both device and circuit levels. In turn, such models should be embedded in standard electronic design automation (EDA) tools allowing for IC design. Those circuit-compatible models could serve to different purposes depending on the Technology Readiness Level (TRL). For low TRL, device models are useful not only for interpreting electrical measurements, but also for designing prototype devices/circuits, and even for device/circuit performance benchmarking against other technologies. If a technology eventually became more mature (higher TRL), a device model would be extremely useful to make the circuit design-fabrication cycle more efficient and complex MMIC designs would be possible.

The current 2DM-based technology is still in its infancy and faces enormous challenges such as the quality of GRMs manufacturing (involving growth and transfer to a suitable substrate), the appropriate integration of 2D-FETs into MMIC design, reproducibility and reliability. Device modelling is an important part of the value chain and is progressing in parallel at a fast pace. The current PhD thesis is focused on the physics-based modelling of 2D-FETs, making especial emphasis on graphene. In order to put graphene technology in context, the next section 1.1 presents a brief overview of the main figures of merit (FoMs) exhibited by outstanding RF FET technologies that have been proposed so far, including the highest FoMs gotten by state-of-the-art GFETs. After that, a survey of general device models is provided in section 1.2, which set the ground for the 2D-FET adapted models presented in chapters 2-4. The chapter ends up with the section 1.3, which presents the thesis outline.

## 1.1 Radio-frequency FETs: state-of-the-art

When operated as an amplifier, a FET does not necessarily need to be switched off. Instead, in most RF amplifier configurations the FET is permanently operated in the on-state and the signal applied to its input appear amplified at the output. The extent to which the input signal is amplified is called gain. Thus the current gain is defined as the RF output current of the transistor divided by the RF input current. Gain is a frequency dependent FoM and decreases with increasing frequency. Two important FoMs of RF transistors are the characteristic frequencies $f_{Tx}$ and $f_{max}$. The cut-off frequency $f_{Tx}$ is the frequency at which the current gain of the transistor drops to unity and the maximum frequency of oscillation $f_{max}$ is the frequency at which the power gain becomes unity. It should be noted that for most RF applications, power gain and $f_{max}$ are even more important than current gain and $f_{Tx}$. As a rule of thumb, the operating frequency should be lower than 20% of the used transistors' $f_{max}$ to guarantee sufficient power gain. Figure 1.1 and Figure 1.2 plot the RF performance of state-of-the-art 2D-FETs ($f_{Tx}$ and $f_{max}$, respectively), which have been benchmarked against other more mature RF technologies. The data presented in those figures have been obtained upon application of a de-embedding procedure.

Regarding $f_{Tx}$, a graphene-based FET of a 67-nm gate length operating at 427 GHz is the highest reported [43]. This number is not that far from the record $f_{Tx}$ exhibited by other competing FETs, namely 688 GHz for a 40-nm GaAs mHEMT [44]. By contrast, 2DM semiconducting FETs, such as $MoS_2$- or phosphorene-based, are still far from reaching 100 GHz. From Figure 1.1, a scaling trend of $1/L$ is observed for all transistor types above 200 nm. Down to about 100 nm gate length, epitaxial or exfoliated graphene based FETs are almost competing with InP HEMTs and GaAs mHEMTs, so it is plausible that by gaining further control of GFET technology, via reduction of the contact resistance and/or output conductance, operation in the THz range could be reached.

In contrast to their impressive $f_{Tx}$ performance, GFETs behave rather poor in terms of the $f_{max}$ as shown in Figure 1.2. The highest $f_{max}$ value reported so far is 200 GHz, corresponding to a GFET of 60-nm channel length [45], which is far from the several hundreds of GHz demonstrated by III-V competitors. For instance, a record $f_{max}$ surpassing 1 THz has been demonstrated by an InP HEMT device of 35-nm channel length [46]. Regarding $MoS_2$- and phosphorene-based FETs, neither $f_{Tx}$ nor $f_{max}$ have demonstrated yet a competitive value, which can be due to a number of reasons. Among them the lack of high mobility transport due to the material quality as well as interface quality with the substrate could be an important bottleneck, so higher technological control is still required to get competitive FoMs.

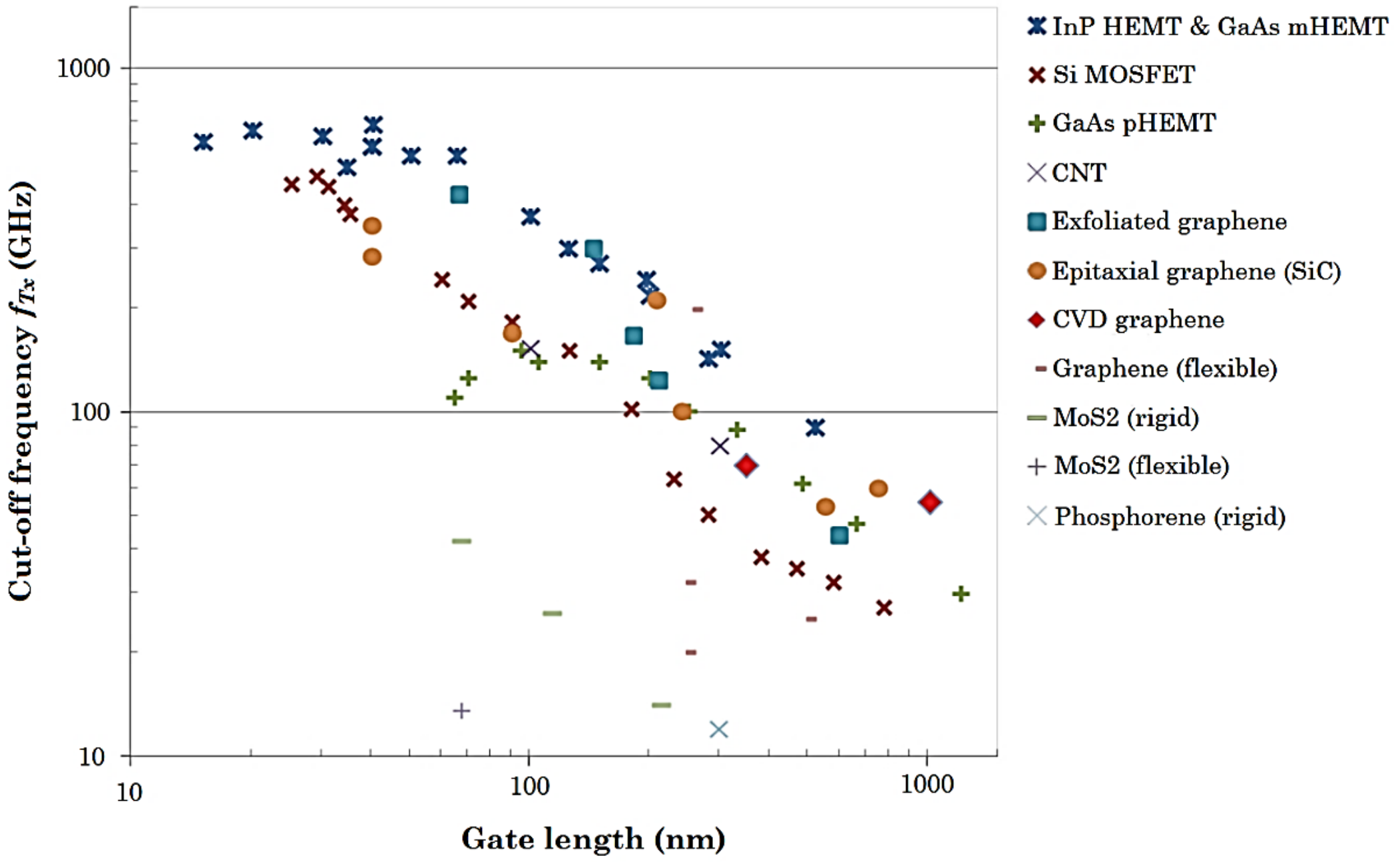

**Figure 1.1** Cut-off frequency of 2D-FETs versus gate length. Also shown is the $f_{Tx}$ performance of the best carbon nanotube FET and that of three classes of conventional RF FETs: InP HEMTs and GaAs mHEMTs (metamorphic HEMT); GaAs pHEMTs (pseudomorphic HEMT); and Si-MOSFETs. (Image taken from [47])

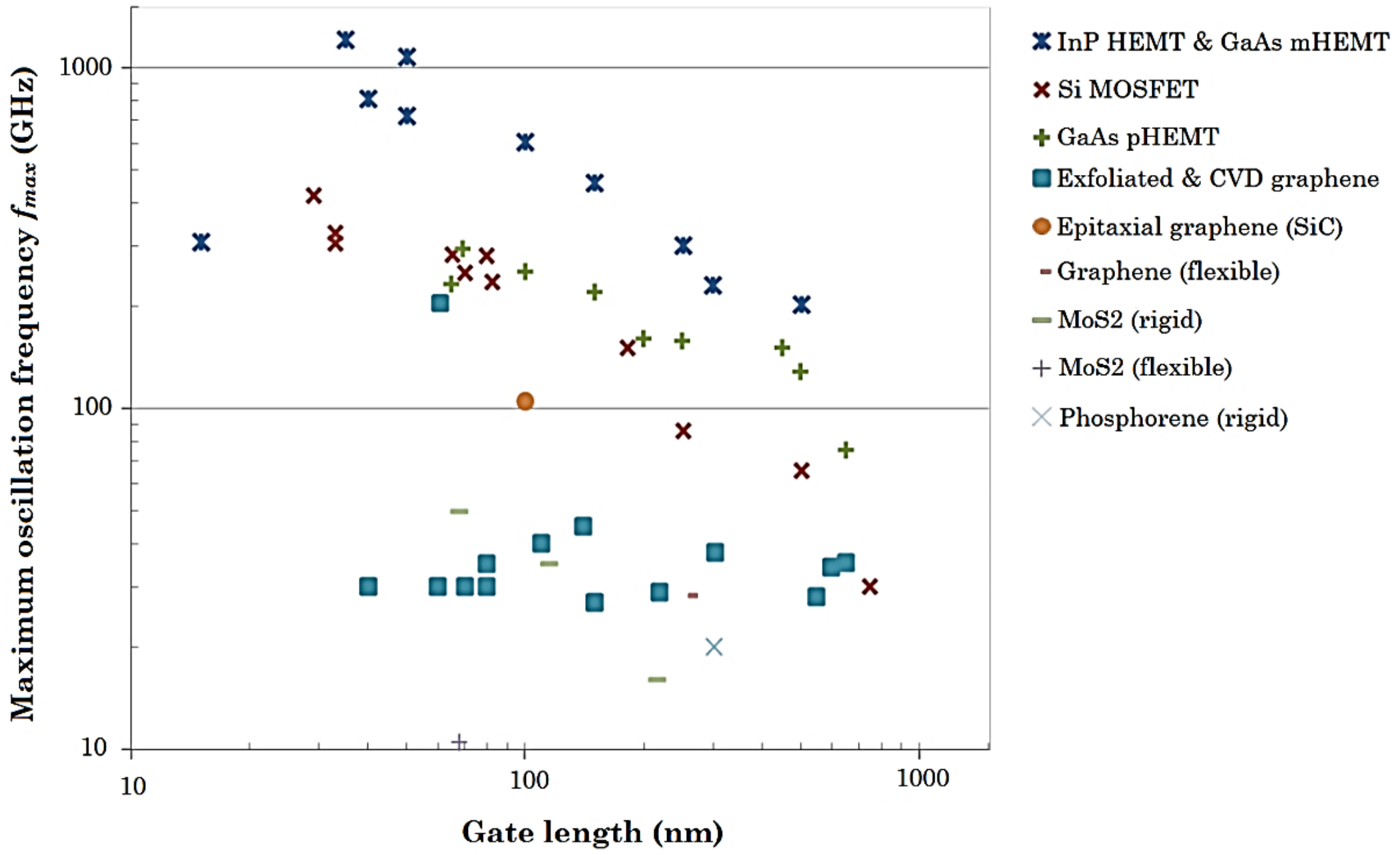

**Figure 1.2** Maximum oscillation frequency of 2D-FETs versus gate length. Also shown is the $f_{max}$ performance of three classes of competing RF FETs: InP HEMTs and GaAs mHEMTs (metamorphic HEMT); GaAs pHEMTs (pseudomorphic HEMT); and Si-MOSFETs. (Image taken from [47])

## 1.2 Types of device models

A survey of the different types of device models is provided in this section for the sake of presenting the main features of each model and, most importantly, the purpose of each one [48]–[50]:

- *Empirical model vs. Physical model*

A purely empirical model relies on just curve fitting. It can use any equation that adequately fits data. Thus, the parameters in such models are the coefficients and exponents used in the curve-fitting expressions, and have no physical significance. Such models can therefore be developed fast, can be formulated in a global form covering any technology, and can be quickly updated. However, a different set of empirical parameters would be needed for each situation, namely, each DC bias, each transistor geometry or even each temperature, since these models have no way to incorporate these effects. Moreover, for example, a drain current curve-fitting expression cannot reliably predict value for biases outside the range in which the model was optimized. No model presented in this thesis belongs to this group.

On the other hand, a physical model is based on device physics which parameters have a physical significance, i.e., the flat-band voltage, the carrier mobility, the oxide thickness, etc. Such models take a long time to develop because of the large amount of physical effects involved in each device. Nevertheless, it is possible to relate the outcome provided by the model to the physical details of the transistor which is very important in IC design because the parameters in such models have a physical significance. Within limits, it would be possible to predict the outcome if the fabrication process parameters were changed. This latter feature is of particular significance in statistical analysis in order to predict ranges of expected performance and yield for given specifications and systematic and random errors of the fabrication process.

- *Small-signal model vs. Large-signal model*

The devices usually operate under time-varying terminal voltages. Depending on the magnitude of the time-varying signals, the dynamic operation can be classified as large-signal operation and small-signal operation. If the time-varying signal is small enough so that the resulting small current and charge variations can be expressed in terms of it using linear relations. This way a non-linear device can be treated as a linear circuit with conductance, inductance and capacitance elements forming a lumped network.

However, if no restrictions on the magnitude of the time-variations are imposed then the device has to be studied under a large-signal dynamic operation, and therefore the evolution of the charges and terminal currents cannot longer be approximate as a linear relation.

A brief mention is worth noticing to a device model which is in between these two categories. A look-up table model is typically in the form of tables containing values of the drain current and small-signal parameters for a large number of combinations of bias voltages. In this way, a large-signal model is built from many small-signal approaches. The values stored can come from measurements, or from physics-based simulations.

- *Numerical model vs. Compact model*

A rigorous way of describing the operation of a 2D-FET is to write the fundamental equations of the 2DM plus all the physical effects affecting a specific technology. These will result in coupled non-linear partial differential equations, one for each of thousands of finite planar elements in the device. That is, in summary, what is usually done to build up a numerical model [48], which usually converges to the solution after a number of iterations. Although such models are invaluable for device analysis and design, the solution can take a long time even for a single transistor. Such an approach is out of the question for general circuit simulation. Much more efficient models are thus needed, which describe the electrical behaviour in an analytical form. In doing so, a compact model represents a device model sufficiently simple to be incorporated in circuit simulators and sufficiently accurate to make the model outcome useful to circuit designers [50].

## 1.3 Thesis outline

The thesis starts with Chapter 2 presenting a small-signal model for 2D-FETs, which is appropriate for circuit simulation. The small-signal parameters can be extracted either from the parameter extraction methodology proposed using the device characterization or from an existing numerical large-signal model of the specific device. Taking advantage of such a small-signal model, the analysis of both stability and RF performance of 2D-FETs is deeply investigated. Then, in Chapter 3 a large-signal model of GFETs is presented in both numerical and compact forms. The compact model allows for circuit simulation so using it, a benchmarking of state-of-the-art GFET-based circuits has been realized. In Chapter 4, a numerical large-signal model of bilayer graphene based FETs has been presented in order to evaluate the impact of bandgap opening on getting better RF performance. All models presented in this thesis are physics-based. Finally, in Chapter 5, the main conclusions have been outlined and future outlook has been given too.

# Chapter 2

# Small-signal model for 2D material based FETs

Research into 2D-FETs is propelling the state-of-the-art of digital and high-frequency electronics both on rigid and flexible substrates [15], [51]–[53]. Ongoing efforts are focused on the demonstration of 2D-FETs outperforming the power consumption of MOSFETs in digital applications and 2D-FETs working at terahertz frequencies exhibiting power gain. In parallel, there is a great deal of interest in developing digital and RF optimized transistors on flexible substrates. A number of advances in those directions have been made in a short time and even a number of simple circuits have been demonstrated [54], [55].

2D-FETs are now operating within the mm-wave range showing intrinsic cut-off frequencies ranging from tenths to hundreds of gigahertz, and maximum oscillation frequencies up to tenths of gigahertz [45], [56]–[58] (see Figure 1.1 and Figure 1.2). Consequently, there is a demand for accurate device models for optimizing the device operation; benchmarking of device performances against other existing technologies; and bridging the gap between device and circuit levels.

To fulfil this demand, a small-signal model of a 2D-FET is proposed in this chapter. Importantly, the parameters of the model could be either directly extracted from the characterization of the device under test (DUT) by means of the $S$-parameters or fed from a numerical physics-based large-signal model of the 2D-FET. In the former case, a different set of small-signal parameters extracted from a characterization of the DUT would be needed for each DC bias or even each temperature. However, in the latter case, this is solved given that the small-signal parameters could be obtained for different biases and temperatures straight from numerical simulations, provided that a numerical physics-based large-signal model was available for a particular technology. For example, the numerical large-signal models for graphene-based FETs presented in sections 3.4 and 4.1 could be used for this purpose. Moreover, a small-signal model, like the one that is being proposed here, where the parameters are extracted from $S$-parameters measurements, is very useful for fast prototyping, which is of upmost importance when dealing with the first stages of new technologies (low TRL).

On the other hand, a compact physics-based large-signal model is more suitable for mature technologies. However, it takes a long time to be developed because of the large amount of physical effects involved in each device and the complexity of including them in a compact way. For example, in the context of 2D-FETs, the quality of the 2D material is crucial, which means including, i.e., the effect of the impurities and defects of the channel into the physical model in a compact way to reproduce the electrical behaviour of 2D-FETs and to make reliable circuit designs based on such devices.

When considering analogue and RF electronic applications, FET terminals are polarized with a DC bias over which an AC signal is superimposed. The amplitude of the AC signal is usually small enough so the $I$-$V$ characteristic can be linearized around the DC bias [48]. This way a non-linear device can be treated as a lumped network, which constitutes the basis of a small-signal equivalent circuit. In this chapter, a specific model that works for 2D-FETs (see Figure 2.1) is formulated, which encompasses both graphene-based FETs and 2D semiconductor based FETs. Specifically, the focus is on modelling the device part between source and drain, containing the 2D layer, the gate oxide and the metal contacts. This part is called the intrinsic part and is the part mainly responsible for transistor action. The rest of the device and surroundings will constitute the extrinsic part and it is responsible for parasitic effects, which can limit the overall performance. Such an extrinsic network could be included as a subcircuit connected to the intrinsic part.

This chapter first provides the description of a small-signal equivalent circuit, which guarantees charge conservation, in section 2.1. It then continues with the analysis of the RF performance of 2D-FETs. In doing so, explicit expressions for the RF FoM calculation based on the charge-conserving small-signal model are provided in section 2.2, comparing its outcome against other calculations reported in the literature. In section 2.3, a methodology to extract the small-signal parameters from $S$-parameter measurements is proposed. Importantly, the approach allows extracting the series combination of the source/drain contact and access resistances which is of upmost importance when dealing with low-dimensional FETs. This methodology has been applied to an exemplary RF GFET. Section 2.4 is devoted to the investigation of scalability of GFET RF performance. This has been done in parallel with device stability analysis, which requires of the previous introduction of tools used by microwave engineers. The chapter ends with section 2.5 providing the main conclusions.

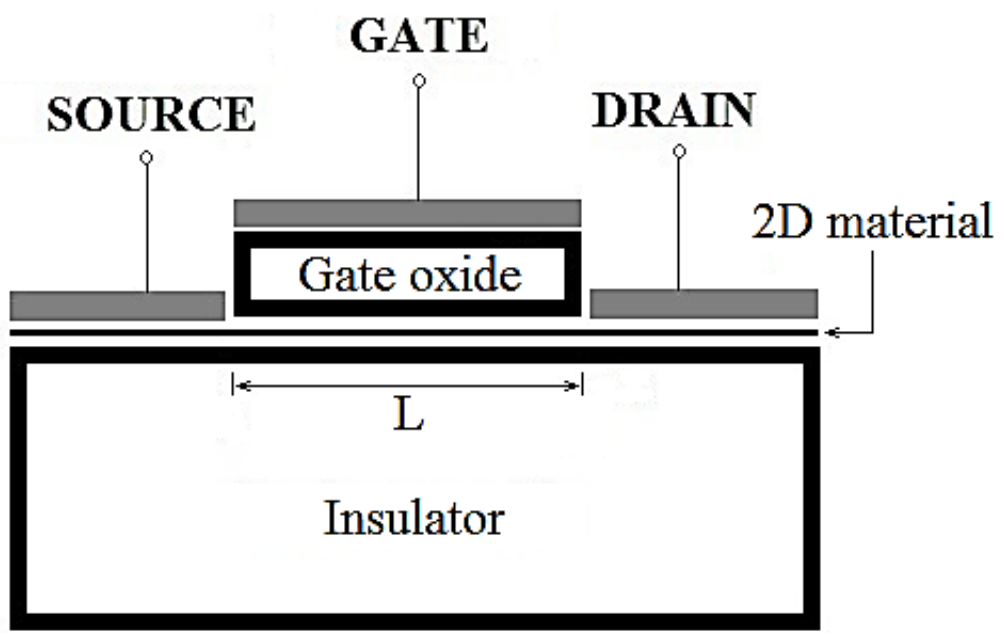

**Figure 2.1** Cross-section of a three-terminal 2D-FET. A 2D material sheet plays the role of the active channel with channel length of $L$. The modulation of the carrier population in the channel is achieved via a top-gate stack consisting of a dielectric and corresponding metal gate.

## 2.1 Charge-conserving small-signal equivalent circuit

The dynamic operation of a device operating under time-varying terminal voltage excitations is influenced by the capacitive effects, becoming indeed essential for circuit design to derive reliable models encompassing such capacitive effects. Several intrinsic capacitance models for FETs have been developed along the years. Basically, they can be categorized into two groups: (i) Meyer [59] and Meyer-like capacitance models and (ii) charge-based capacitance models. The advantages and shortcomings of the two groups of models have been widely discussed and both of them have been implemented in circuit simulators [49], [60].

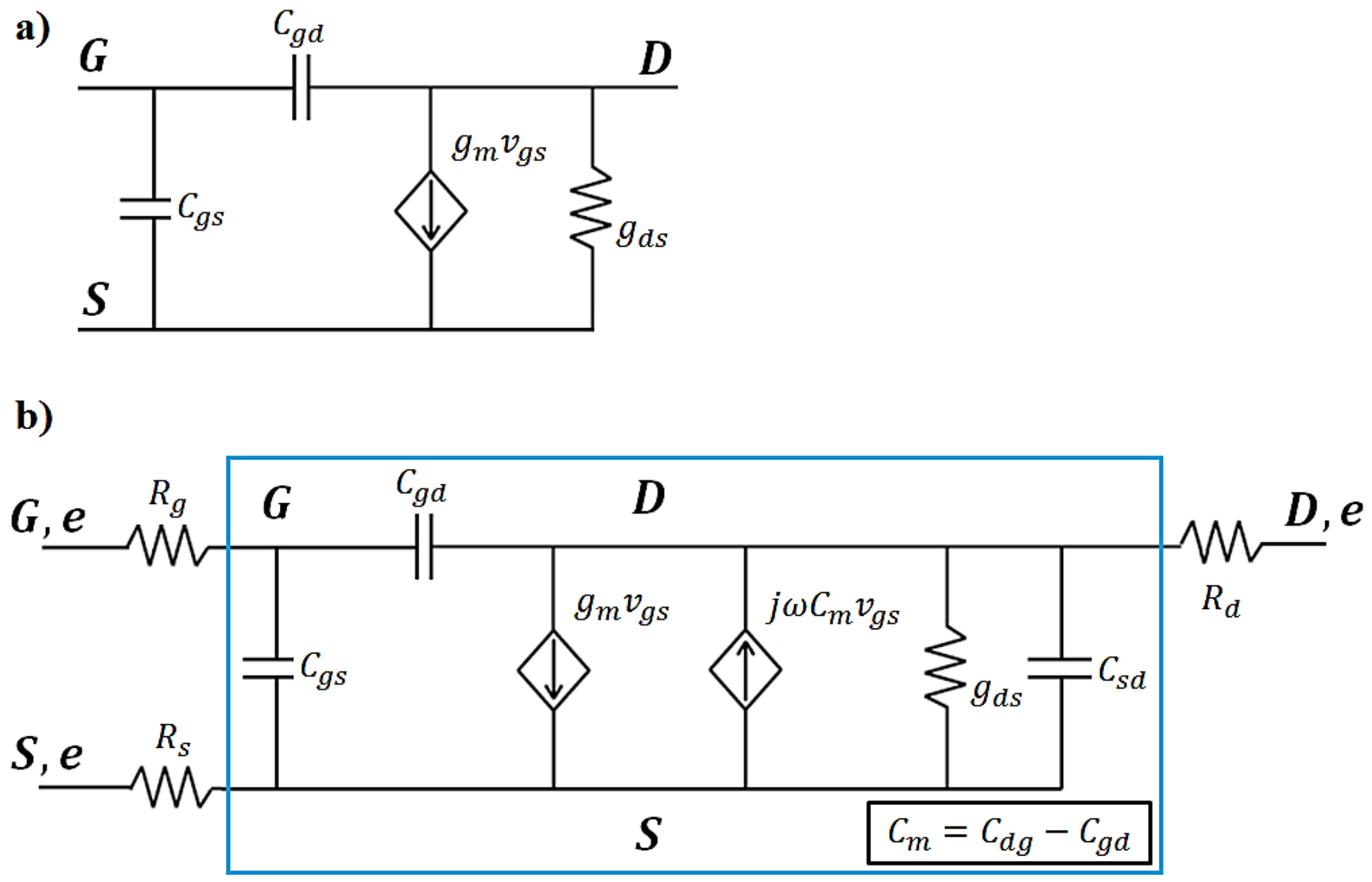

**Figure 2.2** a) Meyer-like intrinsic small-signal model for a three-terminal FET. b) Small-signal model that guarantees charge conservation. The equivalent circuit of the intrinsic device is framed in blue. The small-signal elements are: $g_m$ transconductance, $g_{ds}$ output conductance and $C_{gs}$, $C_{gd}$, $C_{sd}$ and $C_{dg}$ intrinsic capacitances. The physical meaning of the elements is explained in section 3.4 for a GFET. $R_g$ is the gate resistance and $R_d$, $R_s$ account for the series combination of the contact and access resistances of the drain and source respectively.

So far, the small-signal equivalent circuits proposed for 2D-FETs are directly imported from Meyer-like capacitance models [52], [57], [58], [61]–[63], which are widely used because of its simplicity and fast computation. This kind of models can be represented with the equivalent circuit shown in Figure 2.2a. They assume that the intrinsic capacitances of a FET are reciprocal, thus, they cannot ensure charge conservation which is of upmost importance not only for accurate device modelling and circuit simulation but even more for proper parameter extraction [64]–[66]. Instead, the charge-conserving small-signal model shown in Figure 2.2b is proposed [48], which is suitable for HF analysis. However, it should be underlined that both Meyer and charge-based modelling approaches assume the so-called *quasi-static-operation* approximation, where the fluctuation of the varying terminal voltages is assumed to be slow, so the stored charge could follow the voltages variations. Such an approximation is found to be valid when the transition time for the voltage to change is less than the transit time of the carriers from source to drain. This approximation works well in many FET circuits, but it could fail for some cases, especially with long-channel devices operating at high switching speeds, when the load capacitance is very small, and for digital circuits [49], [60].

Based on the above-mentioned assumption, let us begin with the derivation of the $Y$-parameters of the intrinsic part of the small-signal model, which is depicted inside a blue frame in Figure 2.2b. Such an equivalent circuit has been considered as a two-port network connected in a common source configuration, as shown in Figure 2.3.

The intrinsic $Y$-parameters ($Y_i$) can be written as:

$$Y_i(\omega) = \begin{pmatrix} y_{11,i} & y_{12,i} \\ y_{21,i} & y_{22,i} \end{pmatrix} \begin{cases} y_{11,i} = i\left(C_{gd} + C_{gs}\right)\omega \\ y_{12,i} = -iC_{gd}\omega \\ y_{21,i} = g_m - iC_{dg}\omega \\ y_{22,i} = g_{ds} + i\left(C_{gd} + C_{sd}\right)\omega \end{cases} \tag{2.1}$$

where $\omega = 2\pi f$ and $f$ is the frequency of the AC signal and ports 1 and 2 refer to the gate-source and drain-source ports, respectively.

Consequently, the $Z$-parameters of the equivalent circuit can be expressed as:

$$Z(\omega) = \left[Y_i(\omega)^{-1} + R\right] \quad \text{where} \quad R = \begin{pmatrix} R_g + R_s & R_s \\ R_s & R_d + R_s \end{pmatrix} \tag{2.2}$$

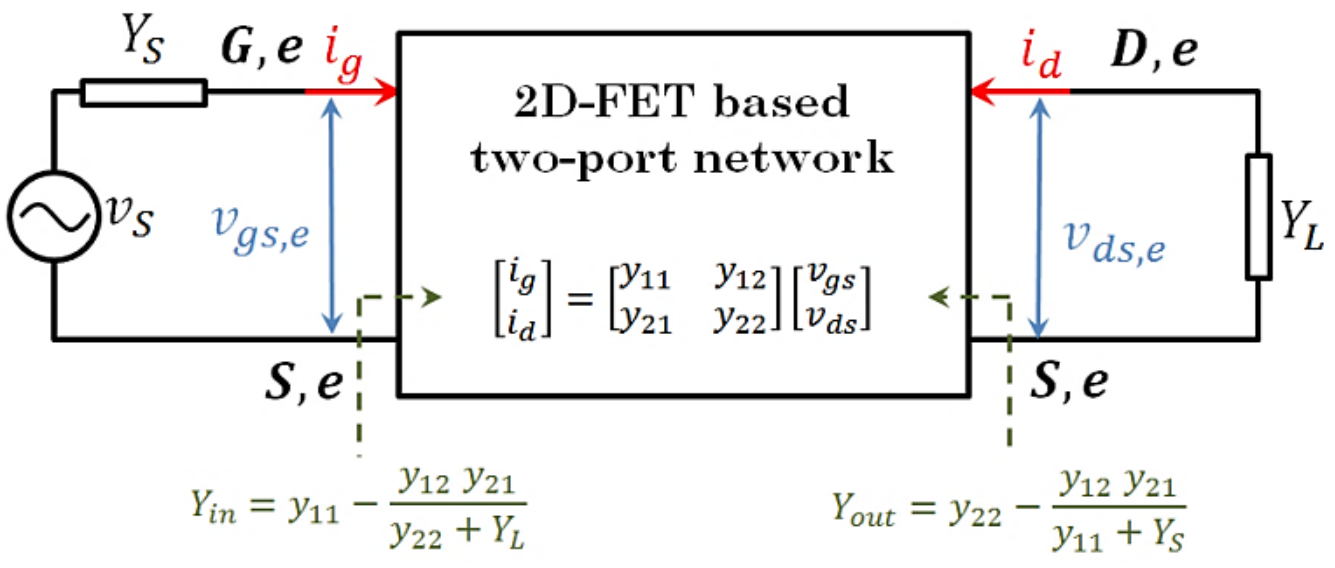

**Figure 2.3** 2D-FET conceptualized as a two-port network, characterized by its $Y$ matrix, connected to source and load admittances.

## 2.2 RF performance of 2D-FETs

Whenever investigating a new technology for electronic applications, it is of primary importance to get the FoMs and compare them against the requirements of the ITRS [10]. Considering the target of HF electronics, the cut-off frequency ($f_{Tx}$) and the maximum oscillation frequency ($f_{max}$) are the most widely used FoMs. The $f_{Tx}$ is defined as the frequency for which the magnitude of the small-signal current gain ($h_{21}$) of the transistor is reduced to unity [67]:

$$h_{21}(\omega) = -\frac{y_{21}}{y_{11}} \rightarrow \left|h_{21}(2\pi f_{Tx})\right| = 1 \tag{2.3}$$

where the $Y$-parameters entering in (2.3) come from the impedance matrix calculated in (2.2):

$$Y(\omega) = \begin{pmatrix} y_{11} & y_{12} \\ y_{21} & y_{22} \end{pmatrix} = Z(\omega)^{-1} \tag{2.4}$$

On the other hand, the $f_{max}$ is defined as the highest possible frequency for which the magnitude of the power gain ($U$, Mason's invariant) of the transistor is reduced to unity [67].

$$U(\omega) = \frac{\left|y_{12} - y_{21}\right|^2}{4\left(\mathrm{Re}[y_{11}]\mathrm{Re}[y_{22}] - \mathrm{Re}[y_{12}]\mathrm{Re}[y_{21}]\right)} \rightarrow U(2\pi f_{\max}) = 1 \tag{2.5}$$

Significant discrepancies between the model proposed here and other models regarding the evaluation of the RF FoMs of 2D-FETs have been found. The reasons for that are the following: (i) the reported expressions have been obtained after assuming a small-signal equivalent circuit based on the Meyer-like capacitance approach, similar as the one depicted in Figure 2.2a; and (ii) approximations usually made for conventional technologies might not be appropriate for 2D-FETs. For instance, for conventional FETs working in the saturation region, the drain edge of the device is depleted of mobile charge carriers, so $C_{gd}$ can be neglected respect to $C_{gs}$. So, in order to keep the accuracy in evaluating the FoMs to the highest level, new explicit expressions with no approximations have been obtained to compute the RF FoMs based on the equivalent circuit presented in Figure 2.2b. In doing so, the definitions of both $f_{Tx}$ and $f_{max}$ given by (2.3) and (2.5) have been applied to obtain (2.7) and (2.9), respectively. Explicit expressions for the intrinsic RF FoMs have also been provided in (2.6) and (2.8), respectively, considering zero contact and access resistances.

### 2.2.1 Assessment of the RF performance of a GFET

In order to assess the new expressions (2.7) and (2.9) to estimate the RF FoMs, the small-signal parameters of a prototype GFET described in Table 2.1 have been obtained by numerical calculations based on the large-signal model presented in section 3.4. The gate bias dependence of the transconductance and output conductance is depicted in Figure 2.4a-b and Figure 2.5a-b, for a drain bias $V_{DS}$ = 0.5 V and $V_{DS}$ = 3 V, respectively, the latter representative of the GFET biased in the negative differential resistance (NDR) region ($g_{ds}$ < 0). The gate bias dependence is expressed as the quantity $V_{GS} - V_{Dirac}$, where $V_{Dirac}$ is related to the gate bias which the graphene channel presents the minimum conductivity ($V_{GS} = V_{Dirac}$). The electronic properties of graphene and their impact into the static and dynamic response of GFETs are presented in Chapter 3. The intrinsic capacitances for $V_{DS}$ = 0.5 V are shown in Figure 2.4c. Predictions of the $f_{Tx}$ and $f_{max}$ have been got using different expressions found in the literature, specifically the ones provided in [17], [57], [62], [67], [68]. Results are presented in Figure 2.4d-e and Figure 2.5c-d. Notice that the RF FoMs are quite sensitive to $V_{GS}$ close to $V_{Dirac}$ bias.

**Table 2.1** Input parameters describing a prototype GFET. The numerical large-signal model used as well as the physical meaning of the parameters is explained in section 3.4

| Input parameter | Value | Input parameter | Value |
|---|---|---|---|
| $T$ | 300 K | $L$ | 1 µm |
| $\mu$ | 2000 cm²/Vs | $W$ | 10 µm |
| $V_{g0}$ | 0 V | $L_t$ | 12 nm |
| $\Delta$ | 0.08 eV | $\varepsilon_t$ | 9 |
| $R_s$, $R_d$ | 20 Ω | $R_g$ | 5 Ω |

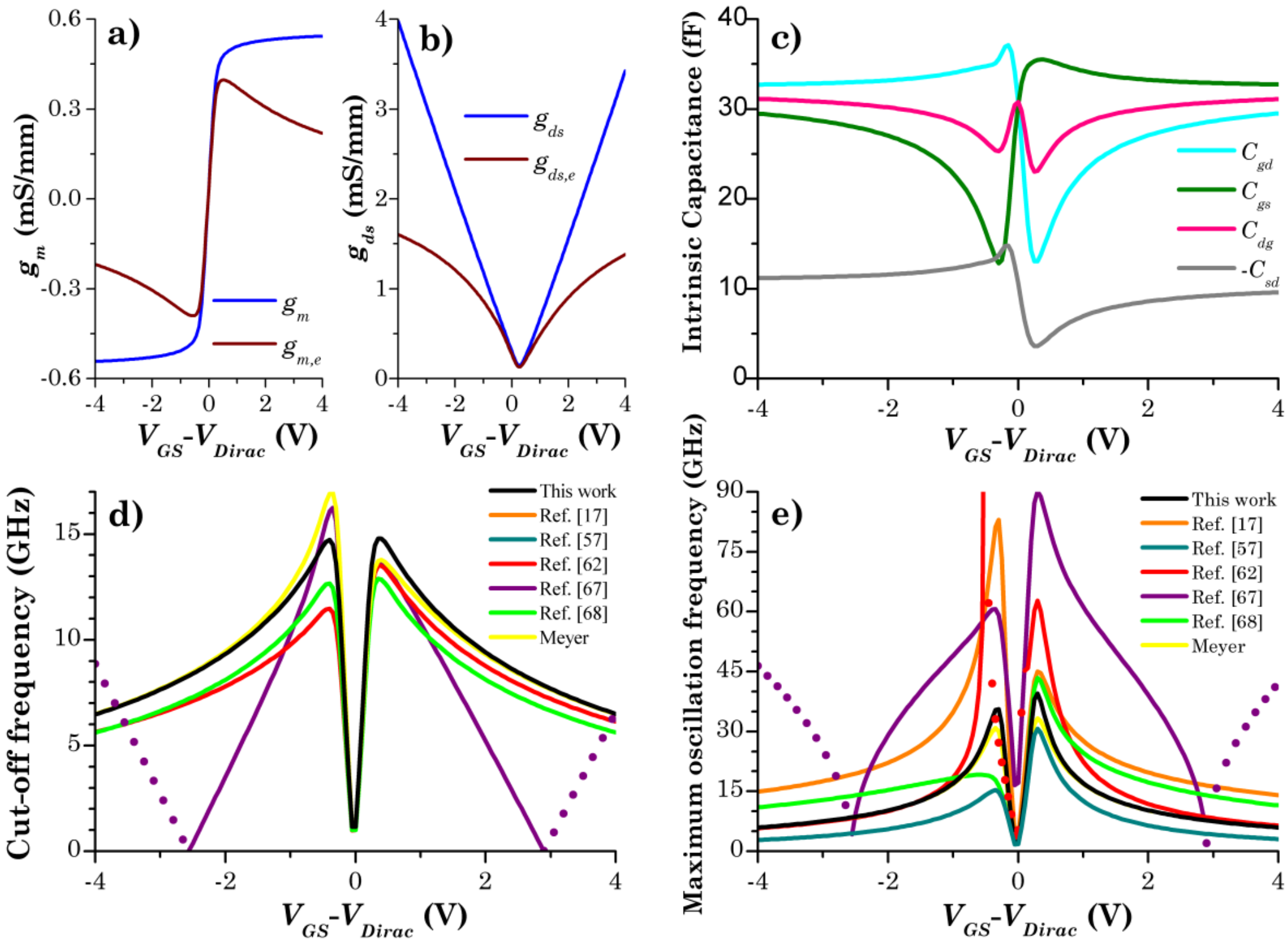

**Figure 2.4** Gate bias dependence of the small-signal parameters and RF FoMs of the GFET described in Table 2.1 for a drain bias $V_{DS}$ = 0.5 V. The closed circles represent the absolute value of the frequency, where the calculated values are real negative or imaginary. a) Intrinsic ($g_m$) and extrinsic ($g_{m,e}$) transconductance; b) intrinsic ($g_{ds}$) and extrinsic ($g_{ds,e}$) output conductance; c) intrinsic capacitances ($C_{gd}$, $C_{gs}$, $C_{dg}$, $C_{sd}$); d) cut-off frequency ($f_{Tx}$); and e) maximum oscillation frequency ($f_{max}$). The $f_{Tx}$ calculation of [17], [57], [68] in d) is the same.

$$f_{T,i} = \frac{|g_m|}{2\pi\sqrt{\left(C_{gs}+C_{gd}\right)^2 - C_{dg}^2}} \tag{2.6}$$

---

$$f_{Tx} = \frac{\sqrt{-\left(c_1+c_2\right)}}{2\pi\sqrt{c_3}}$$

$$C_{gg} = C_{gd} + C_{gs}$$

$$c_1 = -C_{dg}^2 + \left(C_{gg} + \left(C_{gg}g_{ds} + C_{gd}g_m\right)R_d\right)^2 + 2\left(C_{gg}\left(\left(-C_{dg}+C_{gg}\right)g_{ds} - C_{sd}g_m\right) + \left(C_{gg}g_{ds}+C_{gd}g_m\right)^2 R_d\right)R_s \tag{2.7}$$

$$c_2 = -\sqrt{2c_3g_m^2 + \left(-C_{dg}^2 + \left(C_{gg} + \left(C_{gg}g_{ds} + C_{gd}g_m\right)R_d\right)^2 + 2\left(C_{gg}\left(\left(-C_{dg}+C_{gg}\right)g_{ds} - C_{sd}g_m\right) + \left(C_{gg}g_{ds}+C_{gd}g_m\right)^2 R_d\right)R_s\right)^2}$$

$$c_3 = 2\left(C_{dg}C_{gd} - C_{gg}\left(C_{gd}+C_{sd}\right)\right)^2 R_d\left(R_d + 2R_s\right)$$

---

$$f_{\max,i} = \frac{|g_m|}{2\pi\sqrt{4R_g\left(C_{gd}+C_{gs}\right)\left(C_{gs}g_{ds} + C_{gd}\left(g_{ds}+g_m\right)\right) - \left(C_{dg}-C_{gd}\right)^2}} \tag{2.8}$$

---

$$f_{max} = \frac{\sqrt{-\left(a_1+a_2\right)}}{4\pi\sqrt{a_3}}$$

$$C_{gg} = C_{gd} + C_{gs}$$

$$R_{gds} = R_gR_d + R_dR_s + R_gR_s$$

$$a_1 = -C_{dg}^2 + 2C_{dg}\left(C_{gd} + 2g_{ds}\left(C_{gd}R_d - C_{gs}R_s\right)\right) + b_1 + b_2 + b_3;$$

$$a_2 = -\sqrt{8a_3{g_m}^2 + a_1^2} \tag{2.9}$$

$$a_3 = 2\left(C_{dg}C_{gd} - C_{gg}\left(C_{gd}+C_{sd}\right)\right)^2 R_{gds}$$

$$b_1 = 4C_{gd}\left(C_{sd}g_mR_d + C_{gs}\left(g_mR_g + 2{g_{ds}}^2R_{gds} + g_{ds}\left(2R_g + 2g_mR_{gds} + R_s\right)\right)\right)$$

$$b_2 = 4C_{gs}\left(-C_{sd}g_mR_s + C_{gs}g_{ds}\left(R_g + R_s + g_{ds}R_{gds}\right)\right)$$

$$b_3 = {C_{gd}}^2\left(-1 + 4g_{ds}R_g + 4\left(g_m\left(R_d+R_g\right) + \left(g_{ds}+g_m\right)^2 R_{gds}\right)\right)$$

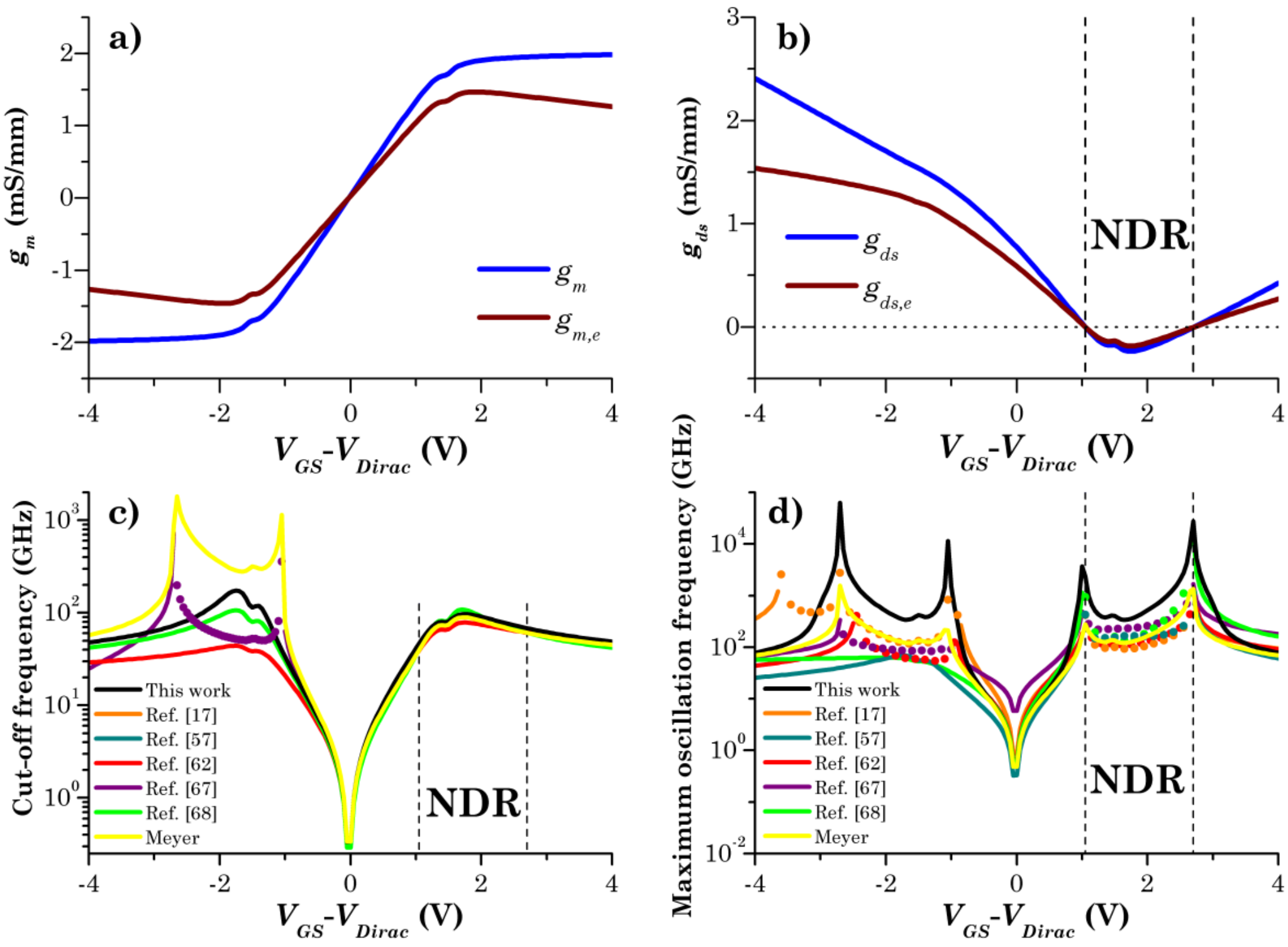

**Figure 2.5** Gate bias dependence of the small-signal parameters and RF FoMs of the GFET described in Table 2.1 for a drain bias $V_{DS}$ = 3 V. The closed circles represent the absolute value of the frequency, where the calculated values are real negative or imaginary. a) Intrinsic ($g_m$) and extrinsic ($g_{m,e}$) transconductance; b) intrinsic ($g_{ds}$) and extrinsic ($g_{ds,e}$) output conductance. Notice that there is a region of negative differential resistance in the range of $V_{GS}$ = [1.05 – 2.7] V; c) cut-off frequency ($f_{Tx}$); and d) maximum oscillation frequency ($f_{max}$). The $f_{Tx}$ calculation of [17], [57], [68] in c) is the same.

Both $f_{Tx}$ and $f_{max}$ expressions from [17], [57], [68] can largely underestimate or overestimate the values depending on the gate voltage overdrive. However, results from [67] are far and give gate bias regions where the $f_{Tx}$ and $f_{max}$ expression results in imaginary or real negative values. Regarding $f_{max}$ evaluation the case where a GFET is operated in its NDR region has been assessed, which is a feature of interest in many applications [68]–[73]. As suggested in Figure 2.5d, there is no expression found in the literature which gives a positive real estimation within this gate bias range. The expressions proposed here are exceptions, delivering results that are physically correct. Moreover, the RF FoMs assuming a Meyer-like model as the one depicted in Figure 2.2a have been calculated, by enforcing $C_{dg}$ = $C_{gd}$ and $C_{sd}$ = 0 in equations (2.7) and (2.9). This has been done for the sake of highlighting the differences with the charge-conserving model. Results have been plotted in Figure 2.4d-e and Figure 2.5c-d (yellow lines). Especially in Figure 2.5c one can realize on the importance of assuming a charge-conserving model and consistently estimating the RF FoMs in accordance to it.

## 2.3 Parameter extraction methodology

To release a successful RF circuit based on 2D-FETs, the device should be fabricated, characterized, and modelled before moving to the design, realization and characterization of the circuit. Hence, 2D-FET small-signal modelling plays a fundamental role because of its utility in enabling a quick and reliable optimization of RF circuit design. It allows for minimizing expensive and time-consuming cycles of design and realization of the RF circuit that hopefully should be characterized only once at the end, to verify its real performance with respect to the predicted behaviour.

In doing the above-mentioned, the small-signal elements which constitute the equivalent circuit that models a transistor, must be extracted. Commonly they are obtained from $S$-parameters measurements, which can be straightforwardly measured with a vector network analyser (VNA). However, the extraction of such small-signal elements from $S$-parameter is an ill-conditioned problem [74] because there are usually more small-signal elements than equations. Figure 2.6 shows a typical topology of the complete small-signal equivalent circuit for a microwave transistor. To solve the ill-conditioned problem the equivalent circuit elements are usually divided into two main groups: the intrinsic elements (i.e., corresponding to the ones depicted in Figure 2.2), which are bias dependent, and the extrinsic or parasitic elements, which are assumed to be bias independent. The latter elements, typically 8 elements according to Figure 2.6, represent the contributions arising from the interconnections between the real device and the outside world, namely, $R_{g,ext}$, $R_{d,ext}$, $R_{s,ext}$, $L_{g,ext}$, $L_{d,ext}$, $L_{s,ext}$, $C_{pd}$ and $C_{pg}$. The intrinsic part can be modelled by any of the small-signal equivalent circuits shown in Figure 2.2. In the case of Figure 2.2b, the complete small-signal equivalent circuit will be composed of 17 elements. The solution of this problem is based on decomposing it into two subproblems and then solving them subsequently. In doing so, first, the extrinsic elements should be extracted so the intrinsic elements can be gotten. Thus, to get the extrinsic elements and consequently subtract their unwanted contribution a de-embedding procedure must be carried out.

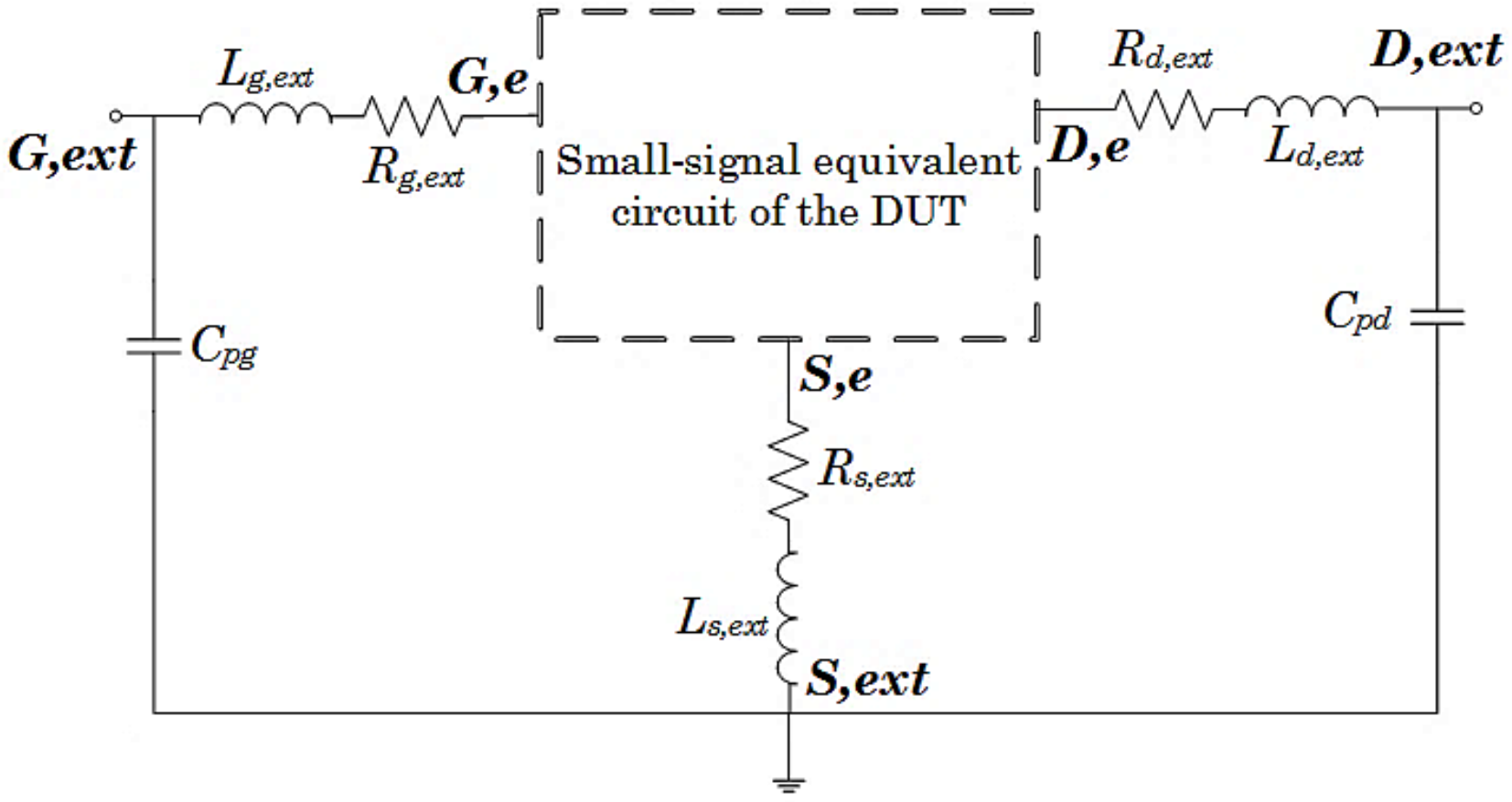

**Figure 2.6** Typical topology of the complete small-signal equivalent circuit for a microwave FET. It is composed of an intrinsic and extrinsic part. The intrinsic part could be either of the networks depicted in Figure 2.2 depending on the capacitance model assumed.

The most common procedure applied to 2D-FETs so far [57], [75]–[79] consists of applying "open" and "short" structures to identical layouts, (see Figure 2.7), one excluding the 2D channel, so to remove the effect of the probing pads, metal interconnections, including the parasitic capacitances and inductances. Since during the de-embedding process the effect of the 2D channel cannot be removed, the parasitic resistance extracted by this method ($R_{d,ext}$ and $R_{s,ext}$) do not include neither the metal contact resistance nor the access resistance [61]. The contact resistance with a 2D material is currently an important bottleneck, together with the lack of perfect current saturation, hampering the realization of power gain at terahertz frequencies [80]–[82]. On the other hand, in many embodiments of the 2D material based transistor an ungated area exists between the drain/source metal and the channel under the gate resulting in additional access resistance, which should be considered.

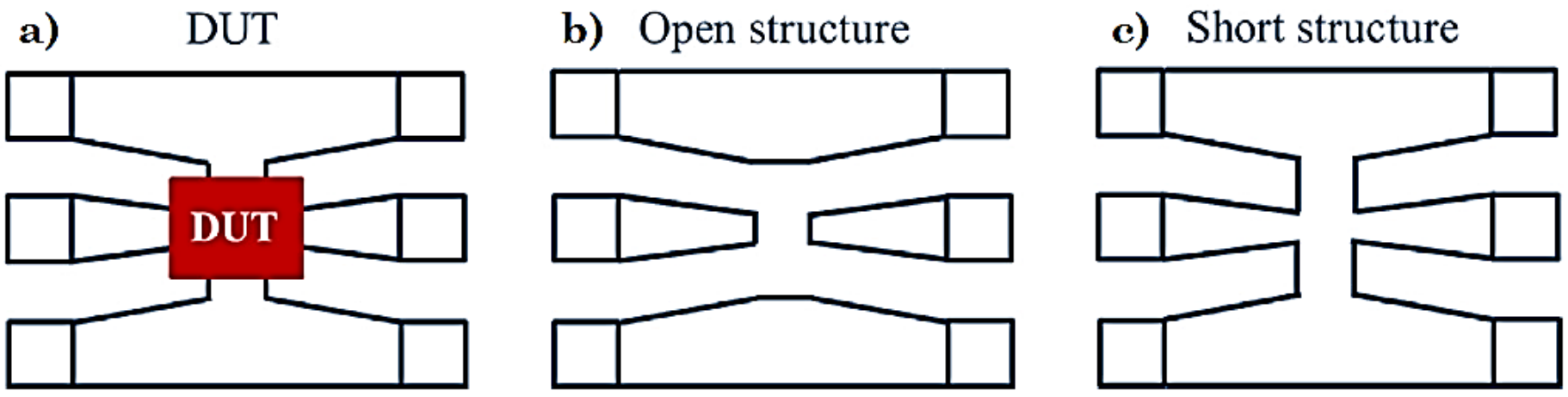

**Figure 2.7** Illustration of two dummy test structures for a) an on-wafer DUT: b) open structure, and c) short structure. (Image taken from [74]).

With the aim of solving this issue affecting 2D-FETs, the effect of the series combination of the drain/source contact and access resistances has been included in the equivalent circuit network, so they can be extracted together with the rest of intrinsic parameters from microwave characterization, i.e. from *S*-parameter measurements. Considering that the equivalent circuit after the de-embedding procedure is the one depicted in Figure 2.2b, the following parameter extraction methodology, suitable to 2D-FETs, is appropriate:

a) Apply “open” and “short” structures to identical DUT’s layouts, but excluding the 2D channel, in order to remove the effect of extrinsic elements [57], [75]–[79].
b) Extract the series combination of the metal contact and access resistances using equation (2.10), where both drain and source resistances have been assumed to be the same, namely: $R_s = R_d = R_c$. Other possibility to estimate these extrinsic resistances is relying on the transfer length method (TLM), which would imply the fabrication and characterization of devices with different channel lengths [83].
c) Direct application of the equations (2.11) - (2.17) to obtain the transconductance ($g_m$), output conductance ($g_{ds}$), gate resistance ($R_g$) and the intrinsic capacitances ($C_{gs}$, $C_{gd}$, $C_{dg}$, $C_{sd}$). These expressions have been derived with no approximations.

As a matter of convenience, equations (2.10) - (2.17) have been expressed in terms of the *Z*-parameters instead of *S*-parameters that had been announced. The equivalence between both kind of parameters is well known and can be found in [84]. It is important to highlight that the above-mentioned extraction approach allows getting the small-signal parameters at any arbitrary bias. This is in contrast to the extraction method reported in [61] that requires biasing a GFET at the minimum conductivity to extract the intrinsic capacitances. So, this procedure is fine when the model is operated close to the Dirac voltage, but discrepancies could arise far from this bias point according to the bias dependence of such intrinsic capacitances observed in Figure 2.4c.

### 2.3.1 Extracting the small-signal equivalent circuit of a GFET

To assess the proposed parameter extraction method, it has been applied to a state-of-the-art GFET, which has been characterized in both DC and RF. The GFET (width $W$ = 12 μm, length $L$ = 100 nm) fabrication process has been described in [85]. Following the extraction method described in section 2.3, the small-signal parameters have been obtained and summarized in Table 2.2. Notice that, due to the non-reciprocity, $C_{dg}$ and $C_{gd}$ are different. Besides, measured and modelled $S$-parameters at $V_{GS,e}$ = 0.2 V and $V_{DS,e}$ = 1 V plotted together in Figure 2.8 are in good agreement. The high-frequency performance of the GFET was characterized using a VNA (Agilent, E8361A) under ambient conditions in the frequency range of 0.25 – 45 GHz. A common calibration procedure of line-reflect-reflect-match was performed before measurements. The de-embedding procedure was implemented to subtract the unwanted contribution of extrinsic elements, as described in [78], [79]. However, the effect of the series combination of the drain/source contact and access resistances could not be de-embedded. The extracted value of these series resistance $R_c = R_s = R_d$ = 215 Ω is in good agreement with the average contact resistance reported by using the TLM technique (around 2200 Ω·μm) for the devices fabricated in [85]. Notice the importance of considering the extraction of these non-negligible resistances after the de-embedding procedure when modelling 2D-FETs.

$$R_c = \frac{\mathrm{Re}[z_{22}]\mathrm{Im}[z_{12}] - \mathrm{Re}[z_{12}]\mathrm{Im}[z_{22}]}{2\,\mathrm{Im}[z_{12}] - \mathrm{Im}[z_{22}]} \tag{2.10}$$

$$g_m = \frac{(2\,\mathrm{Im}[z_{12}] - \mathrm{Im}[z_{22}])(2\,\mathrm{Im}[z_{21}]\mathrm{Re}[z_{12}] - \mathrm{Im}[z_{22}]\mathrm{Re}[z_{12}] - 2\,\mathrm{Im}[z_{12}]\mathrm{Re}[z_{21}] + \mathrm{Im}[z_{22}]\mathrm{Re}[z_{21}] + \mathrm{Im}[z_{12}]\mathrm{Re}[z_{22}] - \mathrm{Im}[z_{21}]\mathrm{Re}[z_{22}])}{(\mathrm{Im}[z_{12}]\mathrm{Im}[z_{21}] - \mathrm{Im}[z_{11}]\mathrm{Im}[z_{22}])\left((-2\,\mathrm{Im}[z_{12}] + \mathrm{Im}[z_{22}])^2 + (-2\,\mathrm{Re}[z_{12}] + \mathrm{Re}[z_{22}])^2\right)} \tag{2.11}$$

$$g_{ds} = \frac{(2\,\mathrm{Im}[z_{12}] - \mathrm{Im}[z_{22}])\left(\mathrm{Im}[z_{11}]\mathrm{Im}[z_{22}](-2\,\mathrm{Re}[z_{12}] + \mathrm{Re}[z_{22}]) + \mathrm{Im}[z_{12}]\left(\mathrm{Im}[z_{22}]\mathrm{Re}[z_{12}] + \mathrm{Re}[z_{21}](2\,\mathrm{Im}[z_{12}] - \mathrm{Im}[z_{22}]) - \mathrm{Im}[z_{12}]\mathrm{Re}[z_{22}]\right)\right)}{\mathrm{Im}[z_{22}](\mathrm{Im}[z_{12}]\mathrm{Im}[z_{21}] - \mathrm{Im}[z_{11}]\mathrm{Im}[z_{22}])\left((-2\,\mathrm{Im}[z_{12}] + \mathrm{Im}[z_{22}])^2 + (-2\,\mathrm{Re}[z_{12}] + \mathrm{Re}[z_{22}])^2\right)} \tag{2.12}$$

$$R_g = \frac{\mathrm{Im}[z_{22}]^2(-\mathrm{Re}[z_{11}] + \mathrm{Re}[z_{12}]) + \mathrm{Im}[z_{12}]\mathrm{Im}[z_{22}](2\,\mathrm{Re}[z_{11}] - \mathrm{Re}[z_{12}] + \mathrm{Re}[z_{21}] - \mathrm{Re}[z_{22}]) + \mathrm{Im}[z_{12}]^2(-2\,\mathrm{Re}[z_{21}] + \mathrm{Re}[z_{22}])}{(2\,\mathrm{Im}[z_{12}] - \mathrm{Im}[z_{22}])\mathrm{Im}[z_{22}]} \tag{2.13}$$

$$C_{gs} = -\frac{\mathrm{Im}[z_{12}] - \mathrm{Im}[z_{22}]}{\omega(\mathrm{Im}[z_{12}]\mathrm{Im}[z_{21}] - \mathrm{Im}[z_{11}]\mathrm{Im}[z_{22}])} \tag{2.14}$$

$$C_{gd} = C_{gs}\frac{\mathrm{Im}(z_{12})}{\mathrm{Im}(z_{22}) - \mathrm{Im}(z_{12})} = \frac{\mathrm{Im}[z_{12}]}{\omega(\mathrm{Im}[z_{12}]\mathrm{Im}[z_{21}] - \mathrm{Im}[z_{11}]\mathrm{Im}[z_{22}])} \tag{2.15}$$

$$C_{dg} = \frac{4\,\mathrm{Im}[z_{12}]^2\,\mathrm{Im}[z_{21}] + \mathrm{Im}[z_{22}]\left(\mathrm{Im}[z_{21}]\mathrm{Im}[z_{22}] + (\mathrm{Re}[z_{12}] - \mathrm{Re}[z_{21}])(2\,\mathrm{Re}[z_{12}] - \mathrm{Re}[z_{22}])\right) + \mathrm{Im}[z_{12}]\left(4(\mathrm{Re}[z_{12}]\mathrm{Re}[z_{21}] - \mathrm{Im}[z_{21}]\mathrm{Im}[z_{22}]) - 2(\mathrm{Re}[z_{12}] + \mathrm{Re}[z_{21}])\mathrm{Re}[z_{22}] + \mathrm{Re}[z_{22}]^2\right)}{(\mathrm{Im}[z_{12}]\mathrm{Im}[z_{21}] - \mathrm{Im}[z_{11}]\mathrm{Im}[z_{22}])\left((-2\,\mathrm{Im}[z_{12}] + \mathrm{Im}[z_{22}])^2 + (-2\,\mathrm{Re}[z_{12}] + \mathrm{Re}[z_{22}])^2\right)\omega} \tag{2.16}$$

$$C_{sd} = \frac{-\mathrm{Im}[z_{12}]\left(\mathrm{Im}[z_{22}]\left(\mathrm{Im}[z_{22}]^2 + (2\,\mathrm{Re}[z_{12}] - \mathrm{Re}[z_{22}])(\mathrm{Re}[z_{12}] + \mathrm{Re}[z_{21}] - \mathrm{Re}[z_{22}])\right) + \mathrm{Im}[z_{12}]\left(-4\,\mathrm{Im}[z_{22}]^2 + (2\,\mathrm{Re}[z_{12}] - \mathrm{Re}[z_{22}])(-2\,\mathrm{Re}[z_{21}] + \mathrm{Re}[z_{22}])\right)\right)}{\mathrm{Im}[z_{22}](\mathrm{Im}[z_{12}]\mathrm{Im}[z_{21}] - \mathrm{Im}[z_{11}]\mathrm{Im}[z_{22}])\left((-2\,\mathrm{Im}[z_{12}] + \mathrm{Im}[z_{22}])^2 + (-2\,\mathrm{Re}[z_{12}] + \mathrm{Re}[z_{22}])^2\right)\omega} + \frac{\mathrm{Im}[z_{11}]\,\mathrm{Im}[z_{22}](-2\,\mathrm{Im}[z_{12}] + \mathrm{Im}[z_{22}])^2 - 4\,\mathrm{Im}[z_{12}]^3\,\mathrm{Im}[z_{22}]}{\mathrm{Im}[z_{22}](\mathrm{Im}[z_{12}]\mathrm{Im}[z_{21}] - \mathrm{Im}[z_{11}]\mathrm{Im}[z_{22}])\left((-2\,\mathrm{Im}[z_{12}] + \mathrm{Im}[z_{22}])^2 + (-2\,\mathrm{Re}[z_{12}] + \mathrm{Re}[z_{22}])^2\right)\omega} \tag{2.17}$$

**Table 2.2** Extracted small-signal parameters of the charge-conserving model for the GFET under test at $V_{GS,e}$ = 0.2 V and $V_{DS,e}$ = 1 V

| Element | Value | Element | Value |
|---|---|---|---|
| $C_{gs}$ | 6.5 fF | $g_m$ | 1.55 mS |
| $C_{gd}$ | 9.5 fF | $g_{ds}$ | -6.5 mS |
| $C_{dg}$ | 10.5 fF | $R_g$ | 0.5 Ω |
| $C_{sd}$ | -3.5 fF | $R_d = R_s$ | 215 Ω |

On the other hand, the extrinsic transconductance ($g_{m,e}$) and the extrinsic output conductance ($g_{ds,e}$) can be calculated as following [86]:

$$\begin{aligned} g_{m,e} &= \frac{\partial I_{DS}}{\partial V_{GS,e}} = \frac{g_m}{1 + g_m R_s + g_{ds}\left(R_s + R_d\right)} \\ g_{ds,e} &= \frac{\partial I_{DS}}{\partial V_{DS,e}} = \frac{g_{ds}}{1 + g_m R_s + g_{ds}\left(R_s + R_d\right)} \end{aligned} \tag{2.18}$$

In [85], a $g_{m,e}$ of ~ -100 μS/μm and a $g_{ds,e}$ of ~ 370 μS/μm were reported at $V_{GS,e}$ = 0.2 V and $V_{DS,e}$ = 1 V. They were extracted from the DC transfer characteristics (TCs, $I_{DS}$ vs. $V_{GS,e}$ curve) and from the output characteristics (OCs, $I_{DS}$ vs. $V_{DS,e}$ curve), respectively. These values are in good agreement with the ones calculated by equation (2.18), using the parameters in Table 2.2, which have been obtained following the parameter extraction methodology explained in the former section 2.3.

Finally, Figure 2.9 shows the experimental current gain ($|h_{21}|$) and Mason's invariant ($U$), both obtained from the $S$-parameter measurements depicted in Figure 2.8, compared to the simulated ones obtained from the small-signal model. Both $f_{Tx}$ and $f_{max}$ coming from different approaches have been calculated using the extracted parameters listed in Table 2.2. They have been summarized in Table 2.3, showing a large dispersion of values, being the values from (2.7) and (2.9) the more accurate prediction. Notice that because the intrinsic output conductance is negative many reported formulas give non-physical real negative or imaginary values.

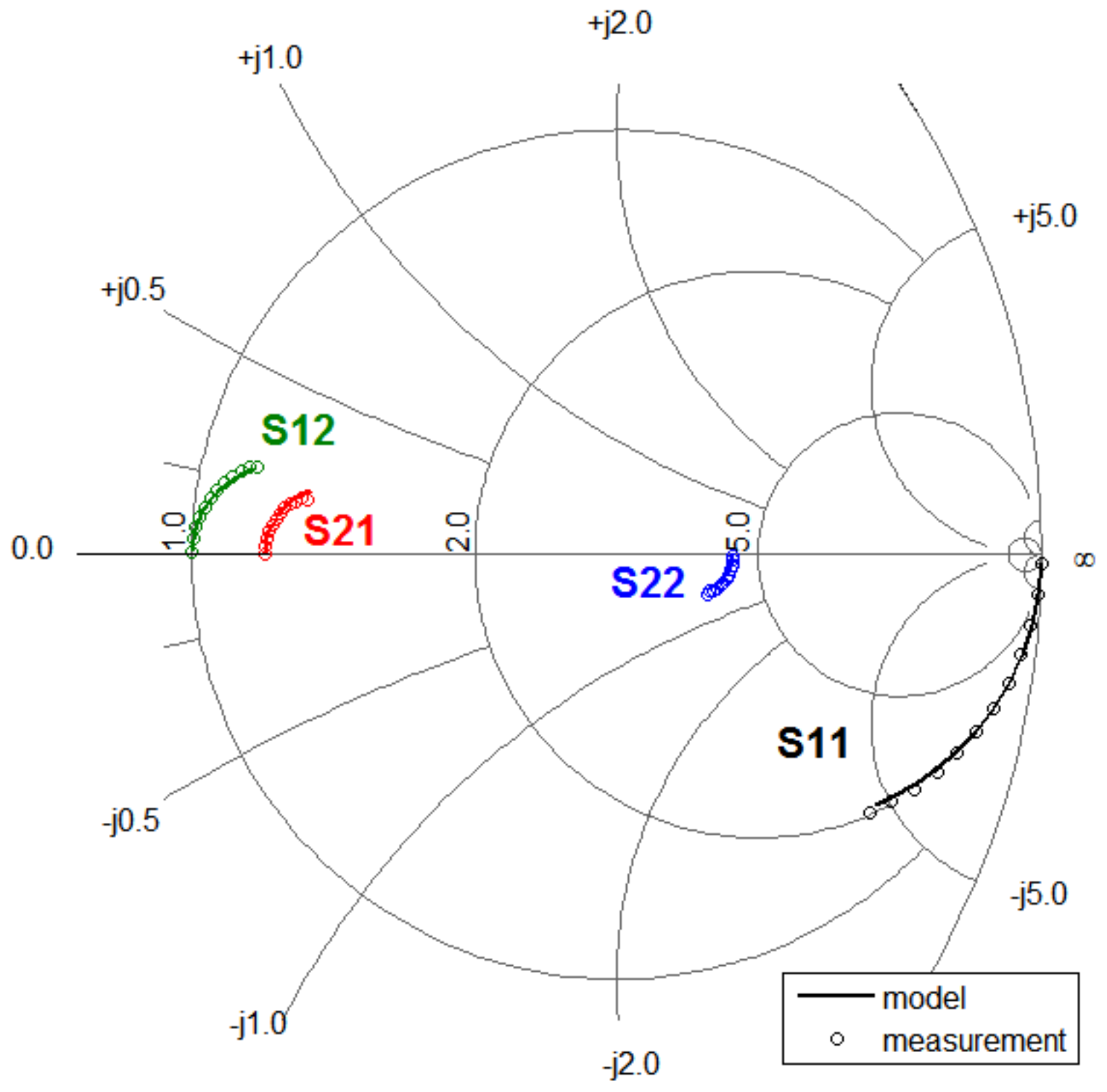

**Figure 2.8** *S*-parameter measurements (circles) and simulations (lines) of the GFET under test assuming a bias $V_{GS,e} = 0.2$ V and $V_{DS,e} = 1$ V.

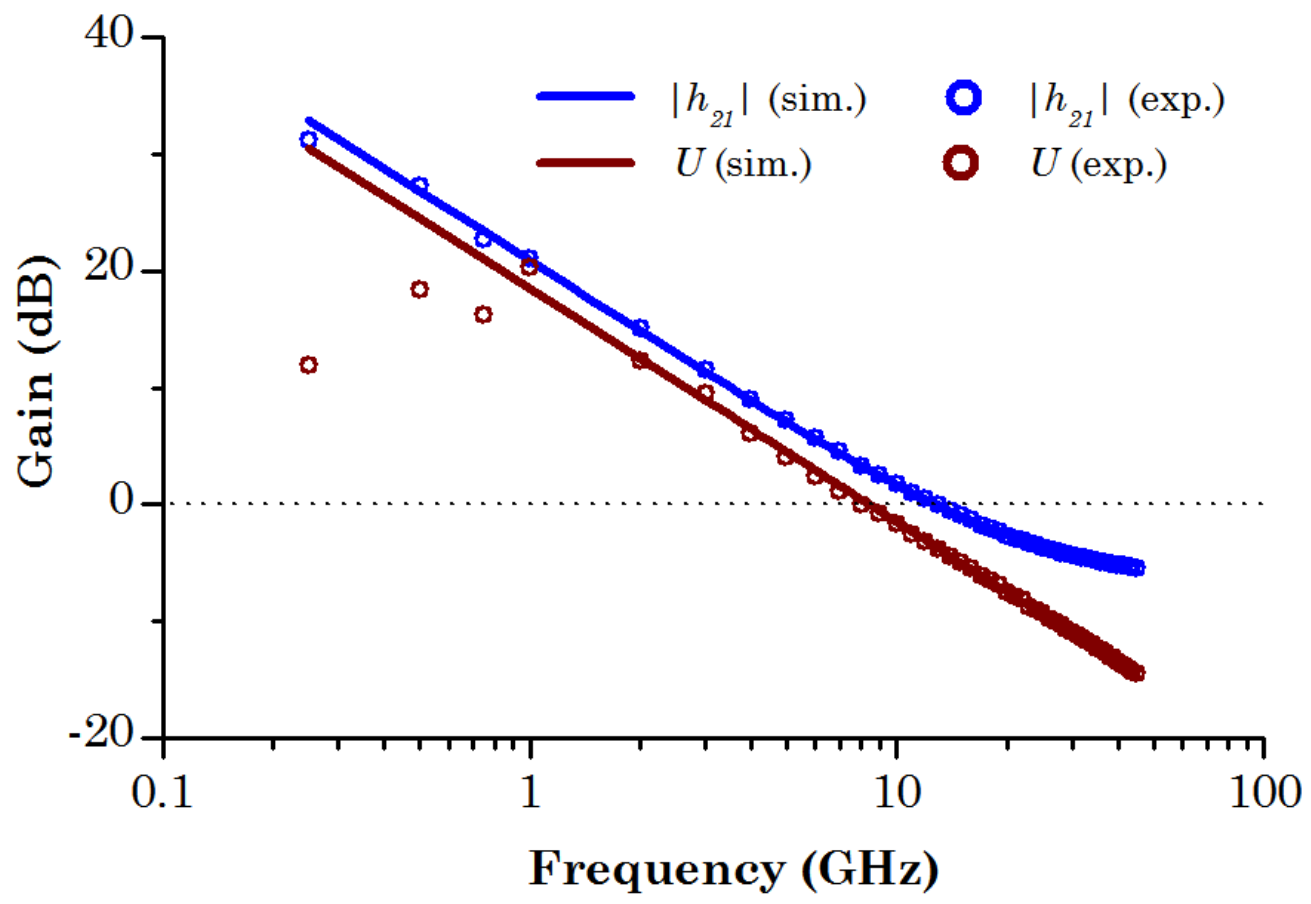

**Figure 2.9** Radio-frequency performance of the GFET under test characterized in Figure 2.8 ($V_{GS,e}$ = 0.2 V and $V_{DS,e}$ = 1 V) with parameters listed in Table 2.2. Measured (symbols) and simulated (solid line) small-signal current gain ($|h_{21}|$) and Mason's invariant ($U$) plotted versus frequency.

**Table 2.3** Estimation of the RF FoMs of the GFET under test (imaginary values are written in italic style)

| | $f_{Tx}$ [GHz] | $f_{max}$ [GHz] |
|---|---|---|
| This work | 11.92 | 8.59 |
| Ref. [17] | -11.02 | *-16.04* |
| Ref. [57] | -11.02 | *4.65* |
| Ref. [62] | 13.69 | 6.75 |
| Ref. [67] | -11.89 | 315.65 |
| Ref. [68] | -11.02 | -25.45 |

## 2.4 Stability of a power amplifier

Signal amplification is one of the most basic and prevalent functions in modern RF and microwave systems. Starting some years ago, there has been a great deal of interest in 2DM-based transistors because of their potential to exhibit power gain in the THz range. Interestingly, 2DMs could also offer mechanical flexibility, so integration on flexible substrates is expected in combination with good RF performances [15]. In the following, a basic power amplifier configuration based on the small-signal model presented in section 2.1 is analysed. A study of the scalability of RF performance of GFETs and a thorough discussion about the device stability will be given.

A general two-port amplifier circuit in terms of the admittance parameters is shown in Figure 2.3. The 2D-FET, represented as a two-port network, is assumed to be connected to the source and load admittances $Y_S$ and $Y_L$, respectively. At the input port, a small-signal AC voltage source $v_s$, of associated admittance $Y_S$, transfers power to the network. A load admittance $Y_L$ is connected at the output to get the transferred power. A small-signal model in form of an admittance matrix $Y$ describes the behaviour of the two-port network.

Taking advantage of the charge-conserving small-signal model presented in section 2.1, microwave techniques can be applied to any design based on a 2D-FET using the *Y*-parameter matrix described in equation (2.4) or any other kind of parameter matrix by transforming it into, i.e., *Z-, S-, h-* or *ABCD*-parameters [84]. In doing so, common amplifier design targets such as getting the maximum gain, or a specified gain combined with low noise figure, could be directly applied. Such designs must be carried out together with a study of the stability issue, which deals with the necessary conditions for a transistor to be stable when acting as a power amplifier.

First of all, the concept of stability of a general two-port amplifier circuit in terms of the *Y*-parameters is recalled. The stability guarantees that no adventitious oscillations can appear at the network for any passive source and load admittances connected to the input and output ports, respectively, by requiring that the reflection coefficient seen looking into the DUT's input and output ports be smaller than one, so to avoid that the injected signal be back reflected towards the source and load with gain greater than one. The unconditional stability of the network can be assessed by means of the *K*-*Δ* test, which is based on the evaluation of the two following factors [84], [87], [88]:

$$K(\omega)=\frac{2\,\mathrm{Re}[y_{11}]\,\mathrm{Re}[y_{22}]-\mathrm{Re}[y_{12}y_{21}]}{|y_{12}y_{21}|}$$
$$\Delta(\omega)=\frac{(Y_0-y_{11})(Y_0-y_{22})-y_{12}y_{21}}{(Y_0+y_{11})(Y_0+y_{22})-y_{12}y_{21}} \qquad (2.19)$$

where $Y_0 = 1 / Z_0$ is the characteristic admittance and $Z_0$ is the characteristic impedance (usually taken as 50 Ω). Note that the stability condition of an amplifier circuit is usually frequency dependent since the input and output impedances as well as the *Y*-parameters describing the device generally depend on frequency. Both conditions $K > 1$ and $|\Delta| < 1$ are necessary and sufficient to ensure device stability. In this context, any passive load and input admittance provide a stable behaviour of the network. Selecting an optimum set of $Y_S$ and $Y_L$, an optimized power gain can be obtained, referred as the maximum available gain (MAG). However, if $-1 < K < 1$, the network is said to be conditionally stable, that is, it becomes stable only for certain combinations of $Y_S$ and $Y_L$. Among those combinations that provide stability, the maximum attainable power gain is known as the maximum stable gain (MSG). Then, the maximum gain can be calculated as following:

$$G_T^{\max}(\omega) = \begin{cases} \text{MAG} = \left|\dfrac{y_{21}}{y_{12}}\right|\left(K - \sqrt{K^2 - 1}\right) & K \geq 1; |\Delta| < 1 \\ \text{MSG} = \left|\dfrac{y_{21}}{y_{12}}\right| & -1 < K < 1 \end{cases} \quad (2.20)$$

where $G_T$ refers to the power gain which represents the ratio of the power delivered to the load to the power available from the source. Again, the $f_{max}$ is defined as highest possible frequency for which the magnitude of the $G_T$ is reduced to unity and its value can be calculated according to (2.9).

### 2.4.1 Scaling of RF GFETs: stability as a limiting factor

The small-signal model presented in section 2.1 is used to investigate the scalability of the RF performance of a GFET via channel length reduction and considering device stability at the same time. Stability is anticipated to play a vital role, especially in short-channel transistors. For such a purpose, the self-consistent model presented in [73] has been used for investigating short-channel transistors. It solves the drift-diffusion (DD) transport equation coupled with the 2D Poisson's equation. Notice that dealing with the 2D electrostatics of a device allows for coping with short-channel effects (SCEs), which significantly reduce the expected $f_{Tx}$ and $f_{max}$ [73]. The main difference between that model and the numerical large-signal model of GFETs presented in section 3.4 is the use of the one-dimensional (1D) Poisson's equation in the latter, which is appropriate only for long-channel transistors not suffering from SCEs.

Then, the prototype GFET described in Table 2.4 has been simulated to obtain the parameters of the small-signal equivalent circuit drawn in Figure 2.2b. The simulated device is a top-gated GFET supported on an hexagonal boron nitride (hBN) substrate, which has been demonstrated to be of upmost technological impact because graphene based devices fabricated on hBN exhibit up to one order-of-magnitude improvement in mobility and carrier inhomogeneities in comparison with conventional oxide dielectrics [89]. State-of-the-art values of the source/drain resistances $R_s \cdot W = R_d \cdot W = 200\ \Omega \cdot \mu m$ have been considered [83]. Regarding the gate resistance, it has been calculated considering a metal gate contacted on both sides of the device [48]. Thus, the gate resistance can be approximated to be inversely proportional to the channel length provided that the channel width is large enough. In this context, a realistic value of $R_g \cdot L = 4.4\ \Omega \cdot \mu m$ could be considered by assuming a prototype 60 nm thick wolfram gate. The increase of $R_g$ with scaling compromises the ultimate $f_{max}$ of GFETs so gate resistance minimization is key in RF applications [90].

The GFET RF performance is, in general, dependent on the bias point [62], [73]. Biases of $V_{DS}$ = 0.6 V and $V_{GS} - V_{Dirac}$ = 2 V have been chosen in order to avoid the region where $f_{max}$ and $f_{Tx}$ are more sensitive to $V_{GS}$ (see Figure 2.4d-e and Figure 2.5c-d).

The impact of the channel length downscaling on the power gain is shown in Figure 2.10. The slope of the MSG is 10 dB/dec and reducing the channel length results in both an increase of the power gain and larger $f_{max}$, i.e., reducing the channel length by a factor of 2 means an increase of the power gain of 6.1 dB, while $f_{max}$ grows from 8.4 to 42.8 GHz.

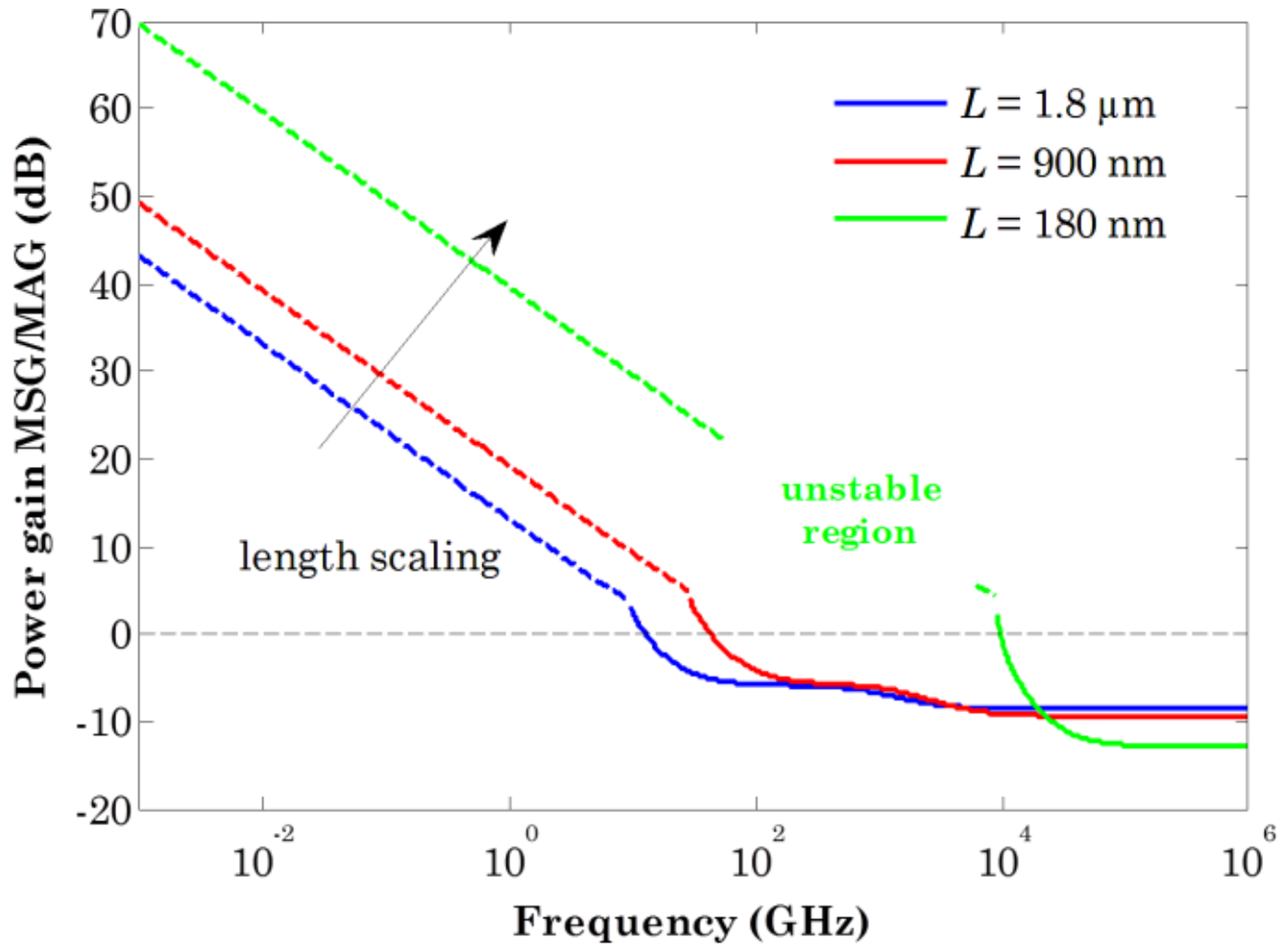

**Figure 2.10** MSG (dashed lines) and MAG (solid lines) of the device described in Table 2.4 for three different channel lengths ($V_{GS} - V_{Dirac}$ = 2 V and $V_{DS}$ = 0.6 V).

**Table 2.4** Input parameters describing a prototype GFET. The physical meaning of the parameters is explained in section 3.4

| Input parameter | Value | Input parameter | Value |
|---|---|---|---|
| $T$ | 300 K | $L$ | 50nm – 18 µm |
| $\mu$ | 7500 cm²/Vs | $W$ | 14 µm |
| $V_{g0}$ | -2.5 V | $L_t$ | 26 nm |
| $\Delta$ | 0.11 eV | $\varepsilon_t$ | 9 |
| $R_s \cdot W$, $R_d \cdot W$ | 200 Ω·µm | $R_g \cdot L$ | 4.4 Ω·µm |

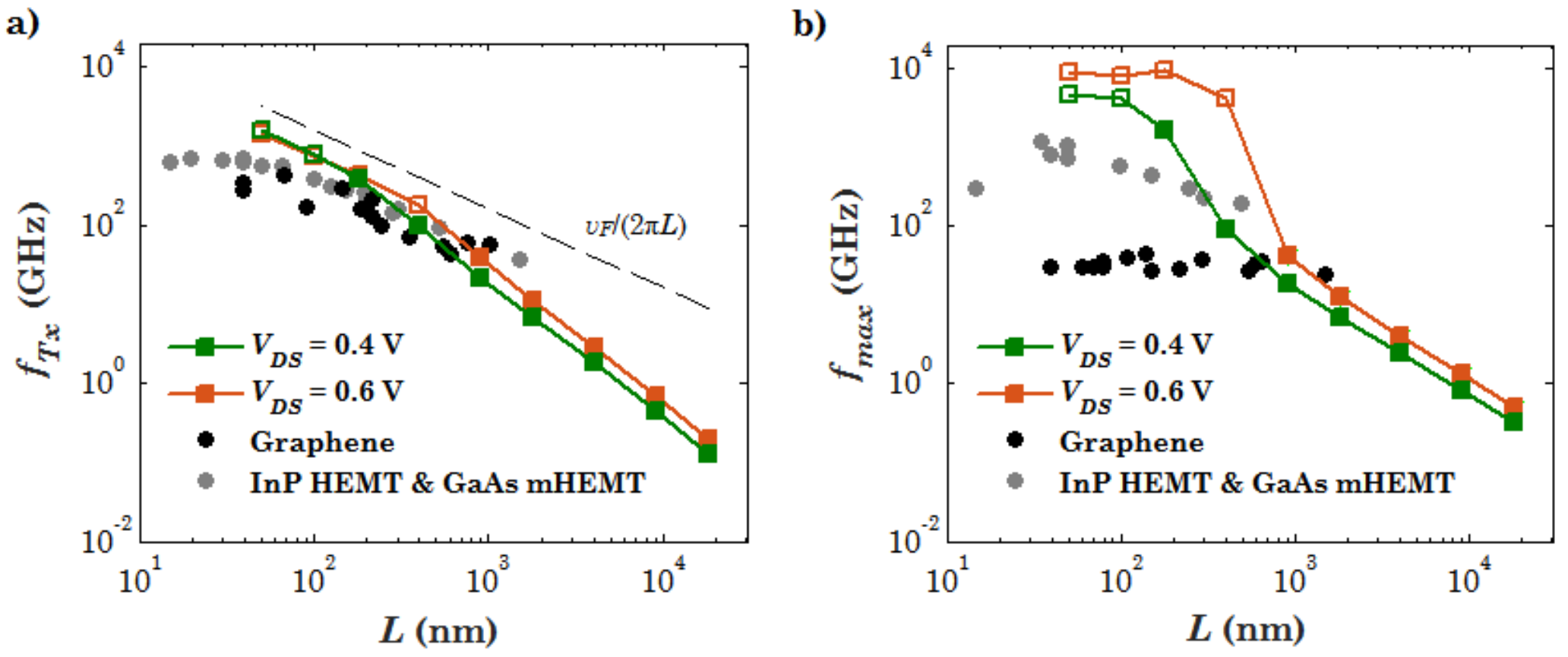

**Figure 2.11** Scaling of a) $f_{Tx}$ and b) $f_{max}$ for two drain voltages (assuming an overdrive gate bias $V_{GS} - V_{Dirac} = 2$ V). Closed and open symbols indicate stability and instability, respectively. The dashed line corresponds to the physical limit of $f_{Tx}$, that is $v_F/(2\pi L)$, where $v_F$ represents the Fermi velocity [91] (the electronic properties of graphene are presented in subsection 3.4.1). This frequency limit comes out from the minimum possible transient time in a graphene channel $L/v_F$. Experimental results from state-of-the-art GFETs on conventional dielectrics (black symbols) and InP / GaAs transistors (red symbols) have also been included for the sake of comparison.

Next, Figure 2.11a-b shows the evolution of $f_{Tx}$ and $f_{max}$ when the channel length is scaled, respectively. Both have been calculated using equations (2.7) and (2.9), respectively. To fully interpret the scaling results of the RF FoMs, Figure 2.12 shows details on the scaling of the small-signal parameters which have been calculated by the simulator presented in [73]. For long-channel lengths ( > 1 µm) $f_{Tx}$ scales as $1/L^2$. This is because the transconductance is proportional to $1/L$ while the intrinsic capacitances, are approximately proportional to $L$. However, for short-channel lengths ( < 1 µm) the scaling law of $f_{Tx}$ approaches $1/L$ because the saturation velocity effect makes the transconductance quite insensitive to $L$. The numbers shown in Figure 2.11a-b are comparable to what has been reported for InP and GaAs high electron mobility transistors (HEMTs), which are the highest reported values for RF transistors [17]. Importantly, it has been found out that the device becomes unstable for short-channel lengths. That issue will be discussed later.

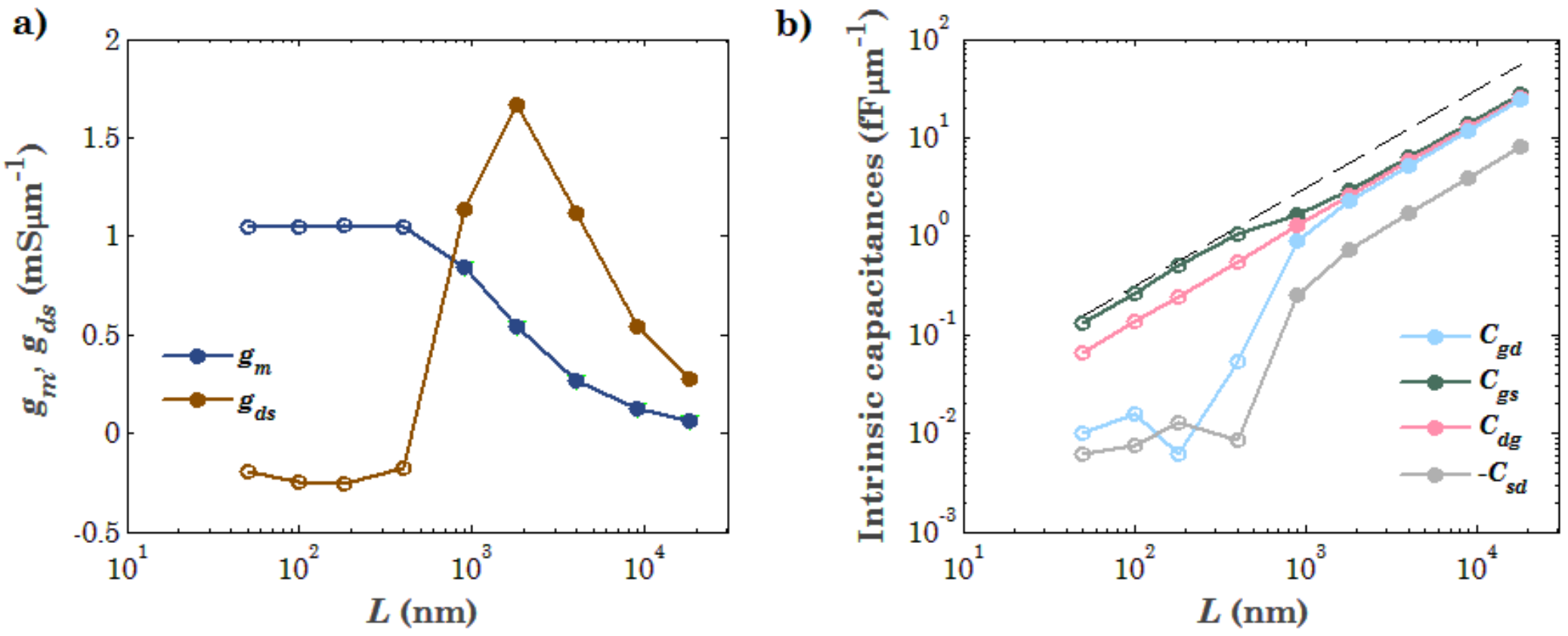

**Figure 2.12** Scaling of a) transconductance and output conductance; and b) intrinsic capacitances per unit width ($V_{GS} - V_{Dirac}$ = 2 V and $V_{DS}$ = 0.6 V). The dashed line in b) represent $C_t \cdot L$ where $C_t = \varepsilon_0\varepsilon_t/L_t$.

Regarding the scalability of $f_{max}$ shown in Figure 2.11b, a different trend from $f_{Tx}$ has been found. Specifically, the scaling law of $f_{max}$ goes as $1/L^n$ with $1 < n < 2$ for long-channel lengths which, in fact, results in a scaling power smaller than $f_{Tx}$ due to the upscaling of $R_g$. However, there is a great increase in $f_{max}$ for short-channel lengths because of current saturation driven by the velocity saturation effect. As a result, the output conductance shown in Figure 2.12a drops and, as a consequence, $f_{max}$ is pushed up. Moreover, the output conductance even reaches the NDR region, which may be the origin of the RF instability [68]. Notice at this point the importance of the charge-conserving model presented in section 2.1 and estimating the RF FoMs in accordance to it. Otherwise, using a Meyer-like small-signal model with parameters coming from a physics-based large-signal model would provide estimations of RF performance without physical meaning, as shown in Figure 2.4d-e and Figure 2.5c-d, especially if the device is biased in the NDR region.

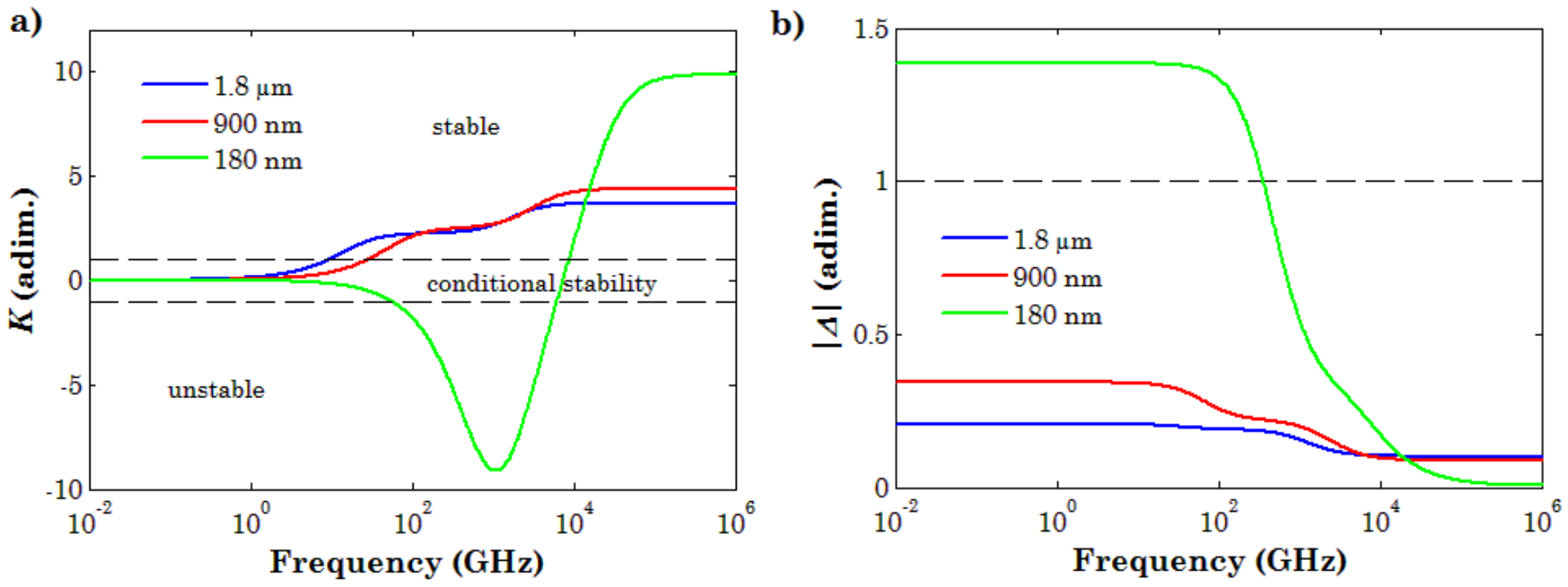

**Figure 2.13** a) $K$ and b) $\Delta$ parameters of the stability test considering the effect of the channel length scaling. Dashed lines separate the regions of instability and stability ($V_{GS} - V_{Dirac}$ = 2 V and $V_{DS}$ = 0.6 V).

Regarding the device stability issue, in Figure 2.13 the stability factors $K$ and $|\Delta|$ have been plotted considering different channel lengths. While longer devices show conditional stability, the factor $K$ corresponding to the short-channel case ($L$ = 180 nm) decreases below -1 for a set of frequencies between ~$10^2$ and $10^4$ GHz. The scaled transistor thus enters in the unstable region, which prevents it from working properly as a power amplifier. Making the device more prone to instability could imply sacrificing some power gain to restore stable RF operation. This can be observed in Figure 2.11a-b where the fact of using a reduced $V_{ds}$ implies that the device stability is extended to lower channel lengths down to 180 nm, although giving a slight decrease in the FoMs. As a result, the choice of the bias point is quite important, not only to maximize $f_{Tx}$, $f_{max}$ and power gain, but also to make sure that the device is working in the stable region.

Finally, the gate series resistance $R_g$ is indeed an important source of RF performance degradation. In this context, Figure 2.14 illustrates how $f_{max}$ decreases with gate resistance. The graph compares the results for devices with different channel lengths, making clear that minimizing the gate resistance produces an important improvement in $f_{max}$. Besides, no relation has been found between the stability of the GFET and the gate resistance.

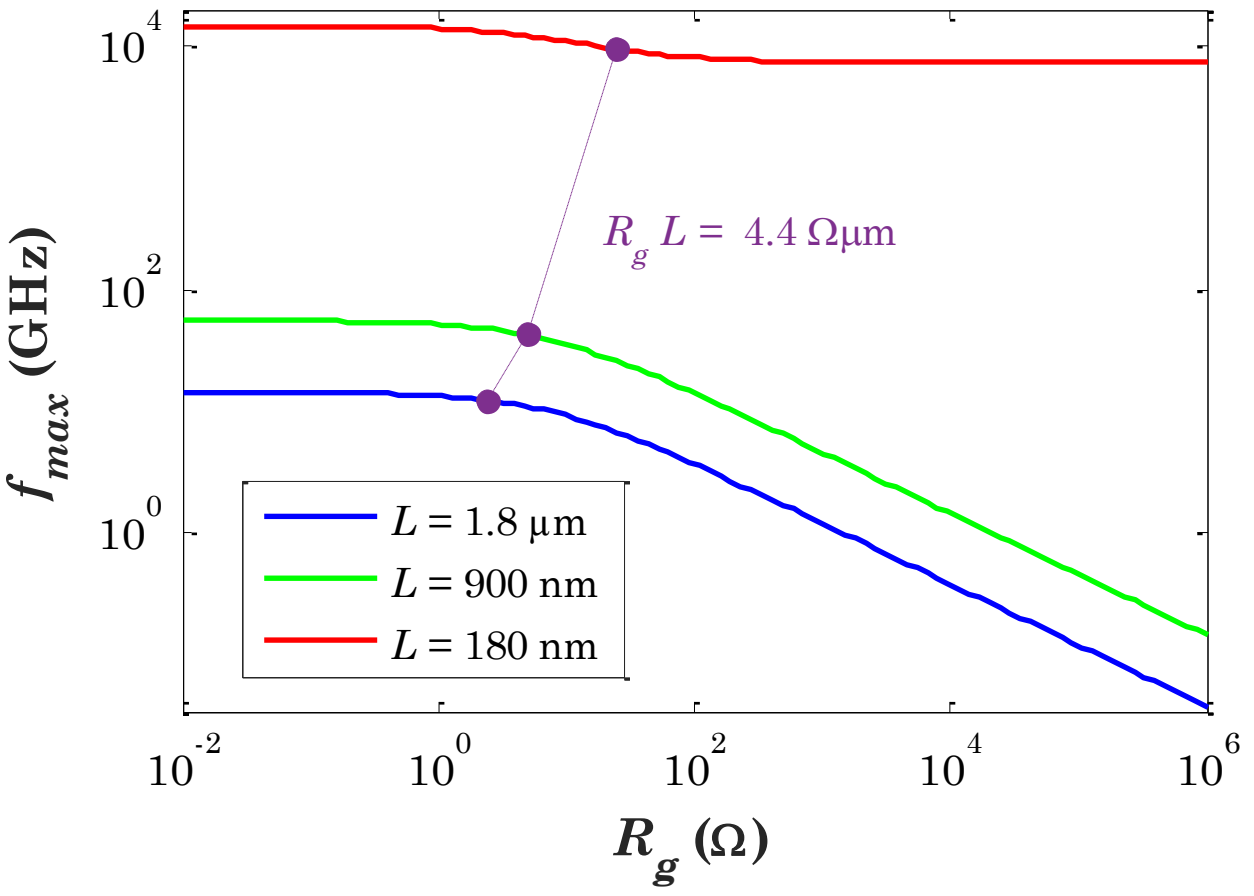

**Figure 2.14** Scaling of $f_{max}$ as a function of gate resistance $R_g$. The circles refer to the point where $R_g$ is equal to 4.4 Ω·µm/$L$. ($V_{GS} - V_{Dirac}$ = 1.5 V and $V_{DS}$ = 0.6 V)

## 2.5 Conclusions

In this chapter, a small-signal model for three-terminal 2D-FETs has been presented. The model formulation is universally valid for any 2DM. Two main features must be highlighted: (i) the small-signal model guarantees charge conservation and (ii) the metal contact and access resistances have been included in the parameter extraction methodology because of the impossibility of removing their effect from a de-embedding procedure.

Explicit and exact expressions for both cut-off and maximum oscillation frequency have been provided consistent with the charge-conserving small-signal model, with no approximations. Such expressions have been compared with other found in the literature finding noticeable discrepancies among them when applied to GFETs, especially when the transistor is biased in the NDR region.

An approach to extract the small-signal parameters (transconductance, output conductance and intrinsic capacitances) and gate resistance from $S$-parameter measurements has been proposed. Additionally, a direct extraction method of the series combination of the metal contact and access resistances has also been provided. The extraction approach has been assessed against RF characterization of a GFET, showing good agreement.

Charge conservation issue is important not only to ensure the model accuracy to predict the FoMs but also to guarantee the compatibility with physics-based large-signal models. Moreover, charge conservation could also be critical when a large-signal model is assembled with a small-signal model, in form of tables containing values of drain current and of small-signal parameters for many combinations of bias voltages. Such a model is the so-called table look-up model presented in section 1.2. Then, by using interpolation functions the values for points in between could be computed.

Besides, the charge-conserving small-signal model has been proposed as a tool for analysing the device stability when it is acting as a power amplifier. In doing so, taking advantage of such a compatibility with numerical physics-based large-signal models, it has been used to make a study of the impact on both RF performance and stability when the channel length of a GFET is scaled. The results show that channel length scaling is a possible way to improve the RF performance, although stability is an important factor that could prevent a device to be usable. In particular, short-channel GFETs could be unstable, so care must be exercised when designing the device. Furthermore, the choice of the bias point is crucial to guarantee a stable operation, as well as increasing the maximum oscillation frequency would require a minimization of the gate resistance.

Chapter 3

# Large-signal modelling of graphene-based FETs

Graphene was theoretically analysed since the late '40s [92] and determined that a single sheet of graphite could not be thermodynamically stable [13]. In 1962, Hofmann and Boehm, who were studying graphene oxide, succeeded to obtain first ultrathin graphitic flakes [93]. Graphene was rediscovered in 2004 by Novoselov and Geim using the so-called scotch-tape method [12]. Due to their work, demonstrating some of the impressive properties of graphene [12], they were awarded the Nobel Prize for Physics in 2010.

Later on its electrical properties were further investigated and the first top-gated graphene transistor was demonstrated by Lemme *et al.* in 2007 [94]. Lots of graphene structures were developed during the last years, and the first IC, a mixer working at frequencies up to 10 GHz, came out in 2011 [30]. From that point onwards, device modelling activities ignited with the aim of helping the design of GFET-based ICs for HF electronics. Following this research line, a complete large-signal model for GFETs is developed in this chapter. It starts with section 3.1, which provides the main features that make graphene a promising material for analogue RF electronics. Section 3.2 presents a brief review about the graphene processing techniques. The state-of-the-art of RF GFETs and a comparison of the main RF FoMs between GFETs and other technologies is provided in section 3.3. Next, a physics-based numerical large-signal model of GFETs is presented in section 3.4. In doing so, this section starts providing a review of the electronic properties of SLG, as well as, the analysis of the electrostatic behaviour of the graphene transistor together with the carrier transport along the graphene channel. In addition, the terminal charge and capacitance description of the device is also presented, which allows for the determination of the device response under dynamic regime. Such a description is proposed under an approach guaranteeing charge conservation, which is of upmost importance to make reliable predictions, as shown in section 2.2. Next, such a numerical model is converted into a compact model in section 3.5, which can be used in standard EDA tools, thus allowing the electrical simulation of arbitrary GFET-based circuits. Finally, a compact model assessment is presented in section 3.6. For such a purpose a number of measured GFET-based circuits, reported by different groups, have been considered.

The physics-based models presented in sections 3.4 and 3.5 only addressed the intrinsic device. Nevertheless, those models are intended to be the kernel of a more complete GFET model that incorporates extrinsic components and additional non-idealities. To end the chapter, the main conclusions are summarized in section 3.7.

## 3.1 Graphene motivation

The crucial property that makes graphene interesting for high speed electronics is the high mobility which record value of $3 \cdot 10^6$ $cm^2V^{-1}s^{-1}$ was measured in suspended samples at low temperature and low carrier densities [95], [96]. Of course, mobility depends on technology and is afflicted by the substrate mismatch. Mobilities up to $2.5 \cdot 10^5$ $cm^2V^{-1}s^{-1}$ have been achieved using an hBN substrate, while in devices based on $SiO_2$ substrate the mobility falls down in the range 1000 – 40000 $cm^2V^{-1}s^{-1}$, as reported in [97]. For comparison Si-MOSFETs show channel mobilities on the order of few hundreds of $cm^2V^{-1}s^{-1}$, while III–V semiconductor transistors present values up to 10000 $cm^2V^{-1}s^{-1}$.

Despite graphene is just one-atom thick it could provide a minimal carrier sheet density in excess of $10^{12}$ cm$^{-2}$ that is enough for FET operations [17]. Furthermore, carrier saturation velocity presents peak values on the order of $10^7$ cm/s [98], [99], and the maximum carrier speed achievable in graphene is theoretically the Fermi velocity (about $10^8$ cm/s). Another important electric property involves the carrier mean free path that is strictly related to scattering phenomena and, therefore, dependent on technology and substrate mismatch. In [100], a mean free path of 70 nm was measured for SLG meanwhile a mean free path of 10 nm has been estimated for exfoliated bilayer graphene (BLG); both at carrier densities of $3 \cdot 10^{12}$ cm$^{-2}$ deposited on a 300 nm $SiO_2$ substrate and at low temperatures. The mean free path is an important property that depends on the graphene quality and plays an essential role on the transport phenomena.

On the other hand, numerous applications demand the development of large-area, flexible and conformal electronics such as wearable electronics. 2DMs, in general, such as graphene or bilayer graphene, could be an ideal choice for future flexible electronics. They tend to have excellent mechanical properties, can be prepared in polycrystalline form over large areas and can be transferred to arbitrary substrates making them mechanically compatible with flexible device fabrication. At the same time, again, the transport properties can be orders of magnitude higher than for materials used at present, such as organic semiconductors, thus enabling higher frequency at low power.

## 3.2 Graphene processing

A high-quality, scalable, Si-CMOS compatible and economical graphene process is the first requirement in order to produce graphene electronics. The scotch-tape method [12] is an example of mechanical exfoliation technique, which offers high-quality samples although it is not suitable for industrial production. Samples of individual crystals can reach millimetre range what makes this technique useful just for the study of fundamental properties.

Looking at those processes that are scalable, a low cost – low quality option is offered by liquid-phase exfoliation (LPE) technique, suitable for flexible electronics. A good cost - quality trade-off is offered by SiC thermal decomposition, which major drawbacks involves the SiC high cost and the high temperature process (in the range 1200 - 1600 ºC) that makes this technique not compatible with a standard Si-CMOS process. The best solution looks to be the chemical vapour deposition (CVD), which offers high-quality graphene sheets that can be transferred into any substrate such $SiO_2$ and hBN, hence making the process Si-CMOS compatible. Recently a roll-to-roll CVD process has produced a 100 m long high-quality graphene sheet [21]. Many issues have to be solved in order to make CVD widely used, but it looks as the most promising option for the future [15], [96], [97], [101].

## 3.3 State-of-the-art of graphene-based FETs

The gapless nature of SLG is the main obstacle to its application in logic/digital applications. The conduction and valence bands (CB and VB) of graphene touch each other in a point, which presents zero available states, which could be used as the off-state in a MOS device. However, it is not achievable in practice due to some puddles originated by the inevitable disorder that causes a minimal conductivity [102]. From this analysis, it is clear how a GFET cannot give a robust off-state, which is a fundamental requirement for a very large scale integration (VLSI) digital design. Instead, FETs based on single layer graphene seem more promising for analogue/RF applications, where the transistor is operated in the on-state.

Regarding the RF performance demonstrated by GFETs so far, $f_{Tx}$ up to 427 GHz [56] and $f_{max}$ of 200 GHz [45] have been reported. Figure 1.1 and Figure 1.2 show those FoMs, together with other competing RF transistors. The largest $f_{max}$ achieved is considered low and still lie behind III-V and Si-based transistors [52]. This is in part because of the absence of a bandgap in SLG, which prevents a proper current saturation. Thus, introducing such a bandgap would be desirable. In this regard, the feasibility of using BLG to open a bandgap and therefore achieving current saturation is investigated in Chapter 4.

## 3.4 Numerical modelling of GFETs

Taking advantage of the new physics behind the graphene requires basic understanding of its electrical properties. So, this section begins with a review of the electronic properties of graphene as a previous step towards the main goal of presenting a numerical physics-based large-signal model of the drain current, charge and capacitance of a GFET.

The physical framework for GFET modelling is a field-effect model and DD carrier transport incorporating saturation velocity effects. Using it as a basis, approaches for the calculation of charge and capacitance based on the Ward – Dutton partition scheme are derived. Such capacitance model is a charge-based capacitance model which guarantees charge conservation (see section 2.1). However, most of the GFET capacitance models hitherto found in the literature are directly based upon Meyer assumption and, therefore, may incorrectly interpret and predict the frequency performance of these devices, as demonstrated in section 2.2. Examples of compact Meyer-like capacitance models of three-terminal devices based on DD theory have been proposed by Rodríguez *et al.* [103], Zebrev *et al.* [63], Champlain [62], or Frégonèse *et al.* [104]. On the other hand, Habibpour *et al.* have proposed a semi-empirical large-signal GFET model based on a small set of fitting parameters, including the intrinsic capacitances $C_{gs}$, $C_{gd}$ and $C_{ds}$, which are extracted from $S$-parameters and DC measurements [61]. However, the intrinsic capacitances are not bias dependent, so the model can be inaccurate depending on the selected bias.

A key application related to graphene-based FETs is the development of ambipolar electronics based on the symmetric $I$-$V$ transfer characteristics. To take advantage of the ambipolarity it is essential (i) controlling the device polarity and (ii) tuning properly the ambipolar voltage (referred as Dirac voltage) of the GFET in a circuit. The inclusion of a back-gate thus is necessary for getting that tunability, which motivates the study of a general four-terminal device. Examples of this are: (i) the polarity-controllable graphene inverter and voltage controlled resistor [40], [105]; and (ii) the graphene-based frequency tripler [25] that has been demonstrated with a properly adjusted threshold voltage separation of two graphene FETs connected in series by a back-gate bias.

### 3.4.1 Electronic properties of graphene

A single layer of graphene consists of carbon atoms arranged with a 2D honeycomb crystal structure as shown in Figure 3.1. The honeycomb structure consists of the hexagonal Bravais lattice with a basis of two atoms, labelled *A* and *B*, at each lattice point [91], [106].

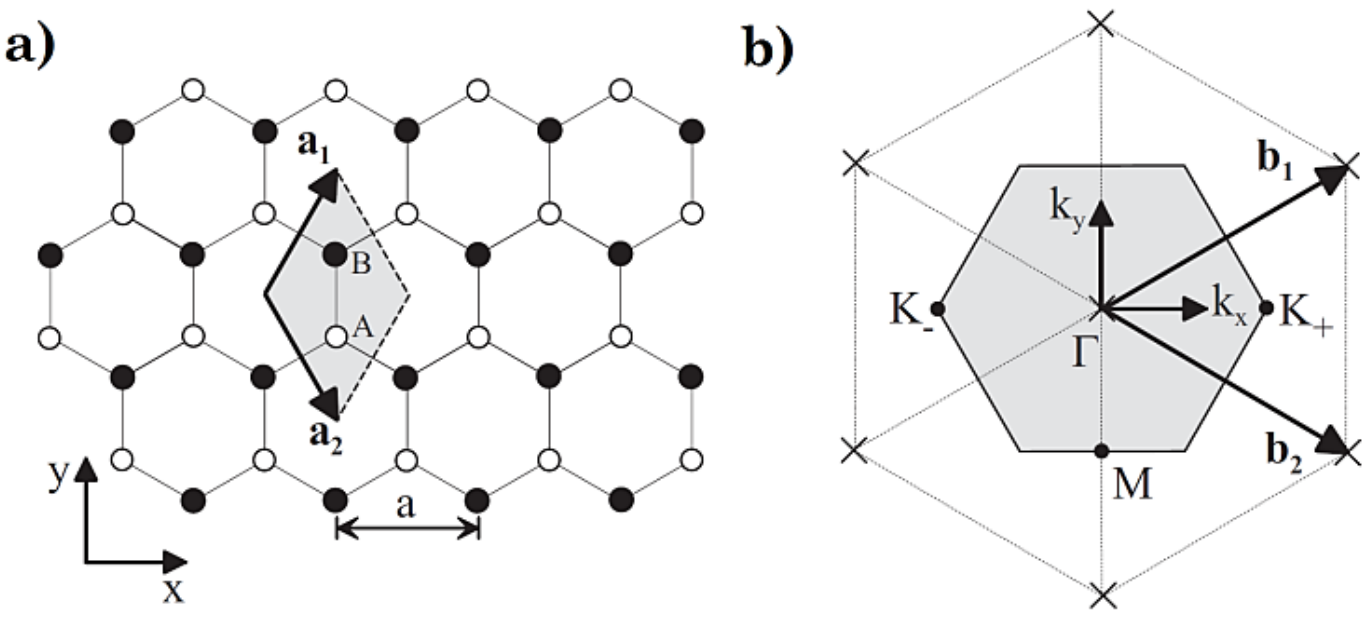

**Figure 3.1** a) Plan view of the crystal structure of graphene. Atoms *A* and *B* are represented as white and black circles, respectively. The shaded rhombus indicates the conventional unit cell; $a_1$ and $a_2$ are primitive lattice vectors of length equal to the lattice constant *a*. b) Reciprocal lattice of graphene with lattice points indicated as crosses; $b_1$ and $b_2$ are primitive reciprocal lattice vectors. The shaded hexagon is the first Brillouin zone with Γ indicating the centre, and $K_+$, $K_-$ showing two non-equivalent corners. (Image taken from [107])

Each carbon atom has six electrons, of which two are core electrons and four are valence electrons. The latter occupy 2*s*, $2p_x$, $2p_y$, and $2p_z$ orbitals. In graphene, the orbitals are $sp^2$ hybridized, meaning that two of the 2*p* orbitals, the $2p_x$ and $2p_y$ that lie in the graphene plane, mix with the 2*s* orbital to form three $sp^2$ hybrid orbitals per atom, each lying in the graphene plane and oriented 120° to each other. They form σ bonds with other atoms, shown as straight lines in the honeycomb crystal structure in Figure 3.1a. The remaining $2p_z$ orbital for each atom lies perpendicular to the plane, and, when combined with the $2p_z$ orbitals on adjacent atoms in graphene, forms a $\pi$ orbital, meaning that the tight-binding model can include only one electron per atomic site, in a $2p_z$ orbital.

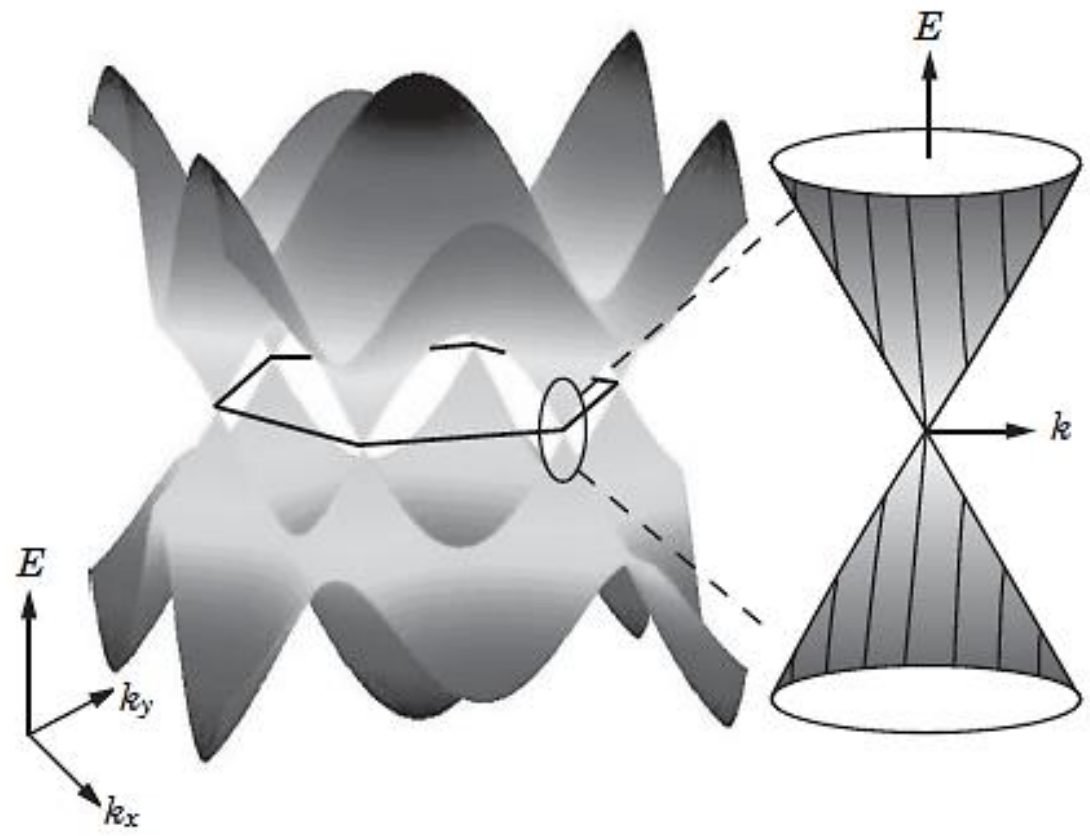

**Figure 3.2** The low-energy band structure of single layer graphene with conduction and valence bands touching at six corners of the Brillouin zone. The zoomed region presents the linear shape of the low-energy dispersion relation.

Therefore, to compute the electronic band structure, the tight-binding model has been applied to graphene, taking into account one $2p_z$ orbital on the two atomic sites in the unit cell, $A$ and $B$, and assuming that the nearest-neighbour hopping is parameterized by coupling $\gamma_{AB} \equiv \gamma_0$ and it leads to the plane velocity or Fermi velocity $v_F = (3a\gamma_0/2\hbar)$, where $\hbar$ is the reduced Planck's constant. The resulting effective Hamiltonian for SLG at low-energy in the vicinity of the valleys $K_+$ and $K_-$ can be written as:

$$\mathcal{H} = \begin{pmatrix} \epsilon_A & v_F \pi^\dagger \\ v_F \pi & \epsilon_B \end{pmatrix} \tag{3.1}$$

where $\pi = \xi p_x + ip_y$, $\pi^\dagger = \xi p_x - ip_y$, $p = (p_x, p_y)$ is the momentum measured with respect to the $K$ point, $\xi$ =+1(-1) labels valley $K_+$ ($K_-$). Parameters $\epsilon_A$, and $\epsilon_B$ describe the on-site energies on the two atomic sites, that are equal in the most general case $\epsilon_A = \epsilon_B = 0$ meaning that the zero of energy is set to be equal to the energy of the $2p_z$ orbital. The energy eigenvalues are given by:

$$E(p) = \pm v_F p \tag{3.2}$$

where ± refer to the CB and VB, respectively. So, because of the high lattice symmetry the band structure for graphene at low energies has the linear conical shape shown in Figure 3.2. This is a remarkable difference from the usual parabolic energy-momentum relation in conventional semiconductors. In graphene, the CB and VB touch each other in one point, dubbed Dirac point (DP) or charge neutrality point (CNP), at the six corners of the two-dimensional hexagonal Brillouin zone and create the zero bandgap.

The two-dimensional density of states (2D-DOS) at low energy can also be derived from (3.2), resulting in:

$$DOS_{2D}\left(E\right)=\frac{2\left|E\right|}{\pi\hbar^{2}{v_{F}}^{2}} \tag{3.3}$$

From the derived 2D-DOS both the $n$ and $p$ (-type) carrier concentration can be calculated as:

$$\begin{aligned} n &= \frac{\upsilon}{2}\int_{E_c}^{+\infty} DOS_{2D}\left(E\right) f\left(E-E_F\right) dE \\ p &= \frac{\upsilon}{2}\int_{-\infty}^{E_\upsilon} DOS_{2D}\left(E\right)\left[1-f\left(E-E_F\right)\right] dE \end{aligned} \tag{3.4}$$

where $f$ is the Fermi-Dirac distribution, $\upsilon = 2$ is the band degeneracy; $E_F$ refers to the Fermi energy; $E_c$ refers to the CB and $E_\upsilon$ refers to the VB edge.

### 3.4.2 Electrostatics of GFETs

The cross-section of a dual-gate graphene-based device considered is the one depicted in Figure 3.3a. The graphene sheet plays the role of the active channel between the source and the drain. To get the electrostatic behaviour, the 1D Gauss law's equation is solved along the $y$-axis. Direction $x$ extends from source to drain along the channel length ($L$). 1D Gauss law's equation then takes the following form [67]:

$$\nabla \cdot E = \frac{\rho_{free}}{\varepsilon} \tag{3.5}$$

where $\rho_{free}$ is the free charge density, $\varepsilon$ is the permittivity of the medium and $E$ is the electric field which is defined by a scalar electric potential field, $E = -\nabla \varphi$, as well as $\varphi$ is directly related to the local position of the Dirac energy $E_D = -q\varphi$, and $q$ is the elementary charge. Upon application of such a 1D Gauss's law equation to the double-gate stack shown in Figure 3.3b, the following expression connecting the external voltages, charge densities, oxide capacitances and carrier concentration can be gotten:

$$\begin{aligned} C_t\left(V_g - V_{g0} + V_c\right) &= -\sigma_1 \\ C_b\left(V_b - V_{b0} + V_c\right) &= -\sigma_2 \end{aligned} \tag{3.6}$$

where $C_t = \varepsilon_0\varepsilon_t/L_t$ and $C_b = \varepsilon_0\varepsilon_b/L_b$ are the top and bottom oxide capacitances, respectively; $V_g$-$V_{g0}$ and $V_b$-$V_{b0}$ are the top- and back-gate voltage overdrive; and $V_{g0}$ and $V_{b0}$ are the flat-band voltages. These quantities comprise work-function differences between the gates and the graphene channel and possible additional charge due to impurities or doping [108]; $-V_c$ is the voltage drop across the graphene and it is directly related to the local position of the chemical potential $E_F - E_D = -qV_c$; and $\sigma_1$ and $\sigma_2$ are the charge densities enclosed at the top- and back-gate stacks, respectively. Therefore, the overall net mobile sheet charge density, $Q_{net} = \sigma_1 + \sigma_2 = q(p\text{-}n)$, is expressed as:

$$Q_{net} = -C_t\left(V_g - V_{g0} + V_c\right) - C_b\left(V_b - V_{b0} + V_c\right) \tag{3.7}$$

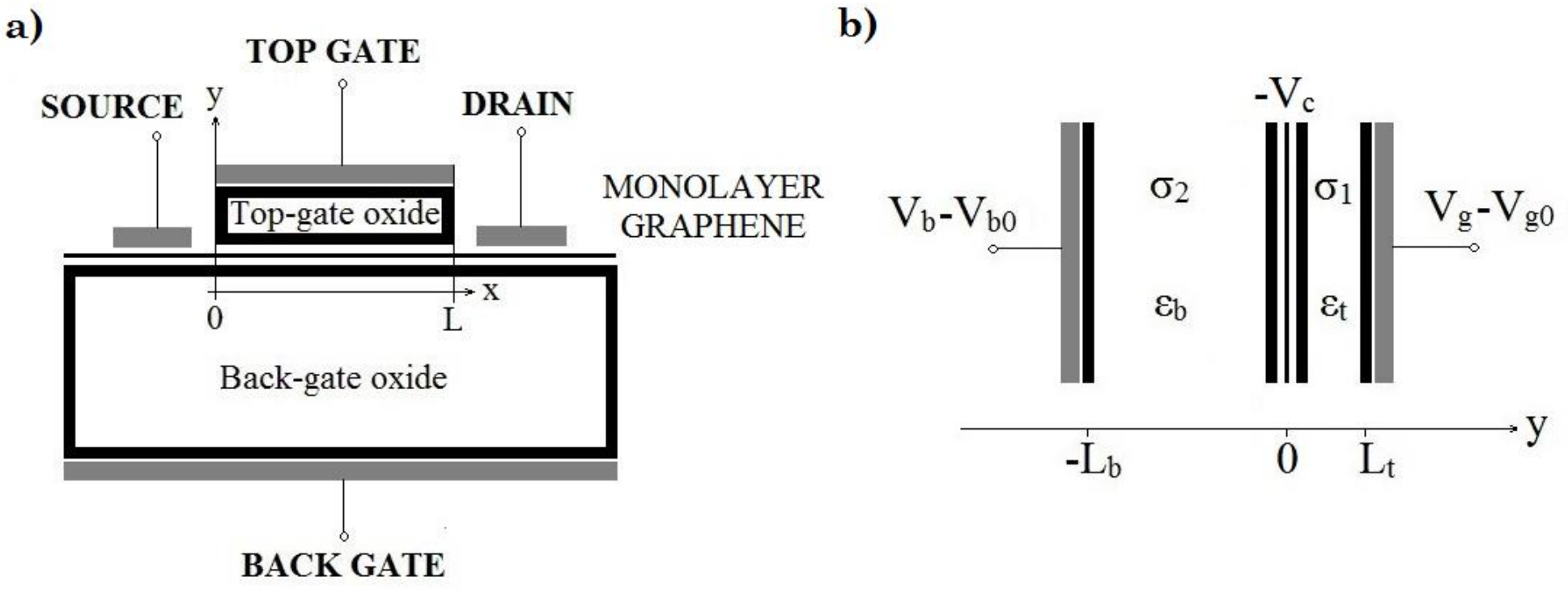

**Figure 3.3** a) Cross-section of a GFET. A graphene sheet plays the role of the active channel. The electrostatic modulation of the carrier concentration in the 2D sheet is achieved via a double-gate stack consisting of top- and back-gate dielectrics and corresponding metal gates. b) Scheme of the SLG-based capacitor showing the relevant physical and electrical parameters, charges and potentials.

**Table 3.1** Input parameters of a prototype SLG-based capacitor

| Input parameter | Value | Input parameter | Value |
|---|---|---|---|
| $T$ | 300 K | $L_t$ | 26 nm |
| $L$ | 10 µm | $L_b$ | 20 µm |
| $W$ | 5 µm | $\varepsilon_t$ | 15 |
| $V_{g0}$ | 0.85 V | $\varepsilon_b$ | 3.9 |
| $V_{b0}$ | 0 V | | |

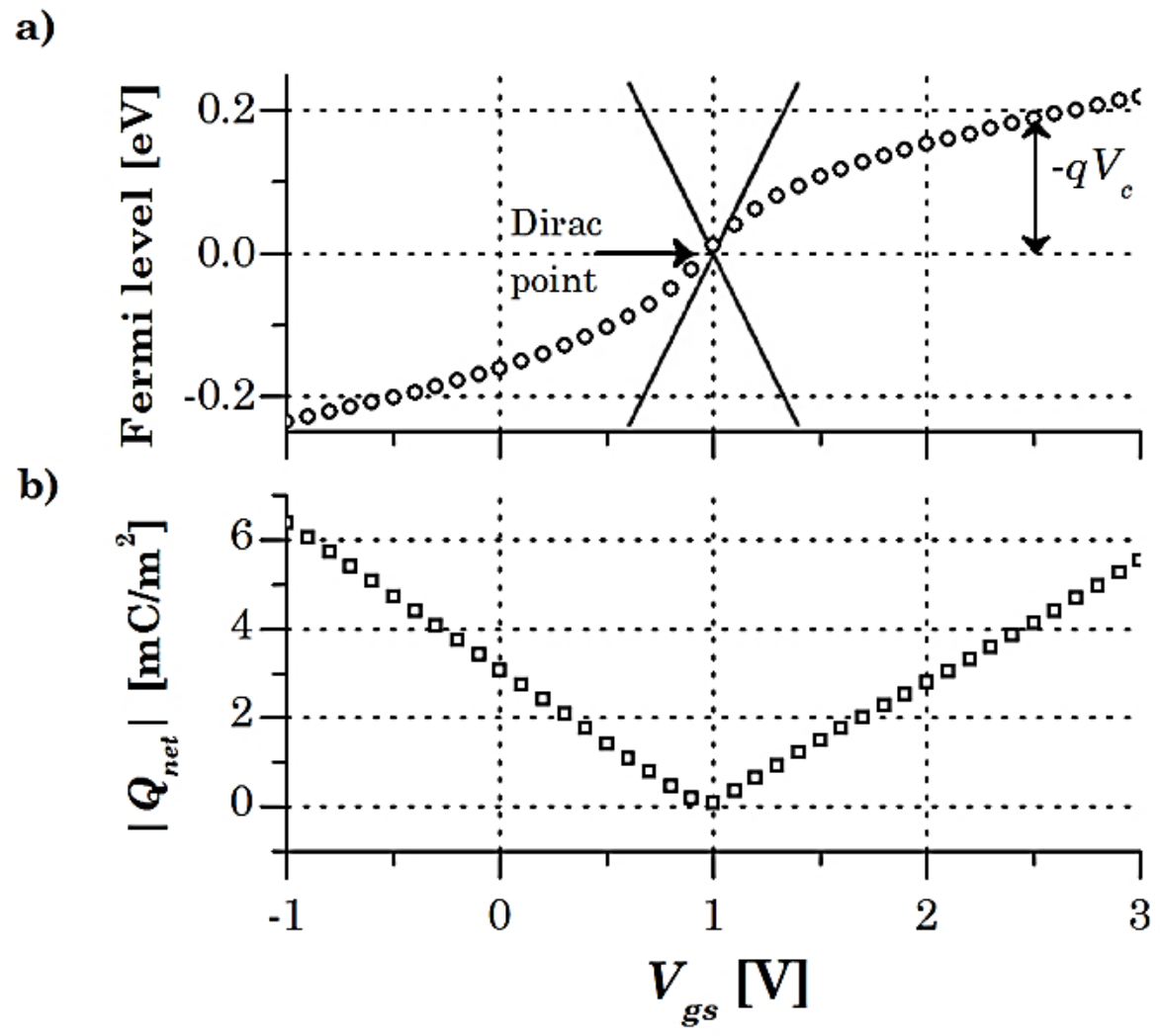

**Figure 3.4** a) Fermi level $E_F$ and b) overall net mobile sheet charge density $|Q_{net}|$ of the graphene-based capacitor shown in Figure 3.3b and described in Table 3.1 versus top-gate bias. A back-gate bias of $V_b$ = 0 V is considered. The Fermi level crosses the DP at a bias $V_g = V_{Dirac}$ and thus the minimum conductance is achieved. The voltage drop across the graphene, labelled as $V_c$, gives the shift of the Fermi level respect to the DP.

Gate voltage electrostatically modulates the carrier concentration in graphene. Figure 3.4 illustrates how the top-gate bias tunes the carrier density and, ultimately, the Fermi energy. The simulated device is described in Table 3.1. To understand the electrostatics of graphene, a positive overdrive top-gate bias is assumed to be applied, which moves the Fermi level from the equilibrium point (referred as the DP) to a new level into the CB, as shown in Figure 3.4. Hence the material becomes n-doped and, if an electric field is applied between source and drain, there will be an electron flux in the form of an electric current. Same effect happens applying a negative overdrive top-gate bias that moves down the Fermi level into the VB making the channel p-doped. Again, a hole flux will appear, provided that an electric field is applied between the source and drain. This behaviour is called ambipolarity since the symmetrical band structure around the DP implies that electrons and holes have the same properties in pure graphene.

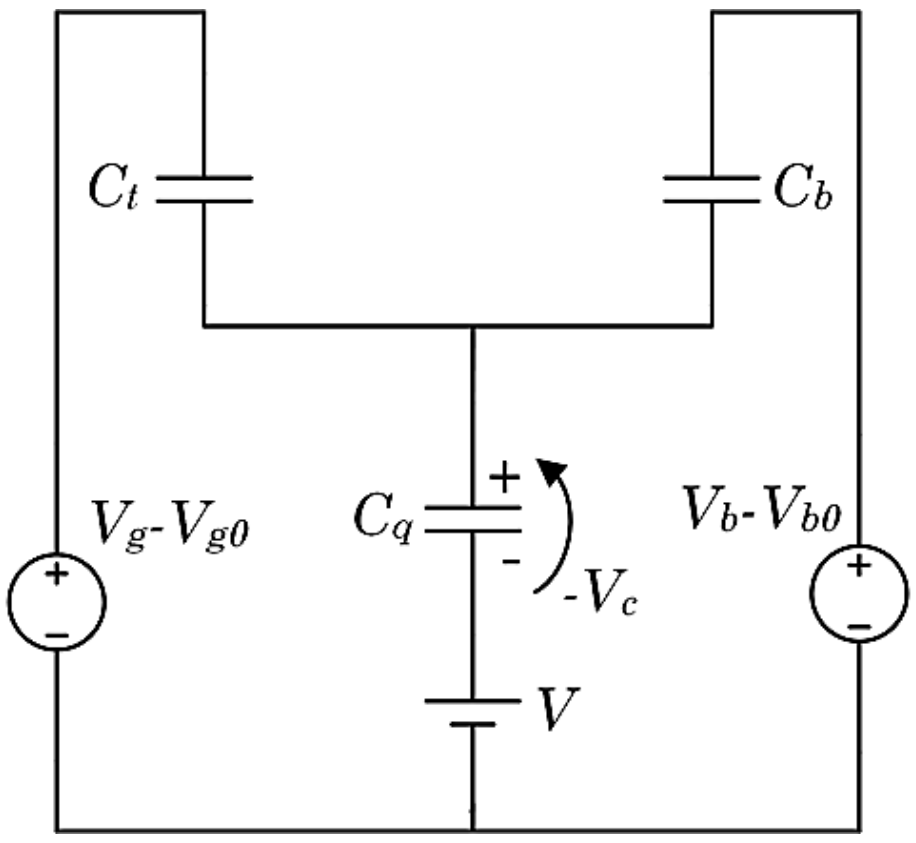

**Figure 3.5** Equivalent capacitive circuit of a GFET.

The electrostatics of a GFET can be also represented using the equivalent capacitive circuit depicted in Figure 3.5, which has been derived from (3.7) but replacing $V_g$ and $V_b$ by $V_g - V(x)$ and $V_b - V(x)$, respectively, where $V(x) = -E_F/q$ is the quasi-Fermi level along the graphene channel. This quantity must fulfil the following boundary conditions: (1) $V(x) = V_s$ at the source end, $x = 0$; (2) $V(x) = V_d$ at the drain end, $x = L$. The potential $-V_c$ in the equivalent circuit represents the shift of the Fermi level (SFL) respect to the Dirac energy or, equivalently, the voltage drop across the quantum capacitance $C_q$, which is pretty the same concept that the surface potential in conventional silicon transistors. This quantity is usually defined as $C_q = dQ_{net}/dV_c$ and it has to do with the 2D-DOS of graphene. Figure 3.6 shows a scheme of these potentials. In nanoscale devices, where the oxide thicknesses could be small and the corresponding geometrical capacitances large, it could play a dominant role in defining the overall gate capacitance [109], [110]. Quantum capacitance of graphene is presented in Figure 3.7. Applying circuit laws to the equivalent capacitive circuit, the following straightforward relation is obtained:

$$V(x)=\frac{C_t}{C_t+C_b}\left(V_g-V_{g0}+V_c\right)+\frac{C_b}{C_t+C_b}\left(V_b-V_{b0}+V_c\right)+\frac{Q_{net}(V_c)}{C_t+C_b} \tag{3.8}$$

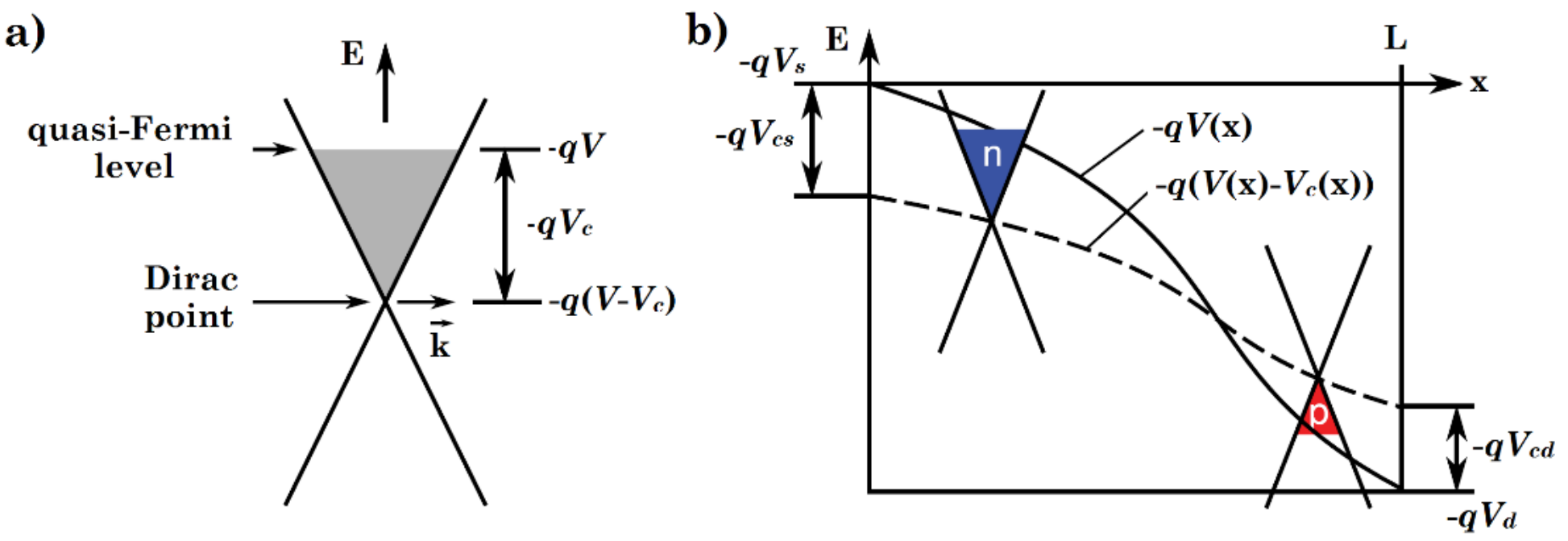

**Figure 3.6** a) Scheme of the energy dispersion relation of graphene, showing the energies defined in this section. $E_F$ = $-qV$ is the quasi-Fermi-level energy, $E_D$ = $-q(V-V_c)$ = $-q\varphi$ is the Dirac energy (where the conduction band and the valence band touch each other). The difference between electrochemical and electrostatic potentials is called the chemical potential $V_c = V-\varphi$, which is directly related to the carrier concentration in graphene. b) Schematic of the band diagram of the intrinsic device [108]: Energy $E$ versus position $x$. The quasi-Fermi-level $-qV(x)$ and the Dirac energy $-q(V(x)-V_c(x))$ are shown. $V_d$ and $V_s$ are the drain and source biases, respectively, and $V_{cd}$ and $V_{cs}$ are the channel potentials at the drain and source side, respectively. Two Dirac cones illustrate the mixed n/p-type channel of this example. (Image taken from [111]).

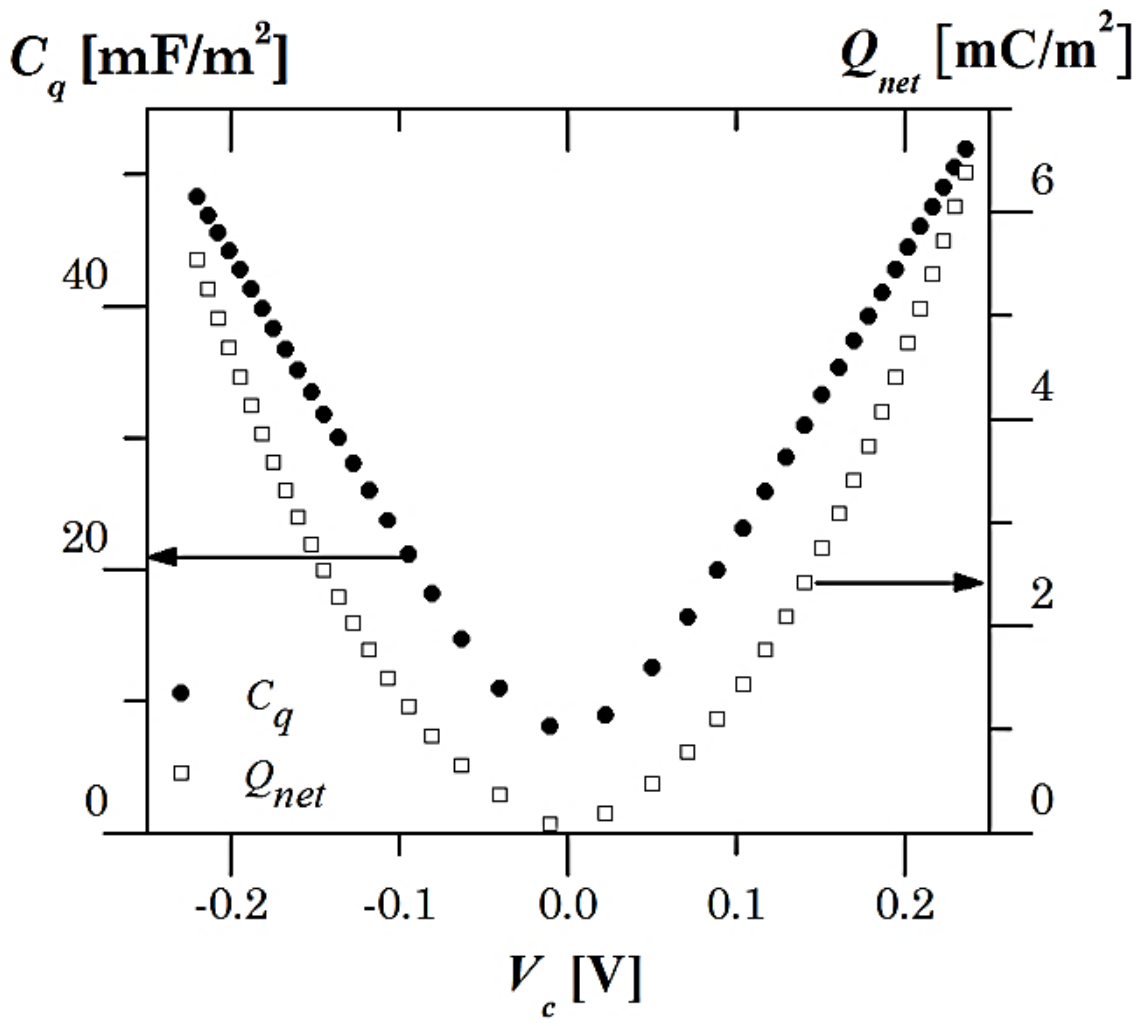

**Figure 3.7** Quantum capacitance and overall net mobile sheet charge density versus the voltage drop across to the quantum capacitance. A back-gate bias of $V_b$ = 0 V is applied. These results of graphene quantum capacitance are consistent with experimental data reported in [110].

The gate bias corresponding to the CNP, the so-called $V_{Dirac}$, can be obtained from (3.8), making $V_c = 0$, $Q_{net} = 0$ and $V = V_{ds}/2$ [71], [112]:

$$V_{Dirac} = V_{g0} + \left( \frac{C_t + C_b}{C_t} \right) \frac{V_{ds}}{2} - \frac{C_b}{C_t} \left( V_b - V_{b0} \right) \tag{3.9}$$

If one of the gate capacitances is larger than the other, i.e. $C_t >> C_b$, then the following well-known rule is found:

$$V_{Dirac} = V_{g0} + \frac{V_{ds}}{2} \tag{3.10}$$

### 3.4.3 Drift-diffusion transport model of GFETs

The carrier transport is strictly related to its mean free path (MFP or $\lambda$); the determination of the MFP in graphene is not trivial due to the strong dependence of the graphene sheet quality. Under practical conditions for common dielectric substrates, room temperature and ambient environment, MFPs of less than a hundred nm have been registered [100]. However, the MFP limiting factors are still under debate [113].

The DD theory usually employed to simulate electronic devices is still applicable while the transistor gate length is larger than the MFP ($L >> \lambda$). Otherwise the carrier transport is mastered by quantum ballistic physics. The latter scenario is out of the scope of this thesis. Therefore, standing to the common MFPs values, for channel lengths about 300 nm the DD theory is still applicable with accuracy, while in the sub-50-nm range ballistic transport must be considered. For channel lengths values between 50 nm and 300 nm, transistors work under the so-called “quasi-ballistic regime” where the DD description is not that accurate due to the weak scattering condition. Nevertheless, even in this condition, a recent study has shown how the current-voltage characteristics of nanoscale devices are still well described by DD models if mobility and saturation velocity are treated as fitting parameters [114].

As most prototype devices present channel lengths greater than the MFP ($L >> \lambda$), the drain-to-source current of a GFET has been modelled under the framework of DD transport [115]:

$$I_{ds} = WQ_{tot}(x)\mu_g(x)\frac{dV}{dx} \tag{3.11}$$

where $W$ is the channel width, $Q_{tot}(x) = Q_t(x)+\sigma_{pud}$ is the free carrier sheet density along the channel at position $x$, $Q_t(x) = q[p(x)+n(x)]$ is the transport sheet charge density, $\sigma_{pud} = q\Delta^2/\pi\hbar^2 v_F^2$ is the residual charge density due to electron-hole puddles [100], with $\Delta$ being the inhomogeneity of the electrostatic potential; $V$ represents the quasi-Fermi level which has been assumed to be the same for both electrons and holes because the generation/recombination times for carriers in graphene are very short (1-100 ps) [116]–[118] and therefore electron and hole quasi-Fermi levels cannot deviate too much from each other [108]; and $\mu_g(x)$ is the mobility considered to be the same for both electrons and holes. The model includes saturation velocity in the form:

$$\mu_g(x) = \frac{\mu}{\left[1+\left(\frac{\mu}{v_{sat}(x)}\left|\frac{d\varphi}{dx}\right|^{\beta}\right)^{1/\beta}\right]} \tag{3.12}$$

where $\mu$ is effective low-field mobility for both electrons and holes and it has been assumed to be independent of the applied electric field, carrier density, or temperature; $v_{sat}$ is the saturation velocity; $-d\varphi/dx$ is the electric field; and $\beta$ is the adimensional saturation coefficient considered to be the same for both kinds of carriers. A soft saturation model ($\beta$ = 1) for the drift carrier velocity has been adopted, consistently with numerical studies of electronic transport in SLG relying on Monte Carlo simulations [119]. Regarding the graphene saturation velocity, the model reported in [120] and shown in Figure 3.8 has been employed. It sets constant $v_{sat}$ below the critical carrier density $\sigma_c$ and carrier density dependent $v_{sat}$ above this threshold:

$$v_{sat} = \begin{cases} \dfrac{2v_F}{\pi} & |Q_{net}| \le q|\sigma_c| \\ \dfrac{2q\Omega}{\pi^2\hbar v_F|Q_{net}|}\sqrt{\dfrac{\pi(\hbar v_F)^2|Q_{net}|}{q}-\left(\dfrac{\hbar\Omega}{2}\right)^2} & |Q_{net}| > q|\sigma_c| \end{cases} \tag{3.13}$$

$$\sigma_c = \frac{1}{2\pi}\left(\frac{\Omega}{v_F}\right)^2$$

where $\hbar\Omega$ is the effective energy at which a substrate optical phonon is emitted.

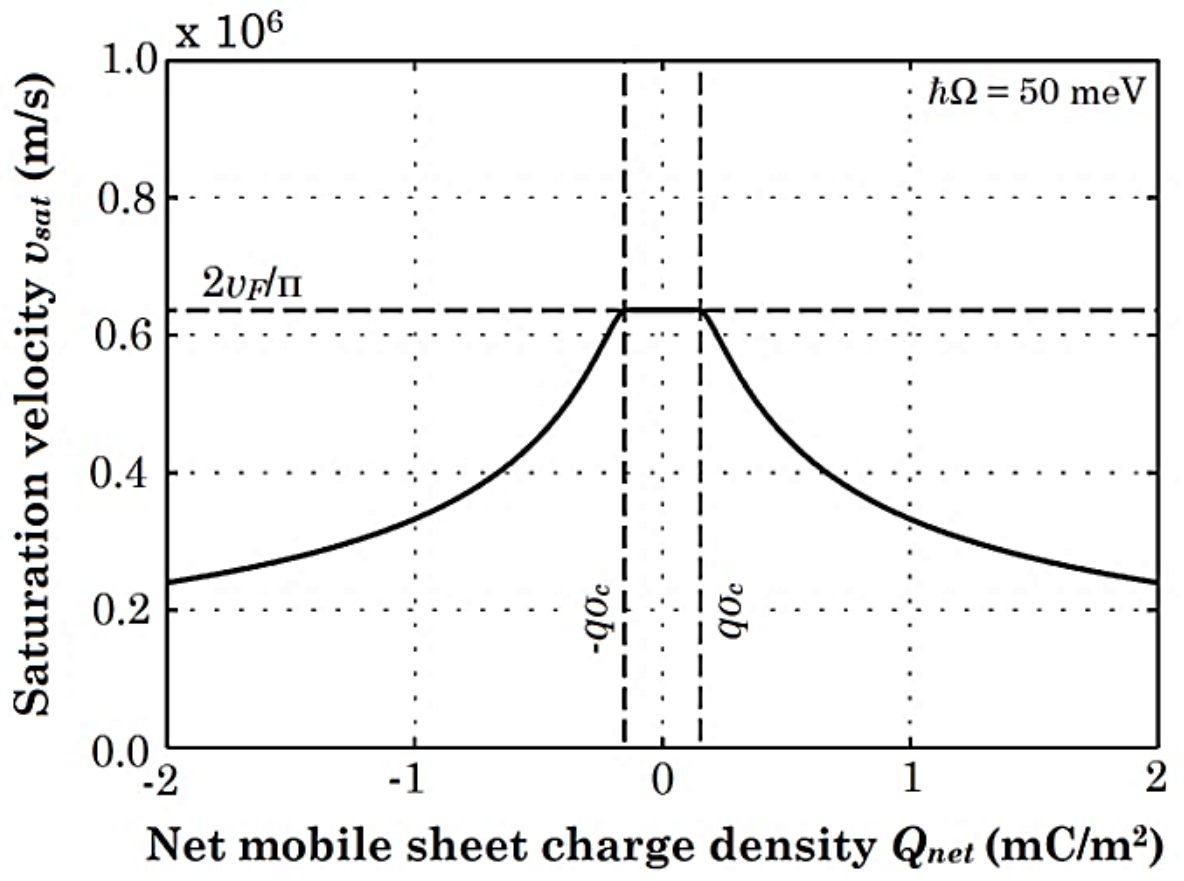

**Figure 3.8** Graphene saturation velocity vs. net mobile sheet charge density [120]. The optical phonon energy $\hbar\Omega$ has been set to 50 meV.

Then, inserting (3.12) into the DD current equation and integrating over the device length, and after assuming there are no generation-recombination processes involved, the drain current can be expressed as:

$$I_{ds} = \frac{\mu W \int_{V_s}^{V_d} Q_{tot} dV}{L + \mu \left| \int_{\varphi_s}^{\varphi_d} \frac{1}{v_{sat}} d\varphi \right|} \tag{3.14}$$

To get the drain current, it is convenient to solve the above integral using $V_c$ as the integration variable, and consistently express $Q_{tot}$ as a function of $V_c$ in the following way:

$$I_{ds} = \mu \frac{W}{L_{eff}} \int_{V_{cs}}^{V_{cd}} Q_{tot}(V_c) \frac{dV}{dV_c} dV_c \tag{3.15}$$

where $L_{eff}$ is the denominator of (3.14) and considered as a correction to the physical channel length to incorporate saturation velocity effects:

$$L_{eff} = L + \mu \left| \int_{V_{cs}}^{V_{cd}} \frac{1}{v_{sat}(V_c)} \left( \frac{dV}{dV_c} - 1 \right) dV_c \right| \tag{3.16}$$

and $V_{cs}$ and $V_{cd}$ are obtained from (3.8), with $V_{cs} = V_c|_{V = V_s}$ and $V_{cd} = V_c|_{V = V_d}$. In addition, the quantity $dV/dV_c$ in (3.15) and (3.16) can also be derived from (3.8) and reads as follows:

$$\frac{dV}{dV_c} = 1 + \frac{C_q(V_c)}{C_t + C_b} \tag{3.17}$$

### 3.4.4 Charge and capacitance models of GFETs

There has been little exploration on the physical behaviour of GFETs under dynamic conditions. Previous GFET capacitance models hitherto found in the literature [62], [121] are directly based upon the Meyer assumption, therefore they assume that the capacitances in the intrinsic FET are reciprocal (as 2-terminal lumped capacitances), which is not the case in real devices, and earlier models based on this assumption cannot ensure charge conservation [65], [66].

On the other hand, charge-based models ensure charge conservation and consider the nonreciprocal property of capacitances in a FET. These features are required especially for RF applications in which the influence of transcapacitances are critical and should be considered. Thanks to some corrections assembled by Ward and Dutton [122] the charge conservation issue was solved at the cost of introducing a capacitive-matrix which adds a bit of complexity. It must be reminded that both Meyer and charge-based modelling approaches assume the so-called *quasi-static-operation* approximation, where the fluctuation of the varying terminal voltages is assumed to be slow, so the stored charge could follow the voltages variations. Such an approximation is found to be valid when the transition time for the voltage to change is less than the transit time of the carriers from source to drain. As a result, an estimation of the maximum frequency valid under the *quasi-static-operation* corresponds to the quotient ($\upsilon_{sat}/2\pi L$), where $\upsilon_{sat}$ is expressed in (3.13) [48]. Further extension of the model to include the non-quasi-static effects is planned for a future work.

So, an accurate modelling of the intrinsic capacitances of FETs requires an analysis of the charge distribution in the channel versus the terminal bias voltages. In doing so, the terminal charges $Q_g$, $Q_b$, $Q_d$, and $Q_s$ associated with the top-gate, back-gate, drain, and source electrodes of a four-terminal device has been considered. For instance, $Q_g$ can be calculated by integrating $Q_{net_g}(x) = C_t(V_{gs}-V_{g0}+V_c(x)-V(x))$ along the channel and multiplying it by the channel width $W$. This expression for $Q_{net_g}(x)$ has been obtained after applying Gauss's law to the top-gate stack, resulting in (3.19). A similar expression can be found for $Q_b$. It is worth noticing that:

$$Q_g + Q_b = -W\int_0^L Q_{net}(x)\,dx \tag{3.18}$$

On the other hand, the charge controlled by both the drain and source terminals can be computed based on Ward-Dutton's linear charge partition scheme, which guarantees charge conservation. The resulting equations are listed next:

$$\begin{aligned}
Q_g &= \frac{WC_t}{C_t + C_b}\left[C_b L\left(V_g - V_{g0} - V_b + V_{b0}\right) - \int_0^L Q_{net}(x)dx\right]\\
Q_b &= \frac{WC_b}{C_t + C_b}\left[C_t L\left(V_b - V_{b0} - V_g + V_{g0}\right) - \int_0^L Q_{net}(x)dx\right]\\
Q_d &= W\int_0^L \frac{x}{L} Q_{net}(x)dx\\
Q_s &= -\left(Q_g + Q_b + Q_d\right)
\end{aligned} \tag{3.19}$$

The above expressions can conveniently be written using $V_c$ as the integration variable, as it was done to model the drain current. Based on the fact that the drain current is the same at any point $x$ in the channel, the following information is gotten from the DD transport model:

$$\begin{aligned}
dx &= \frac{\mu W}{I_{ds}(V_c)} Q_{tot}(V_c)\frac{dV}{dV_c}dV_c - \mu\frac{1}{\upsilon_{sat}(V_c)}\frac{dV}{dV_c}\left|dV_c\right|\\
x &= \frac{\mu W}{I_{ds}(V_c)}\left[\int_{V_{cs}}^{V_c} Q_{tot}(V_c)\frac{dV}{dV_c}dV_c\right] - \mu\left|\int_{V_{cs}}^{V_c}\frac{1}{\upsilon_{sat}(V_c)}\frac{dV}{dV_c}dV_c\right|
\end{aligned} \tag{3.20}$$

A four-terminal FET can be modelled with 4 self-capacitances and 12 intrinsic transcapacitances, which makes 16 capacitances in total. The capacitance matrix is formed by these capacitances where each element $C_{ij}$ describes the dependence of the charge at terminal $i$ with respect to a varying voltage applied to terminal $j$ assuming that the voltage at any other terminal remain constant.

$$\begin{aligned} C_{ij} &= -\frac{\partial Q_i}{\partial V_j} \qquad i \neq j \\ C_{ij} &= \frac{\partial Q_i}{\partial V_j} \qquad i = j \end{aligned} \tag{3.21}$$

where *i* and *j* stand for *g*, *d*, *s*, and *b*.

$$\begin{bmatrix} C_{gg} & -C_{gd} & -C_{gs} & -C_{gb} \\ -C_{dg} & C_{dd} & -C_{ds} & -C_{db} \\ -C_{sg} & -C_{sd} & C_{ss} & -C_{sb} \\ -C_{bg} & -C_{bd} & -C_{bs} & C_{bb} \end{bmatrix} \tag{3.22}$$

Each row must sum to zero for the matrix to be reference-independent, and each column must sum to zero for the device description to be charge-conservative. Note that of the 16 intrinsic capacitances only 9 are independent.

Finally, the dynamic response of a GFET would be calculated as:

$$\begin{aligned} i_g(t) &= C_{gg}\frac{dv_g}{dt} - C_{gd}\frac{dv_d}{dt} - C_{gs}\frac{dv_s}{dt} - C_{gb}\frac{dv_b}{dt} \\ i_d(t) &= -C_{dg}\frac{dv_g}{dt} + C_{dd}\frac{dv_d}{dt} - C_{ds}\frac{dv_s}{dt} - C_{db}\frac{dv_b}{dt} \\ i_s(t) &= -C_{sg}\frac{dv_g}{dt} - C_{sd}\frac{dv_d}{dt} + C_{ss}\frac{dv_s}{dt} - C_{sb}\frac{dv_b}{dt} \\ i_b(t) &= -C_{bg}\frac{dv_g}{dt} - C_{bd}\frac{dv_d}{dt} - C_{bs}\frac{dv_s}{dt} + C_{bb}\frac{dv_b}{dt} \end{aligned} \tag{3.23}$$

### 3.4.5 Extrinsic and parasitic elements

The model presented in this section is an intrinsic large-signal model. In this regard, the modelling of extrinsic effects (source, drain, top-gate and back-gate resistances, overlap and junction capacitances, etc.) is out of the scope of this work. Nevertheless, to reproduce any experimental current-voltage characteristics of GFETs, accounting of the voltage drop at the source/drain contacts is necessary. This quantity must be removed from external biases $V_{ds,e}$, $V_{gs,e}$ in order to get the internal ones $V_{ds}$, $V_{gs}$, respectively. This is simply done by solving the following equations:

$$\begin{aligned} V_{ds} &= V_{ds,e} - I_{ds}(R_d + R_s) \\ V_{gs} &= V_{gs,e} - I_{ds}R_s \end{aligned} \tag{3.24}$$

It is worth noticing that the intrinsic transconductance and the intrinsic output conductance are defined as $g_m = (\partial I_{ds}/\partial V_{gs})\,|\,_{V_{ds}}$ and $g_{ds} = (\partial I_{ds}/\partial V_{ds})\,|\,_{V_{gs}}$, respectively. Then, the relation between these intrinsic and extrinsic small-signal parameters written in (2.18) is straightforwardly obtained from (3.24) [123].

## 3.5 Compact modelling of GFETs

In this section, the numerical large-signal model of GFETs presented in section 3.4 is converted into a compact model. In doing so, the drain current compact model from [111] is taken. Then, a compact model of the intrinsic capacitances is proposed by obtaining an analytical description of them. Both drain current and intrinsic capacitance models are properly combined to obtain both static and dynamic descriptions covering continuously all the operation regions, respectively. What is more, the compact large-signal model of GFETs built is implemented in Verilog-A, a language suited to circuit simulators.

### 3.5.1 Compact drain current model of GFETs

The compact drain current model of a GFET has been extracted from [111]. In the following, the main modelling aspects considered are described:

- *Electrostatics*

The electrostatics is described by (3.8) and also by the equivalent capacitive circuit shown in Figure 3.5. Thus, the net mobile sheet charge density $Q_{net} = q(p\text{-}n)$ in the channel is defined from (3.4):

$$Q_{net} = \frac{2q\left(k_BT\right)^2}{\pi\left(\hbar v_F\right)^2}\left(\mathfrak{F}_1\left(\frac{qV_c}{k_BT}\right) - \mathfrak{F}_1\left(-\frac{qV_c}{k_BT}\right)\right) \tag{3.25}$$

where $p$ and $n$ are evaluated using Fermi-Dirac integrals of first-order. Since there is no closed-form solution for such integrals, they are approximated with a maximum relative error of ~ $10^{-6}$ using elementary mathematical functions [124], [125].

Then, the $Q_{net}$ is stored in the quantum capacitance which was defined as $C_q = dQ_{net}/dV_c$. The following exact solution of this derivative has been implemented into the electrostatic analysis:

$$C_q = \frac{2q^2 k_B T}{\pi\left(\hbar v_F\right)^2} \ln\left[2\left(1+\cosh\left(\frac{qV_c}{k_B T}\right)\right)\right] \tag{3.26}$$

An iterative Verilog-A algorithm has been implemented to obtain the channel potential at the source and drain edges. In doing so, $V_{cs}$ and $V_{cd}$ must be obtained, respectively, from (3.8), with $V_{cs} = V_c|_{V = V_s}$ and $V_{cd} = V_c|_{V = V_d}$. Because of the complexity of (3.8), it is not possible to express $V_c$ explicitly. However, a construct has been used (see Figure 3.9) to let the circuit simulator iteratively solve the equation during run-time [126].

$$\begin{aligned} \mathrm{LHS}\left(V_c\right) &= \left(C_t + C_b\right)V_c + Q_{net}\left(V_c\right) \\ \mathrm{RHS}\left(V\right) &= -\left(V_g - V_{g0} - V\right)C_t - \left(V_b - V_{b0} - V\right)C_b \end{aligned} \tag{3.27}$$

In doing so, the equation's left-hand side LHS($V_c$) is equated with its right-hand side RHS($V$), both written in (3.27), by assigning the respective values to current sources connected in series. The simulator forces the two currents to be equal, and $V_c$ is then obtained by reading it out as the voltage drop over one of the current sources.

At this point, the electrostatics would be solved and the values of $V_{cs}$ and $V_{cd}$ would be available. So, not only the analytical expression of the drain current but also the analytical expressions of the intrinsic capacitances will be formulated depending on these quantities.

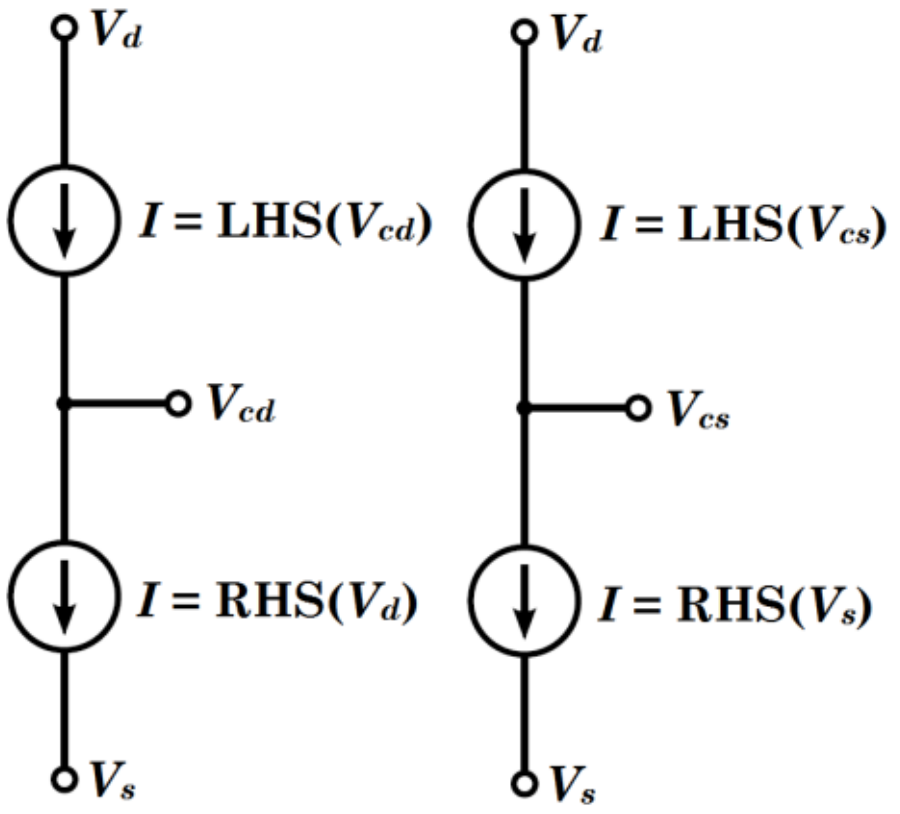

**Figure 3.9** Verilog-A construct [126] to obtain the channel potentials $V_{cd}$ and $V_{cs}$ by iteratively evaluating (3.8). (Image taken from [111])

- *Drift-diffusion transport*

Next step is to derive an analytical equation for the GFET drain current $I_{ds}$ expressed in (3.15). In doing so, the compact modelling of $L_{eff}$, $Q_{tot}$ and the relation $dV/dV_c$ is considered.

So, first, under the condition of symmetrical electron and hole mobilities, the transport sheet charge density $Q_t = q(p+n)$ is expressed as a quadratic polynomial [127]:

$$Q_t = \frac{2q\left(k_B T\right)^2}{\pi\left(\hbar v_F\right)^2}\left(\mathfrak{F}_1\left(\frac{qV_c}{k_B T}\right) + \mathfrak{F}_1\left(-\frac{qV_c}{k_B T}\right)\right) = \frac{q\pi\left(k_B T\right)^2}{3\left(\hbar v_F\right)^2} + \frac{q^3 V_c^2}{\pi\left(\hbar v_F\right)^2} \tag{3.28}$$

The polynomial's constant term represents the thermal charge density at the DP. Notice that the residual charge density due to electron-hole puddles $\sigma_{pud}$ must be included, then the total transport sheet carrier density considered in (3.15) is $Q_{tot} = Q_t + \sigma_{pud}$.

Then, an accurate square-root-based approximation [128] is used into (3.17):

$$C_q = \frac{2q^2 k_B T \ln(4)}{\pi\left(\hbar v_F\right)^2}\sqrt{1 + \left(\frac{qV_c}{k_B T \ln(4)}\right)^2} \tag{3.29}$$

introducing a maximum relative error of 7.97% but allowing expressing the drain current in an analytical form in the end.

Finally, an analytical expression of (3.16) is reported in [111] based on the following approximations [115]:

$$C_q = \frac{2q^3 |V_c|}{\pi(\hbar v_F)^2} \qquad Q_{net} = \frac{C_q V_c}{2} \tag{3.30}$$

Notice at that point that three different approximations for the calculation of graphene quantum capacitance have been used. In Figure 3.10, the relation between the different approximations of $C_q$ vs. $V_c$ is shown. A thorough discussion about the modelling error of considering each expression is reported in [111], as well as, a complete benchmarking of GFETs under static regime.

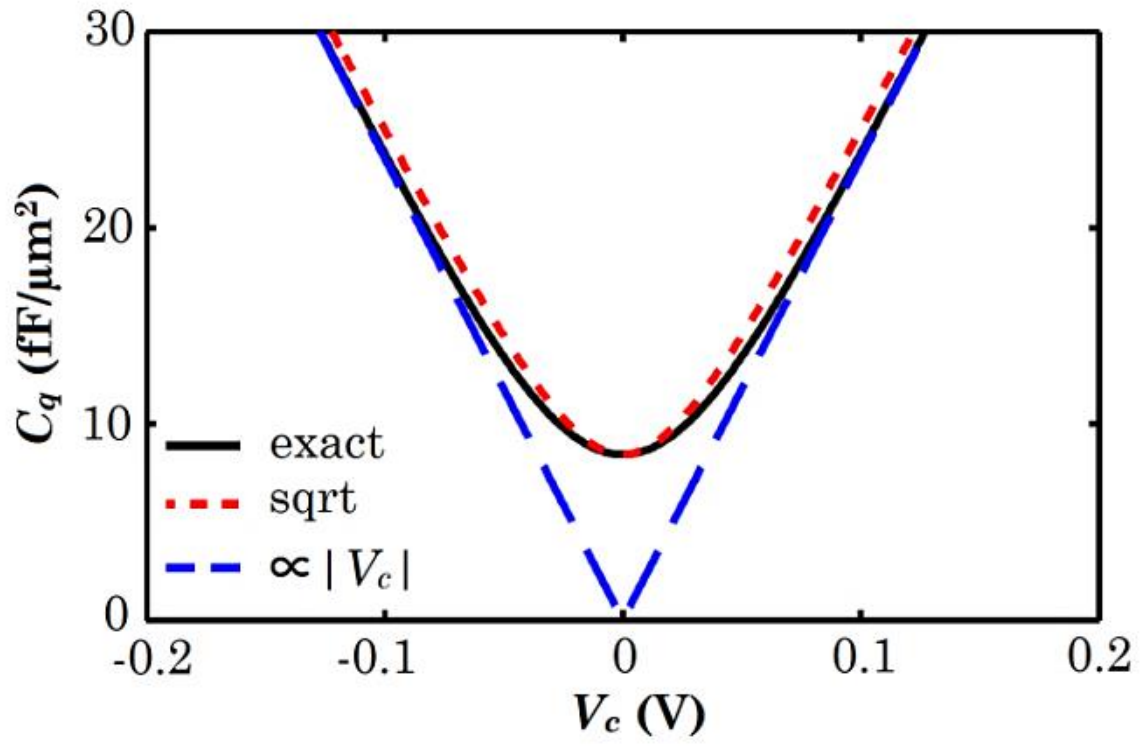

**Figure 3.10** Quantum capacitance $C_q$ versus channel potential $V_c$: exact $C_q$ (3.26); square root-approximation (3.29); and absolute value approximation (3.30). (Image taken from [111])

### 3.5.2 Compact intrinsic capacitance model of GFETs

In order to get a compact description of the dynamic response of GFETs, an analytical form of the intrinsic capacitances must be obtained. Then, it must be implemented in Verilog A and must be combined with the compact drain current model described in the previous subsection. For such a purpose, the scheme of the intrinsic capacitances described in (3.22) must be gotten in an analytical way and it must be dependent on the local potentials $V_{cs}$ and $V_{cd}$, which are calculated from the electrostatics. In this way, the working procedure of the circuit simulator consists of (i) evaluating the electrostatics of the device to obtain such potentials $V_{cs}$ and $V_{cd}$, and then (ii) using them to calculate both the static drain current and the intrinsic capacitances determining the dynamic response.

Because of the complexity of getting an analytical description of the intrinsic capacitances described in (3.22), the approximation in equation (3.30) is assumed. In doing so, first equation (3.13) is rewritten as:

$$\begin{aligned} v_{sat} &= \begin{cases} \dfrac{1}{2z_1z_2} & |V_c| \le \dfrac{\sqrt{2}}{2}\hbar\Omega \\ \dfrac{1}{z_2V_c^2}\sqrt{V_c^2 - z_1^2} & |V_c| > \dfrac{\sqrt{2}}{2}\hbar\Omega \end{cases} \\ z_1 &= \frac{\hbar\Omega}{2} \\ z_2 &= \frac{\pi}{4z_1v_F} \end{aligned} \tag{3.31}$$

Then, the approximation of the quantum capacitance in (3.30) is again used into (3.17), to calculate consecutively (3.16), (3.15), (3.20), and finally obtaining the scheme of the charge distribution in the channel, described in (3.19), expressed in an analytical way respect to $V_{cs}$ and $V_{cd}$.

Next, the intrinsic capacitances described in (3.21) are gotten using:

$$\frac{\partial Q_i}{\partial V_j} = \frac{\partial Q_i}{\partial V_{cd}}\frac{\partial V_{cd}}{\partial V_j} + \frac{\partial Q_i}{\partial V_{cs}}\frac{\partial V_{cs}}{\partial V_j} \tag{3.32}$$

In the derivation of the capacitances, the following relations, extracted from (3.8) after inserting the approximation (3.30), have been used:

$$\begin{aligned} &\frac{\partial V_{cs}}{\partial V_g} = \frac{\partial V_{cd}}{\partial V_g} = \frac{C_t}{C_b}\frac{\partial V_{cs}}{\partial V_b} = \frac{C_t}{C_b}\frac{\partial V_{cd}}{\partial V_b} = -\frac{C_t}{C_t + C_b + C_q(V_c)} \\ &\frac{\partial V_{cs}}{\partial V_s} = \frac{\partial V_{cd}}{\partial V_d} = \frac{C_t + C_b}{C_t + C_b + C_q(V_c)} \\ &\frac{\partial V_{cs}}{\partial V_d} = \frac{\partial V_{cd}}{\partial V_s} = 0 \end{aligned} \tag{3.33}$$

Moreover, from (3.19) and (3.33) the following relations between the top- and back-gate capacitances can be worked out:

$$
\begin{aligned}
&C_{bd}=C_{gd}\left(\frac{C_b}{C_t}\right) \qquad C_{db}=C_{dg}\left(\frac{C_b}{C_t}\right)\\
&C_{bs}=C_{gs}\left(\frac{C_b}{C_t}\right) \qquad C_{sb}=C_{sg}\left(\frac{C_b}{C_t}\right)\\
&C_{bb}=C_{gg}\left(\frac{C_b}{C_t}\right)^2 \qquad C_{bg}=C_{gb}=-C_{gg}\left(\frac{C_b}{C_t}\right)=-C_{bb}\left(\frac{C_t}{C_b}\right)
\end{aligned}
\tag{3.34}
$$

Basic compact modelling rules have been followed in order to meet the requirements reported in [48], [129] and to guarantee the continuity of the model over any bias condition, temperature or geometry. It is worth noticing, that in order to keep the symmetry and non-singularity at zero drain-source bias, the limits of the intrinsic capacitances at that bias have been calculated. This singularity is well-known [129] and it is produced because of the use of the soft saturation model ($\beta$ = 1) for the drift carrier velocity in (3.12).

Once the charge-based compact intrinsic capacitance description has been obtained, it has been integrated in a circuit simulator together with the drain current model presented in subsection 3.5.1, both written in Verilog-A. The complete large-signal model is available online [130]. The resulting intrinsic large-signal GFET equivalent circuit is depicted in Figure 3.11.

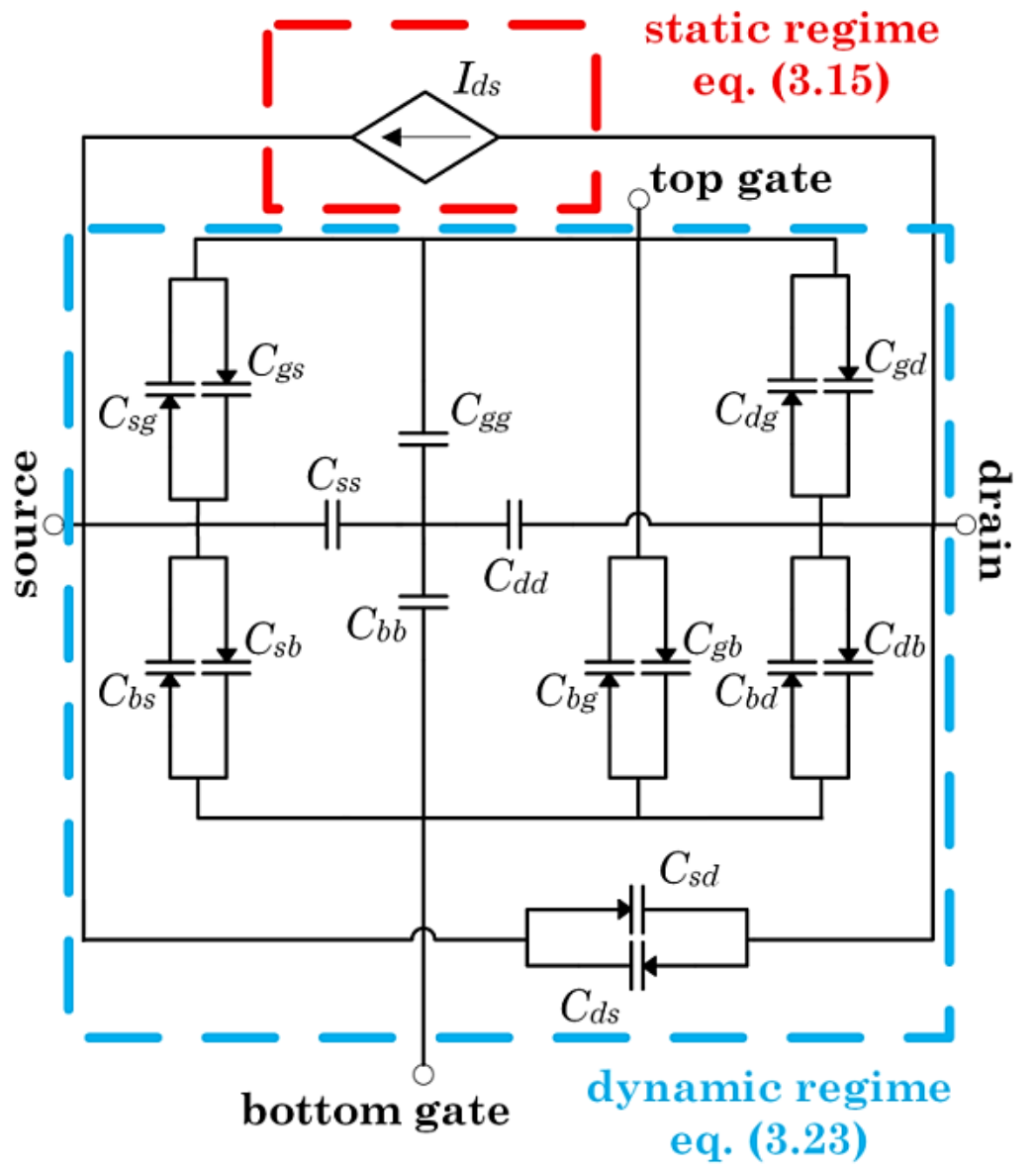

**Figure 3.11** Large-signal GFET equivalent circuit formed by the drain current model and the intrinsic capacitance model, presented in subsections 3.5.1 and 3.5.2, respectively.

The accuracy and assessment of the drain current model have been reported in [111]. However, the accuracy of the compact intrinsic capacitance model presented in subsection 3.5.2 must be checked, especially around the DP because of the use of the quantum capacitance approximation in (3.30) instead of using the exact calculation (3.26). In doing so, in Figure 3.12a, compact calculations of such capacitances have been compared against numerical calculations from the intrinsic capacitances using the large-signal model presented in section 3.4. The prototype GFET considered, is used as a key component for a frequency doubler reported in [24]. That is just an example illustrating how to face the calculation of the transient behaviour or frequency response of the circuit, where it is essential to know how the intrinsic capacitances are related with the terminal voltages, which is exactly what the presented model does.

In doing so, the prototype GFET is described in Table 3.2. It is a double-gated transistor with $C_t/C_b \approx 185$. A set of independent intrinsic capacitances have been plotted in Figure 3.12a-b as a function of $V_{gs}$ and $V_{ds}$, respectively. A thorough discussion of the terminal charges and capacitances for the different operation regions can be found in section 4.2.3 and in [115], and could be directly applied to these results.

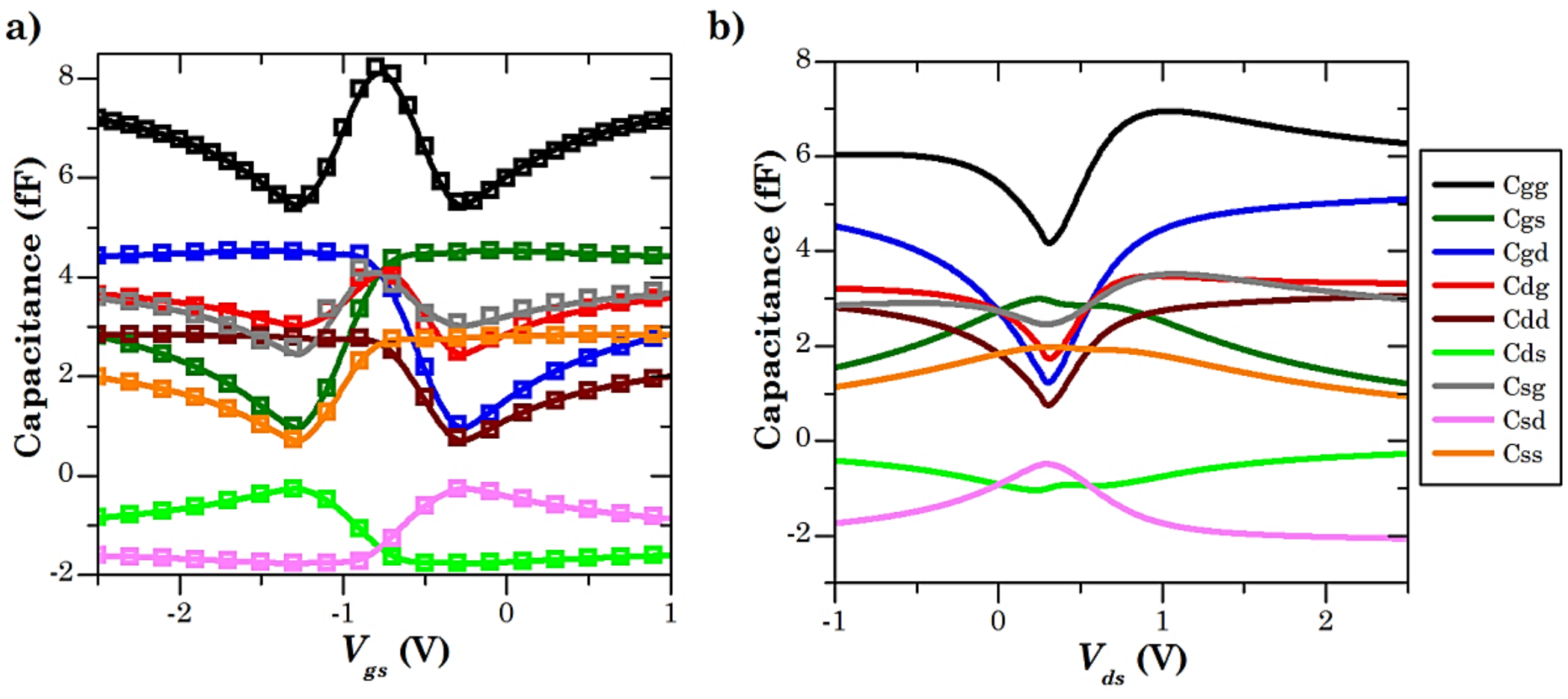

**Figure 3.12** Compact model (solid lines) and numerical (symbols) calculation of the intrinsic capacitances versus a) the gate bias ($V_{ds}$ = 1 V) and b) the drain bias ($V_{gs}$ = 1 V) for the device described in Table 3.2.

**Table 3.2** Input parameters of the GFET used to simulate the device reported in [24].

| Input parameter | Value | Input parameter | Value |
|---|---|---|---|
| $T$ | 300 K | $L$ | 0.5 μm |
| $\mu$ | 1300 $cm^2$/Vs | $W$ | 0.84 μm |
| $V_{g0}$ | -1.06 V | $L_t$ | 5 nm |
| $V_{b0}$ | 0 V | $L_b$ | 300 nm |
| $\Delta$ | 0.140 eV | $\varepsilon_t$ | 12 |
| $\hbar\Omega$ | 0.075 eV | $\varepsilon_b$ | 3.9 |

### 3.5.3 Extrinsic and parasitic elements

The compact model presented in this section is an intrinsic large-signal compact model. In this regard, the extrinsic elements such as: source, drain, top-gate and back-gate resistances, overlap and junction capacitances, probing pads, metal interconnections, including any parasitic capacitances and inductances must be included as lumped elements in the circuit simulator.

## 3.6 Compact model validation: circuit performance benchmarking

The compact large-signal model of the intrinsic GFET is assessed against experimental measurements. For such a purpose it has been embedded in the Cadence Virtuoso Spectre Circuit Simulator [131], which is a widely used general purpose circuit simulator. A Verilog-A version of the compact model is available online at http://ieeexplore.ieee.org in [130], [132].

The benchmarking has been split in two subsections. First, in subsection 3.6.1, the DC and frequency response of a high-frequency voltage amplifier [33] have been assessed. Such a voltage amplifier is a main building block of RF electronics. On the other hand, in subsection 3.6.2, exemplary circuits that take advantage of the graphene ambipolarity as the working principle have been chosen. Specifically, the benchmarking of the DC, transient dynamics, and frequency response of a high performance frequency doubler [24], a radio-frequency subharmonic mixer [28] and a multiplier phase detector [133] have been carried out.

### 3.6.1 High-frequency performance of GFETs

In this subsection, a high-frequency graphene voltage amplifier has been simulated and later compared with experimental results [33]. The GFET consists of a gate stack with an ultrathin high-κ dielectric (4 nm of $HfO_2$, equivalent oxide thickness EOT of 1.75 nm), which has been demonstrated to enhance current saturation [134]. The circuit under test is shown in Figure 3.13, which is a common-source amplifier. The input parameters used for the GFET are described in Table 3.3. The DC TCs and the transconductance are shown in Figure 3.14a. Besides, the DC OCs at various gate biases are depicted in Figure 3.14b.

Figure 3.14c shows the key RF characteristics of the GFET-based voltage amplifier, specifically the current gain and the power gain. The simulated $f_{Tx}$ = 8.7 GHz and $f_{max}$ = 5.4 GHz are in close agreement with the measurements of 8.2 GHz and 6.2 GHz, respectively. Finally, the voltage gain of the amplifier has been assessed (Figure 3.14d). The simulation gives a DC voltage gain of ~ 7.4 dB, which is ~ 20log($g_m g_{ds}^{-1}$), with a 3-dB bandwidth of 6.2 GHz.

**Table 3.3** Input parameters of the GFET-based voltage amplifier reported in [33].

| Input parameter | Value | Input parameter | Value |
|---|---|---|---|
| $T$ | 300 K | $L$ | 500 nm |
| $\mu$ | 4500 $cm^2$/Vs | $W$ | 30 μm |
| $V_{g0}$ | 0.613 V | $L_t$ | 4 nm |
| $\Delta$ | 0.095 eV | $\varepsilon_t$ | 12 |
| $\hbar\Omega$ | 0.12 eV | $R_s \cdot W$, $R_d \cdot W$ | 435 Ω·μm |
| $R_g \cdot L$ | 7 Ω·μm | | |

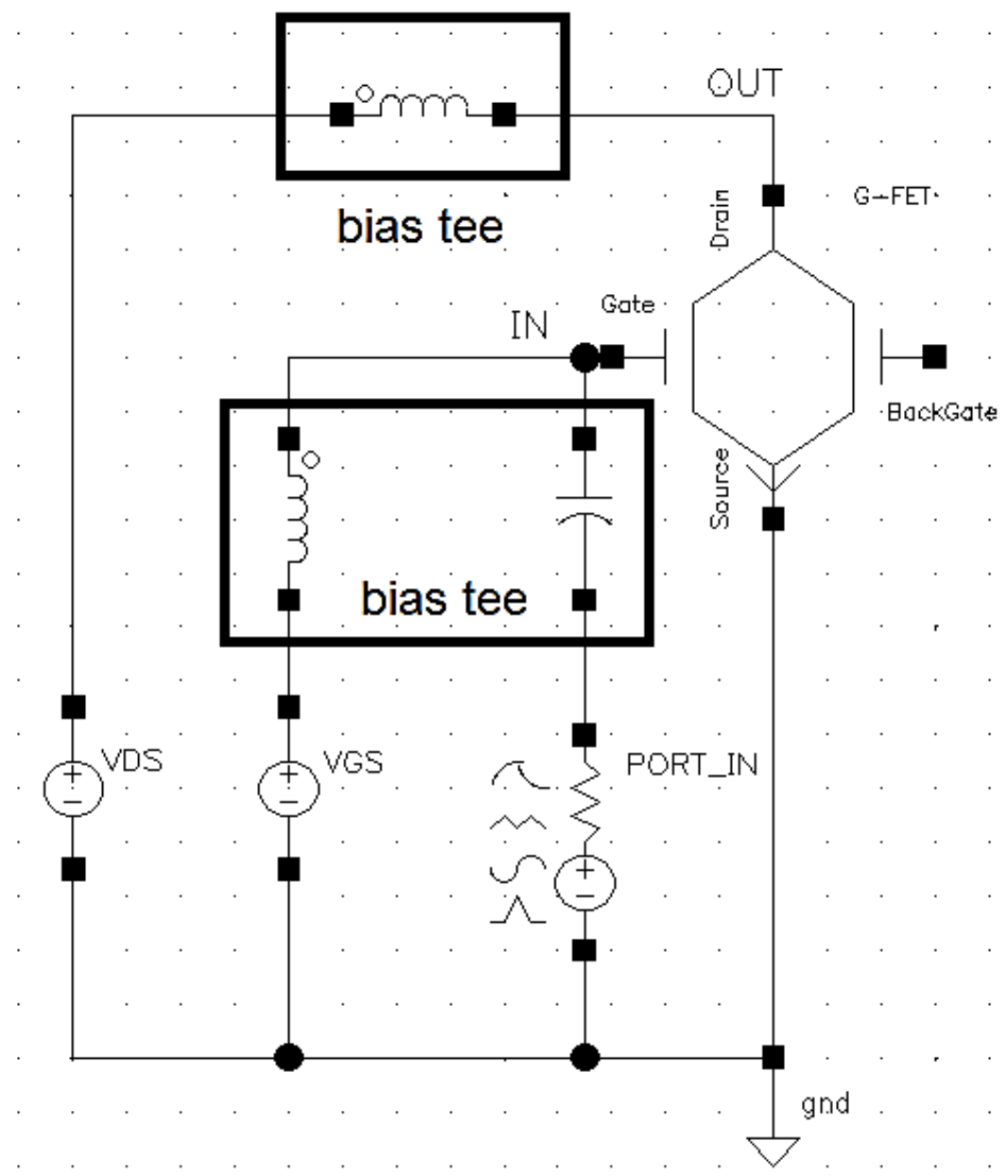

**Figure 3.13** Schematic circuit of the GFET-based voltage amplifier. Bias tees are used for setting the DC bias point. The "G-FET" symbol includes the compact large-signal model implemented in Verilog-A plus the contact ($R_s$, $R_d$) and gate ($R_g$) resistances.

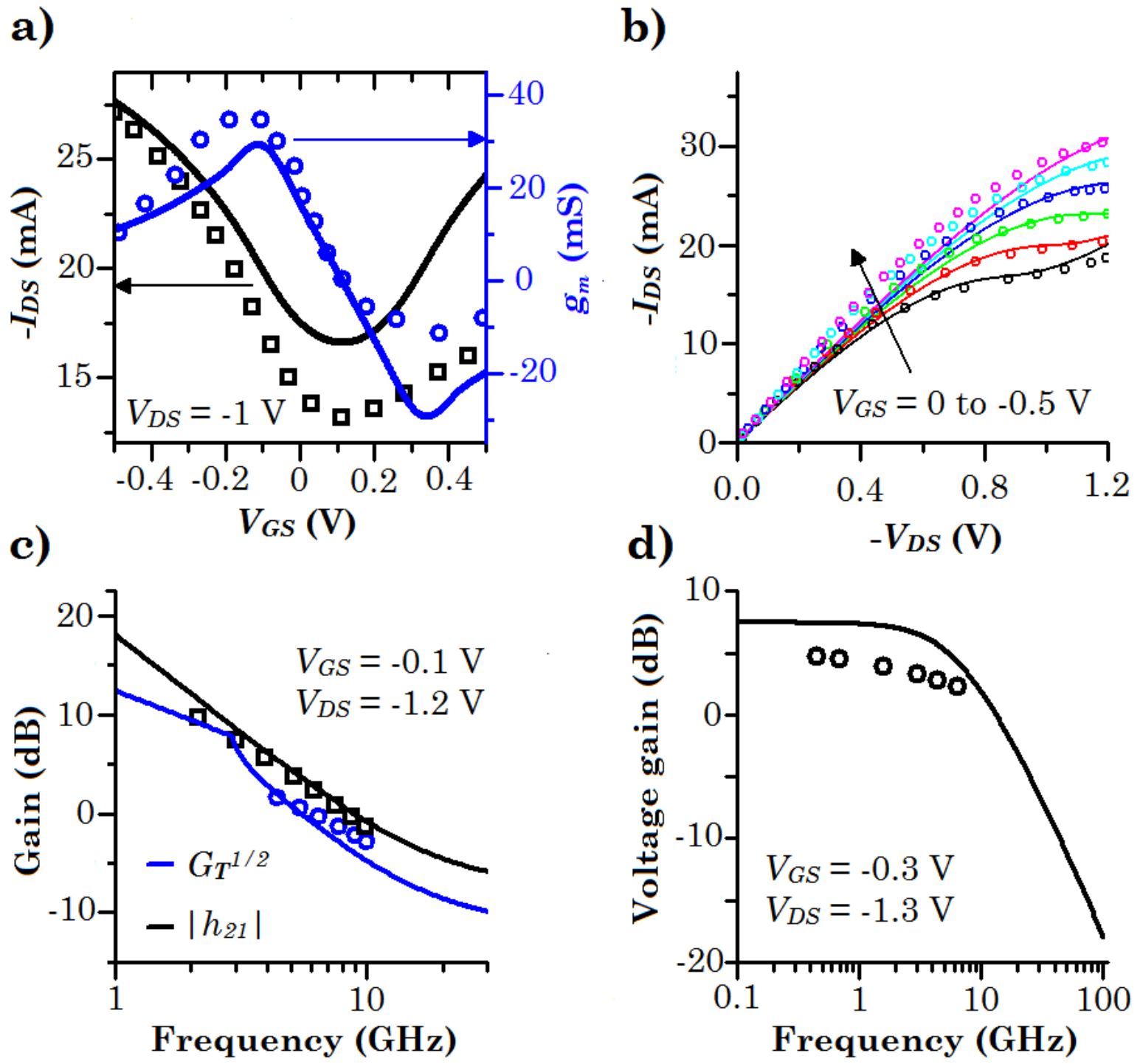

**Figure 3.14** a) DC transfer characteristics and extrinsic transconductance of the GFET-based voltage amplifier. The device is biased at $V_{DS}$ = -1 V. b) DC output characteristics at various gate voltages. c) Power gain ($G_T^{1/2}$) and current gain ($|h_{21}|$) as a function of frequency. $f_{Tx}$ and $f_{max}$ are the frequency at which current gain and power gain becomes unity (0 dB), respectively. d) Frequency response of the amplifier's voltage gain when the input port level is -17 dBm. (Lines correspond to simulations and symbols to experimental data from [33])

### 3.6.2 Graphene-based ambipolar electronics

Ambipolar electronics based on symmetric $I_{ds} - V_{gs}$ relation around $V_{Dirac}$ has attracted lot of attention. The ability to control device polarity allows for (i) simplification of conventional circuits such as frequency multipliers [24]–[26], [43], [135]–[137], RF mixers [27]–[30], [43], [138]–[140], digital modulators [34], [40], [42], phase detectors [133] or active balun architectures [141]; and (ii) new functionalities in both analogue/RF and digital domains. In this subsection, the compact large-signal model has been benchmarked against exemplary ambipolar circuits such as a high performance frequency doubler [24], a radio-frequency subharmonic mixer [28] and a multiplier phase detector [133].

- *Frequency doubler*

The frequency doubler's working principle takes advantage of the quadratic behaviour of the GFET TC, which can be written as:

$$I_{ds} = a_0 + a_2\left(V_{gs} - V_{Dirac}\right)^2 \tag{3.35}$$

where $a_0$ and $a_2$ are appropriate parameters describing the TC. When a small AC signal with an offset $V_{GS} = V_{Dirac}$, namely $V_{in} = V_{GS} + A\sin(\omega t)$, is input to the transistor's gate in the circuit of Figure 3.15, the output voltage $V_{out} = V_{ds}$ results in:

$$V_{out} = V_{DD} - a_0 R_0 - \frac{1}{2}a_2 R_0 A^2 + \frac{1}{2}a_2 R_0 A^2 \cos\left(2\omega t\right) \tag{3.36}$$

where $A$ is the signal amplitude, $\omega = 2\pi f_{in}$ the angular frequency, and $R_0$ a load resistor connected to the drain. The output frequency is double because of the quadratic TC. If the TC was not perfectly parabolic and/or symmetric, which is the practical case, the output voltage would contain, in the former case, other even high order harmonics and, in the latter case, other odd high order harmonics, resulting in harmonic distortion. Examples of frequency doublers can be found in [24], [26], [43], [135]–[137]. Moreover, with a properly adjusted threshold voltage separation of two graphene FETs connected in series, a graphene-based frequency tripler has been demonstrated [25].

Next, the frequency doubler circuit shown in Figure 3.15 is analysed by means of a circuit simulator that includes the large-signal compact model of the GFET. The goal is to benchmark the model's outcome against the experimental data reported in [24]. The input parameters used for the GFET are shown in Table 3.2 with drain and source resistances scaled by the channel width of $R_s \cdot W = R_d \cdot W$ = 1.1 kΩ·µm and gate resistance of $R_g$ = 20 Ω. The DC transfer characteristics and the GFET's transconductance, are shown in Figure 3.16a, with a nearly symmetric shape respect to the Dirac voltage, $V_{Dirac}$ = -1.15 V.

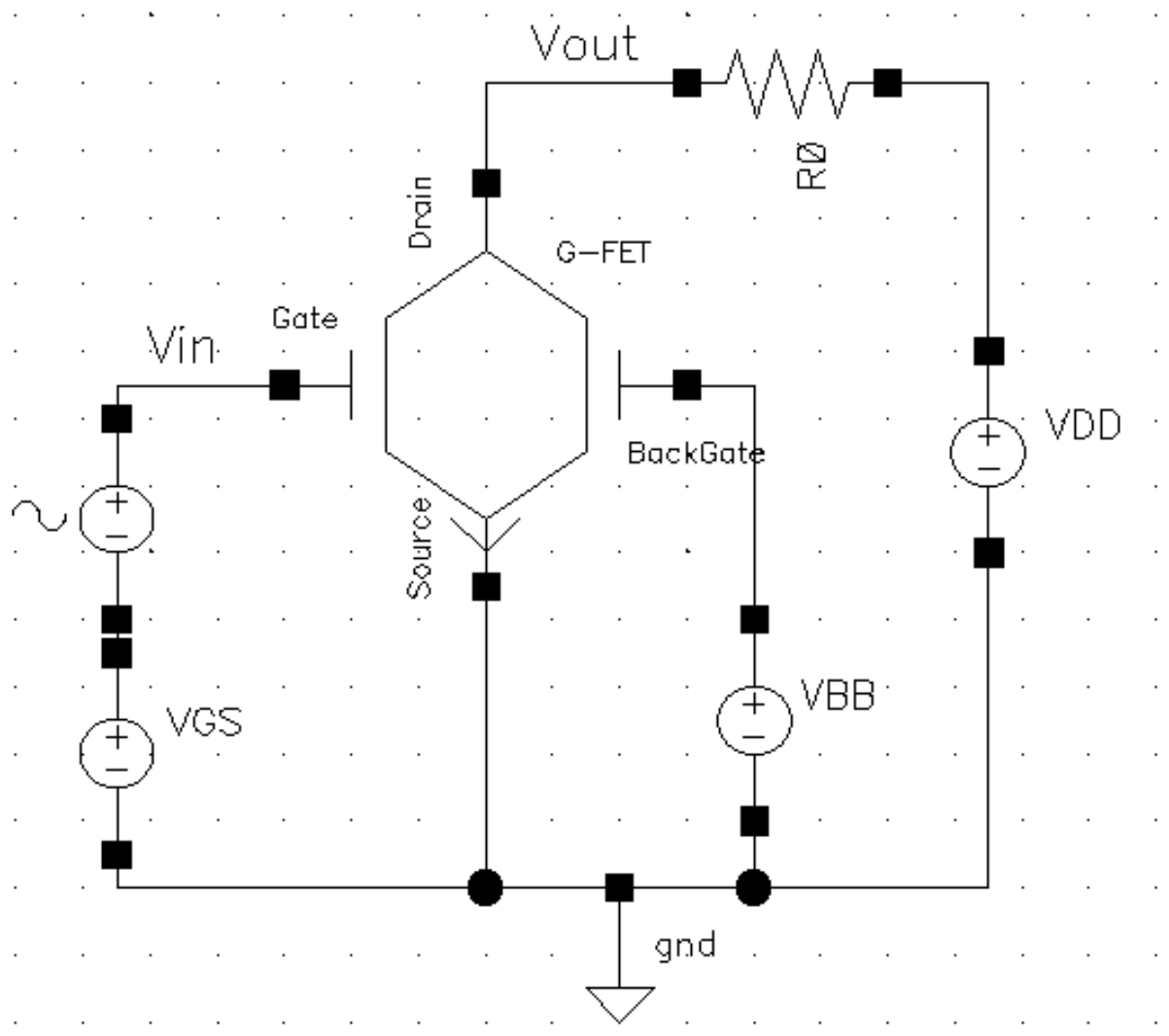

**Figure 3.15** Schematic circuit of the GFET-based frequency doubler. The device is described in Table 3.2.

Using the GFET model, the output waveform has been analysed for different input frequencies, which are shown in Figure 3.16b-d. For the lowest frequency, $f_{in}$ = 10 kHz, the output waveform consists of the doubled frequency with an amplitude ~ $A$/10, with a clear distortion coming from other higher order harmonics (see Figure 3.16b). A Fourier transform of the waveform, shown in Figure 3.17, reveals that 60% of the output RF power is concentrated at the doubled frequency of 20 kHz.

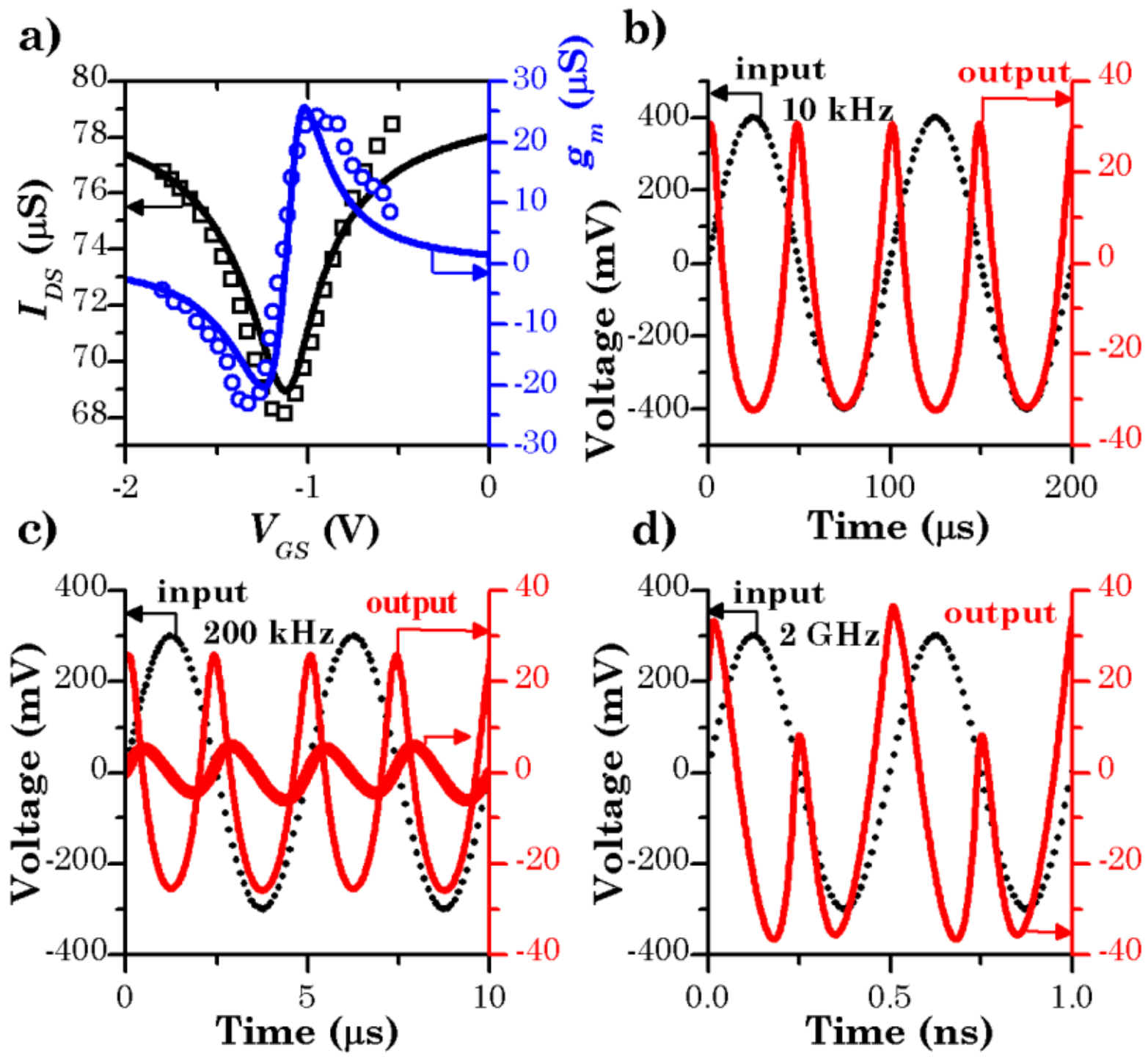

**Figure 3.16** a) DC transfer characteristics and extrinsic transconductance of the GFET-based frequency doubler. The device is biased at $V_{DD}$ = 1 V, $V_{BB}$ = 40 V and $V_{GS}$ = -1.15 V. The description of the device is given in Table 3.2. b) Input and output waveforms considering an input frequency of $f_{in}$ = 10 kHz and amplitude $A$ = 400 mV. c) Input and output waveforms considering an input frequency of $f_{in}$ = 200 kHz and amplitude $A$ = 300 mV. A thicker solid line shows the output waveform when a parasitic capacitance ($C_{pad}$ = 600 pF) is placed between the drain/source and the back-gate, taking into account the effect of the electrode pads reported in [24]. d) Input and output waveforms considering an input frequency of $f_{in}$ = 2 GHz and amplitude $A$ = 300 mV.

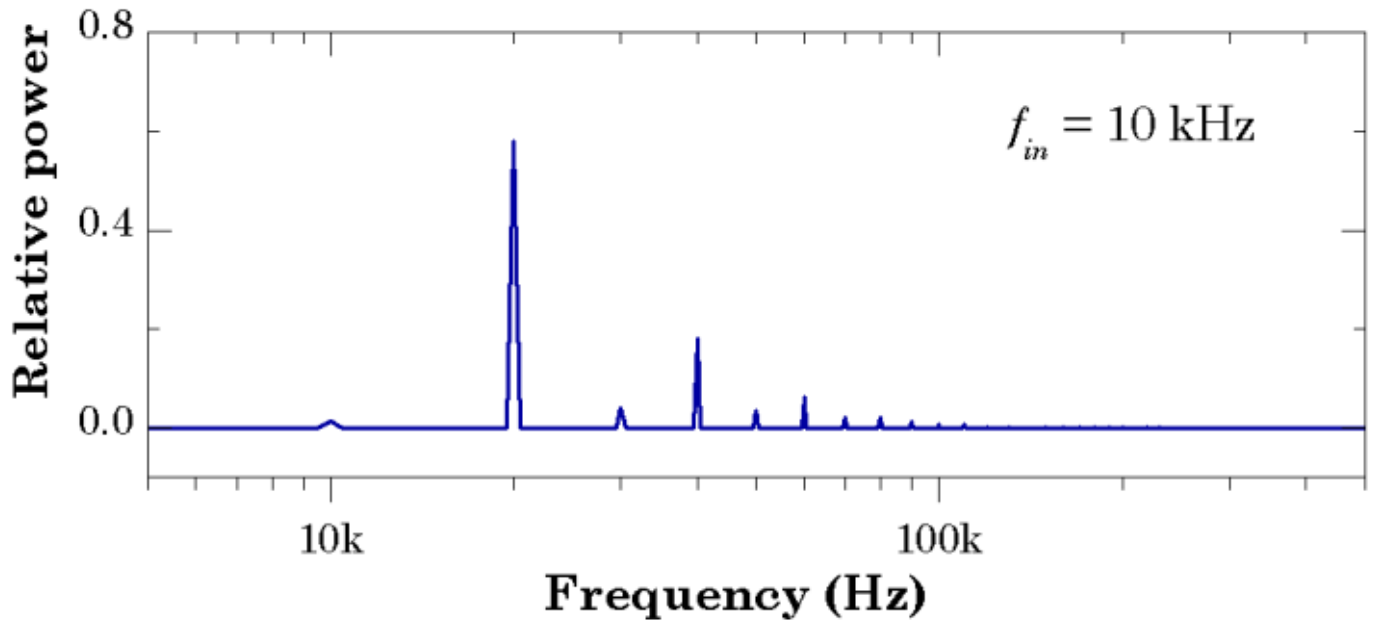

**Figure 3.17** Power spectrum obtained via Fourier transforming the output signal in Figure 3.16b.

When the input signal is increased up to $f_{in}$ = 200 kHz and beyond a severe decay of the output signal amplitude was observed in the experiment, with a voltage gain ~ $A$/100 [24], likely because of the presence of a parasitic capacitance $C_{pad}$ = 600 pF between the GFET source/drain terminals and its back-gate, getting a similar output waveform as in the experiment for an input frequency of 200 kHz (see Figure 3.16c). If the input frequency is further increased up to 2 GHz the output waveform, shown in Figure 3.16d, displays the doubled frequency, although with a greater distortion because the group delay is not constant with the frequency according to Figure 3.18, meaning that the phase is not linear with the frequency. To achieve high efficiency gigahertz frequency multipliers the parasitic capacitances must be diminished. Besides, these non-idealities must be incorporated to the device model to make realistic predictions on the performance of high-frequency circuits.

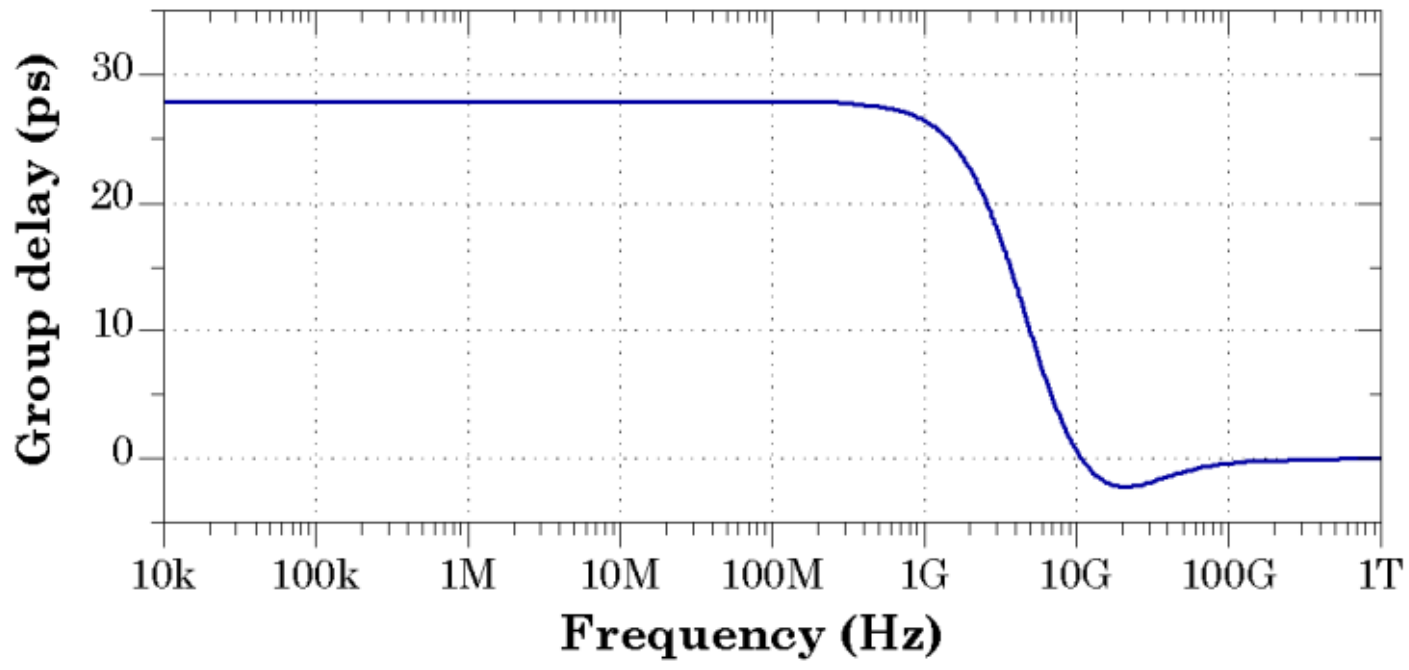

**Figure 3.18** Group delay vs. frequency for the GFET-based frequency doubler.

- *RF mixer*

In telecommunications, a mixer is a non-linear device that receives two different frequencies (the local oscillator LO signal at $f_{LO}$ and the radio-frequency RF signal at $f_{RF}$) at the input port and a mixture of several frequencies appears at the output, including both original input frequencies, the sum of the input frequencies, the difference between the input frequencies (the intermediate frequency IF signal at $f_{IF}$), and other intermodulations [142]. There are basically two operating principles for a FET mixer; either utilizing the change in transconductance, $g_m$, or channel conductance, $G_{ds}$ (= $I_{ds}/V_{ds}$), with the gate voltage. In both approaches a LO signal is applied to the gate to achieve a resulting time-varying, periodic quantity $g_m(t)$ or $G_{ds}(t)$. The former case is referred to as an active transconductance mixer, where the RF signal is applied to the gate, and the latter a resistive mixer, with the RF signal applied to the drain [140].

**Table 3.4** Input parameters of the GFET used to simulate the subharmonic graphene-based mixer reported in [28].

| Input parameter | Value | Input parameter | Value |
|---|---|---|---|
| $T$ | 300 K | $L$ | 1 μm |
| $\mu$ | 2200 cm²/Vs | $W$ | 20 μm |
| $V_{g0}$ | 1 V | $L_t$ | 25 nm |
| $V_{b0}$ | 0 V | $L_b$ | 300 nm |
| $\Delta$ | 0.116 eV | $\varepsilon_t$ | 9 |
| $\hbar\Omega$ | 0.075 eV | $\varepsilon_b$ | 3.9 |
| $R_s \cdot W$, $R_d \cdot W$ | 560 Ω·μm | $R_g \cdot L$ | 10 Ω·μm |

On the one hand, best possible performance from a transconductance mixer is realized by maximizing the variation in $g_m$, which is accomplished by biasing the FET in the saturation region. Examples of graphene-based transconductance mixers can be found in [27], [43]. However, as a consequence of the currently low transconductance in GFETs and the weak current saturation, the so far reported graphene-based transconductance mixers have shown poor performance. Instead, it does seem better to use the resistive mixing concept combined with the unique properties of graphene allowing for the design of subharmonic mixers with a single FET. The mixer operation is based on a sinusoidal LO signal also applied to the gate of the GFET, biased at the Dirac voltage. The idea is to make a frequency doubler operation with the LO signal, but keeping the drain unbiased. Thus, the conductance variation as seen from the drain, $G_{ds}(t)$ would have a fundamental frequency component twice as $f_{LO}$. Therefore, a subharmonic mixer only needs half the LO frequency compared to a fundamental mixer. This property is attractive particularly at high-frequencies where there is a lack of compact sources providing sufficient power [143]. Moreover, subharmonic mixers suppress the LO noise [144], and the wide frequency gap between the RF and LO signals simplifies the LO and RF separation [145]. Examples of resistive mixers without subharmonic operation are reported in [30], [138], [139] and examples of resistive subharmonic mixers can be found in [28], [29], [140]. Besides, due to near symmetrical ambipolar conduction, graphene-based mixers can effectively suppress odd-order intermodulations, which are often present in conventional unipolar mixers and are harmful to circuit operations [146].

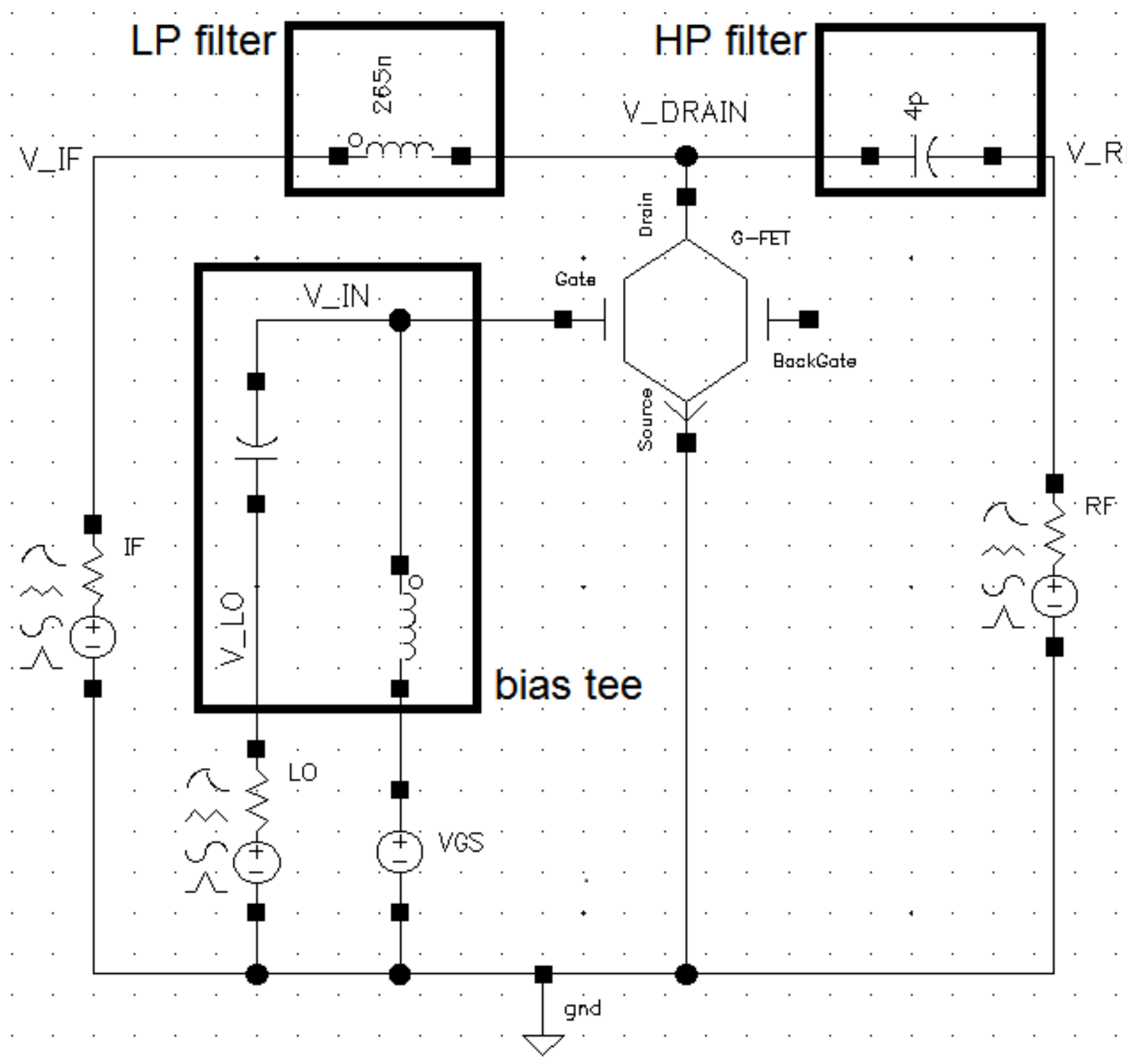

**Figure 3.19** Schematic circuit of the subharmonic resistive GFET mixer. A bias tee is used for setting the DC bias point. The characteristic impedance of the ports is $Z_0 = 50\ \Omega$.

The compact GFET model has been used to simulate the subharmonic resistive mixer circuit shown in Figure 3.19. The goal is to benchmark the model's outcome against the experimental data reported in [28]. The input parameters used for the GFET are shown in Table 3.4. The circuit under test only uses a transistor and no balun is required in that implementation, which makes the mixer more compact, as opposed to conventional subharmonic resistive FET mixers, which require two FETs in a parallel configuration, including a balun for feeding the two out-of-phase LO signals [147], [148]. In the subharmonic mixer, the RF signal is applied to the drain of the GFET through a high-pass filter and the IF is extracted with a low-pass filter, both designed with cut-off frequencies of 800 MHz and 30 MHz, respectively.

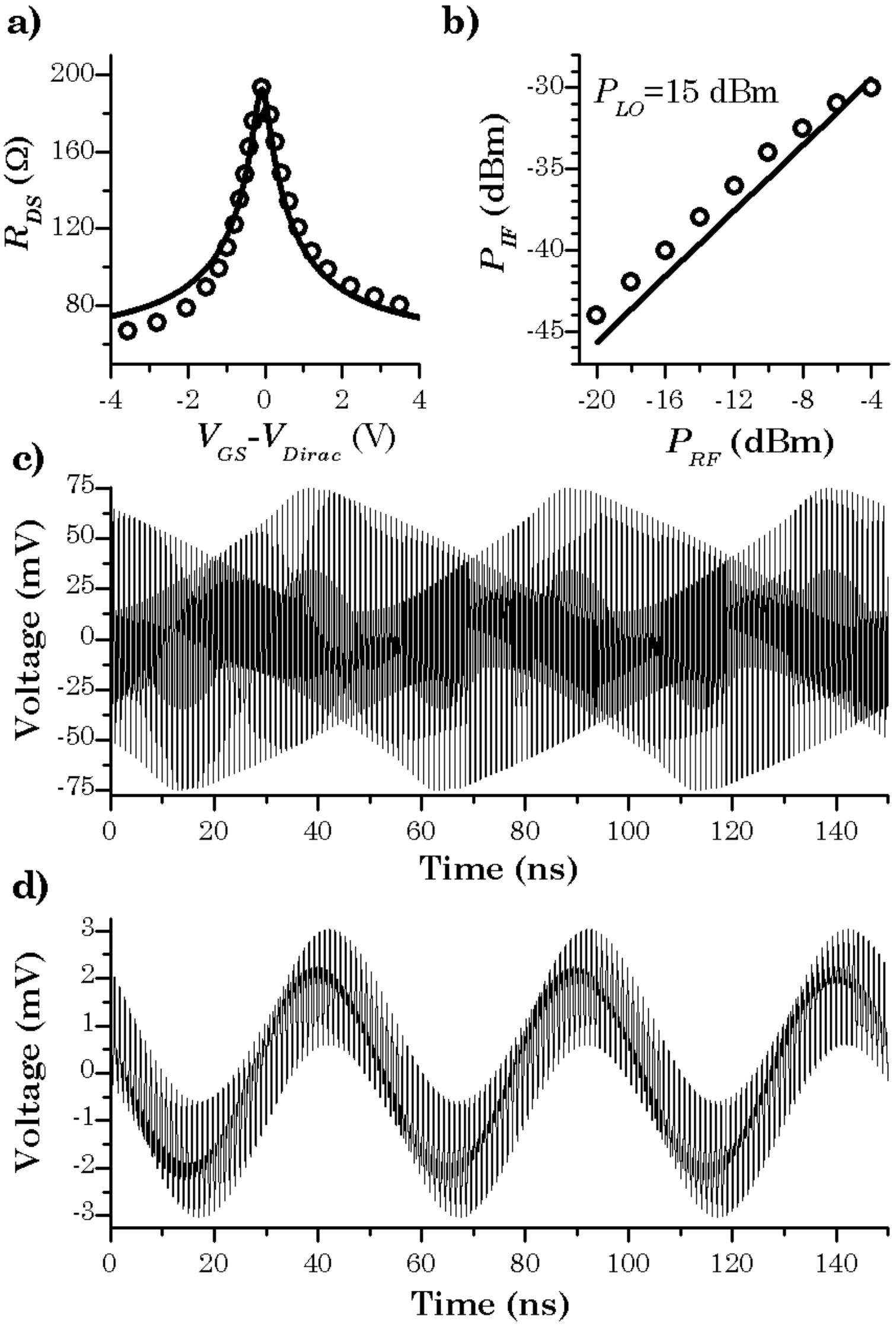

**Figure 3.20** a) Drain-to-source resistance $R_{DS} = 1/G_{DS}$ versus the gate voltage $V_{GS}$, with $R_{DS} = R_d + R_s + R_{ch}$, where $R_{ch}$ is the channel resistance and $R_d$ and $R_s$ are the extrinsic contact resistances at the drain and source sides. Solid lines correspond to simulations and the symbols to the experimental results in [28]. b) IF output power as a function of the RF input power. The device is biased at $V_{GS} = V_{Dirac}$ and $P_{LO}$ = 15 dBm. c) Transient evolution of the signal collected at the drain at $V_{GS} = V_{Dirac}$. The following conditions have been assumed: $P_{LO}$ = 15 dBm and $f_{LO}$ = 1.01 GHz; $P_{RF}$ = -20 dBm and $f_{RF}$ = 2 GHz. d) Transient evolution of the IF signal collected at the IF port under the same conditions as in c). The separation between peaks is 50 ns, which corresponds to $f_{IF} = |f_{RF} - 2f_{LO}|$ = 20 MHz.

The drain-to-source resistance $R_{DS} = 1/G_{DS}$ versus the gate voltage is shown in Figure 3.20a. The device has been biased at $V_{GS} = V_{Dirac}$ = 1 V through a bias tee. The RF signal has been introduced to the RF port connected to the drain and the LO signal has been introduced to the LO port connected to the gate through the bias tee, where the IF signal is collected at the IF port, according to the schematics shown in Figure 3.19. Figure 3.20b depicts the mixer IF output power versus the RF input power, where a near constant conversion loss rate of ~ 25 dB has been obtained. The transient evolution of the signal collected at the drain is shown in Figure 3.20c, as well as the signal collected at the IF port which oscillates as expected frequency of $f_{IF} = |f_{RF} - 2f_{LO}| = 20$ MHz is shown in Figure 3.20d. Finally, the spectrum of the signal collected at the drain is represented in Figure 3.21, being the output power of ~ -49 dBm. Lower levels of odd harmonics are observed as well, which are attributed to the non-perfect symmetry of $R_{DS}$ versus $V_{GS}$.

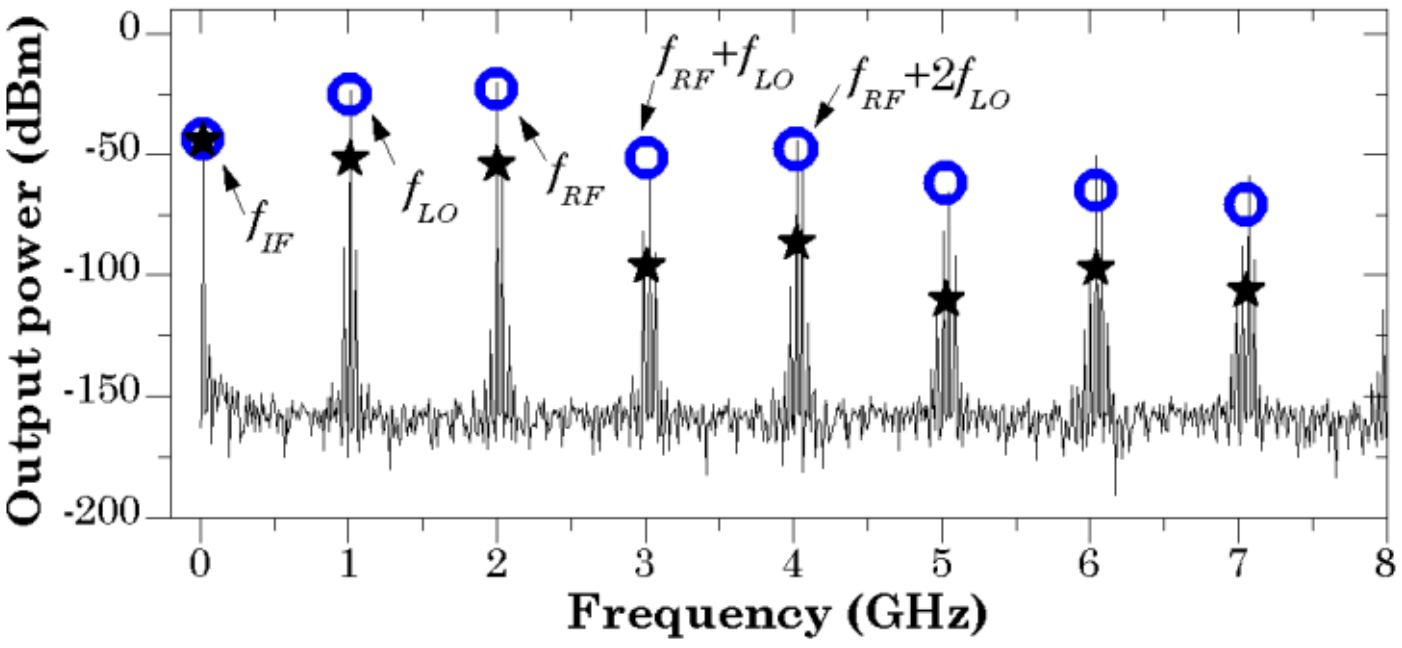

**Figure 3.21** Spectrum (solid lines) of the signal collected at the drain ($P_{LO}$ = 15 dBm and $f_{LO}$ = 1.01 GHz; $P_{RF}$ = -20 dBm and $f_{RF}$ = 2 GHz). The bubbles correspond to the experimental results in [28]; and the stars correspond to the power peaks of the signal collected at the IF port.

- *Multiplier phase detector*

The multiplier phase detector is a vital component of the phase-locked loop, which is one of the most important building blocks in modern analogue, digital, and communication circuits [149].

Upon application of a sinusoidal wave $A_1\sin(\omega t+\theta_1)$ and a square wave $A_2\text{rect}(\omega t+\theta_2)$ to the input of a phase detector, the DC component of the output can be written as the product of the two input signals [133]:

$$A_d = A_1 A_2 \frac{2}{\pi}\sin\left(\theta_1 - \theta_2\right) \simeq K_d \theta_e \tag{3.37}$$

where $K_d$ is the gain of the detector and $\theta_e$ is the phase difference in radians between the input signals. Hence, the relation between the DC component and the phase difference can be utilized for phase detection. A multiplier is generally needed for this process, which complicates the circuit. However, taking advantage of the ambipolarity of a GFET, the simplified circuit structure shown in Figure 3.22 is enough to perform the phase detection.

**Table 3.5** Input parameters of the GFET used to simulate the phase detector reported in [133]

| Input parameter | Value | Input parameter | Value |
|---|---|---|---|
| $T$ | 300 K | $L$ | 1.28 μm |
| $\mu$ | 2100 cm²/Vs | $W$ | 2.98 μm |
| $V_{g0}$ | 0.495 V | $L_t$ | 23 nm |
| $V_{b0}$ | 0 V | $L_b$ | 300 nm |
| $\Delta$ | 0.074 eV | $\varepsilon_t$ | 9.35 |
| $\hbar\Omega$ | 0.075 eV | $\varepsilon_b$ | 3.9 |
| $R_s \cdot W$, $R_d \cdot W$ | 4.3 kΩ·μm | $R_g \cdot L$ | 38.5 Ω·μm |

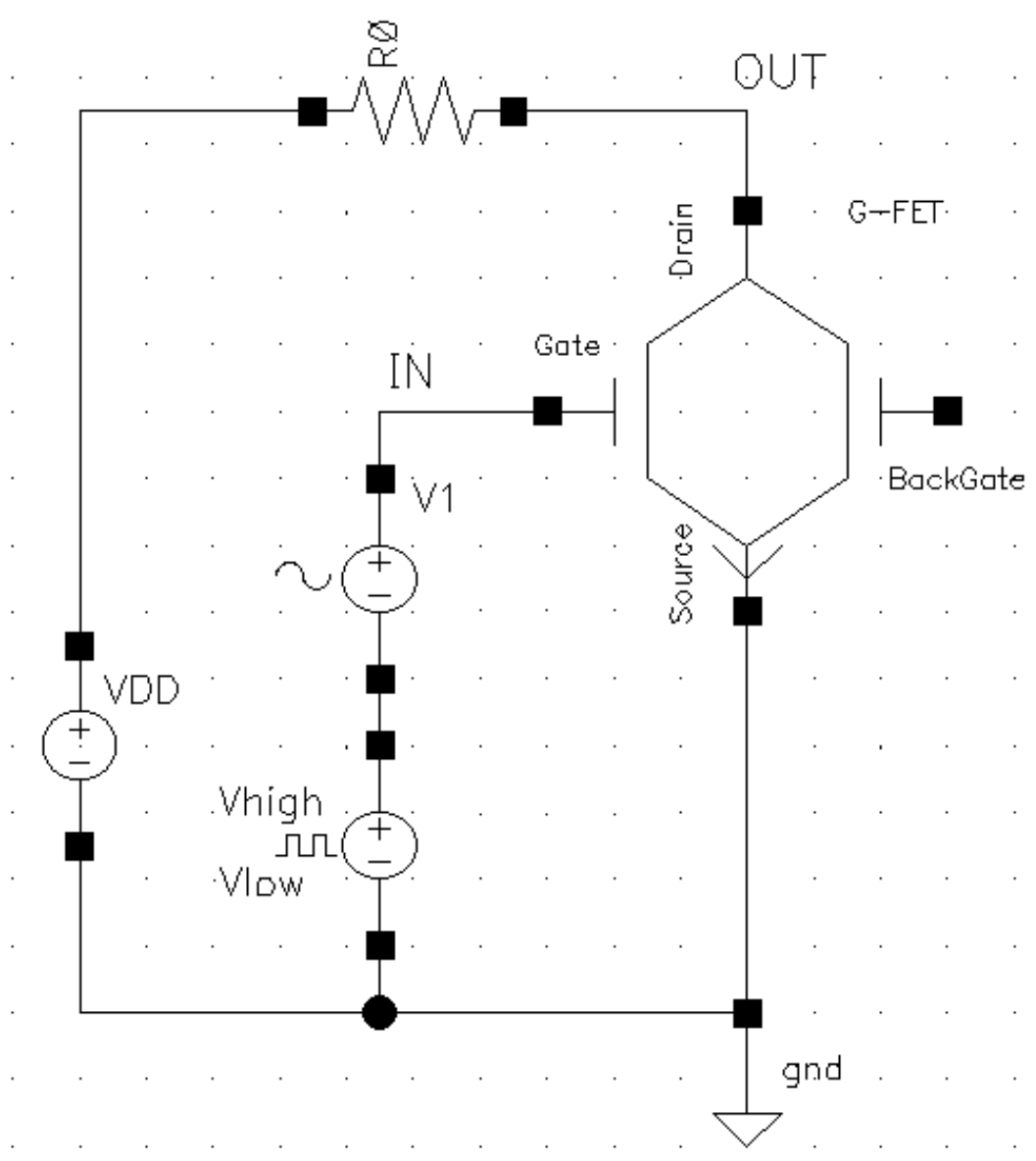

**Figure 3.22** Schematics of the multiplier phase detector based on a single graphene transistor and a load resistor.

Next, the GFET compact model has been used to simulate the phase detector circuit shown in Figure 3.22 with the goal of benchmarking the model outcome against the experimental data reported in [133]. The input parameters used for the GFET are shown in Table 3.5. The DC TCs and GFET's transconductance at $V_{DS}$ = 0.1 V are shown in Figure 3.23a. The device shows a nearly symmetric characteristic around the Dirac voltage ($V_{Dirac}$ = 0.55 V). Then, the GFET is biased at $V_{DD}$ = 1.8 V through a series resistor $R_0$ = 20 kΩ, according to the schematics shown in Figure 3.22. The back-gate has been assumed disconnected, as in [133]. A square-wave signal is used as the gate bias voltage, where the low level ($V_{low}$ = 0.36 V) and high level ($V_{high}$ = 0.82 V), satisfy $V_{low} < V_{Dirac}$ and $V_{high} > V_{Dirac}$. Both levels match with the two $g_m$ peaks so to get the maximum voltage gain. A sinusoidal-wave signal of 0.1 V amplitude oscillates around the two levels of the square-wave signal. Both the signals have 100 kHz of frequency, thus resulting in the following combined gate input signal:

$$v_{IN} = 0.1\sin\left(2\pi 10^5 t + \theta_1\right) + \left[0.46\mathrm{rect}\left(2\pi 10^5 t + \theta_2\right) + 0.36\right] \tag{3.38}$$

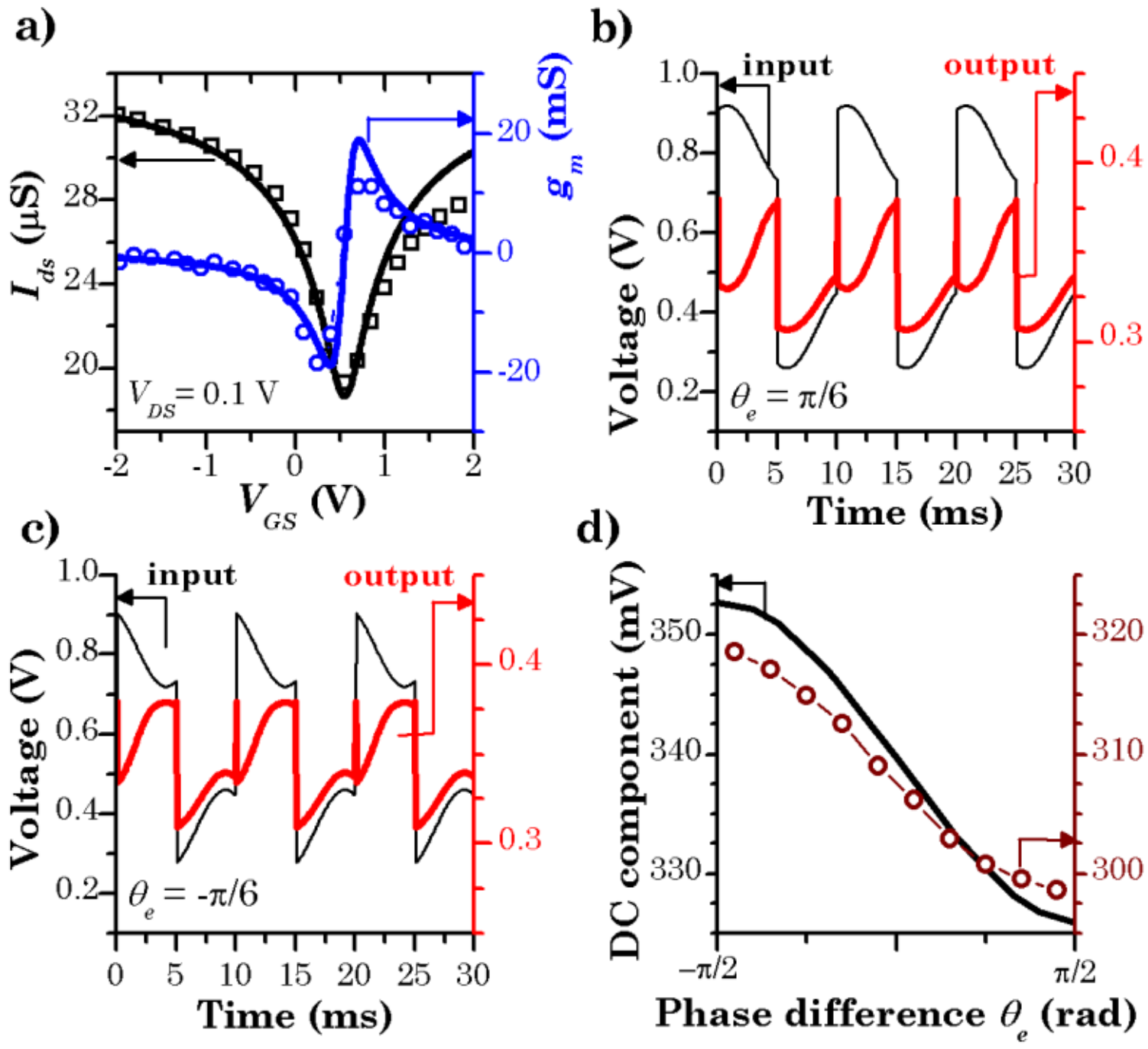

**Figure 3.23** a) Experimental (symbols) and simulated (solid lines) DC transfer characteristics and extrinsic transconductance of the device described in Table 3.5. The drain bias has been set to $V_{DS}$ = 0.1 V. b, c) Simulated input and output waveforms of the phase detector circuit shown in Figure 3.22, biased at $V_{DD}$ = 1.8 V, where a phase difference (b) $\theta_e$ = π/6 and (c) $\theta_e$ = -π/6 has been assumed. The transient responses are quite similar to the data reported in [133]. d) Experimental (symbols) and simulated (solid line) output DC component versus the phase difference $\theta_e$.

In Figure 3.23b-c the transient response of the multiplier phase detector circuit has been depicted, assuming both $\theta_e$ = π/6 and $\theta_e$ = -π/6, respectively, which looks very similar to the experimental results. The circuit corresponds to a common-source amplifier, therefore, the voltage gain could be estimated as $A_v \approx -g_m(g_{ds}^{-1} || R_0)$. It is approximately 0.1, which agrees with the reported value in [133]. Finally, in Figure 3.23d, the output DC component is shown for different $\theta_e$. As the phase difference goes from -π/2 to π/2 rad, the DC component decreases from 353 to 326 mV, which corresponds to a detector gain of $K_d \approx$ -8.6 mV/rad, which can be further improved by combining a reduction of the series resistance, increasing the gate efficiency (increase $g_m$), and pushing the transistor to the saturation region (reducing $g_{ds}$).

## 3.7 Conclusions

This chapter has first introduced the motivation of using graphene for analogue/high-frequency electronics, which is rooted in its ultrahigh carrier mobility and saturation velocity. Given that, several GFET-based circuits working at RF have already been demonstrated, thus modelling is becoming increasingly important to make circuit design and validation more systematic to push up the TRL. For such a purpose, the goal of this chapter has been the development of an intrinsic compact large-signal model of GFETs suitable for conventional circuit simulators. In doing so, a brief review of the electronic properties of single layer graphene has been presented first, followed by the electrostatics analysis of a GFET-based structure. Then, a drain current model and a charge-based intrinsic capacitance model have been proposed assuming a field-effect model and DD carrier transport. Taking all in consideration, a numerical large-signal model of GFETs has been built which allows for performance assessment, benchmarking against other technologies and providing guidance for device design.

However, for the model to be used in ordinary EDA tools making circuit simulations possible, compact modelling techniques have been applied to turn the numerical model into a compact one. Then, the large-signal compact model of GFETs implemented in Verilog-A has been benchmarked against high-performance and ambipolar electronics' circuits. Specifically, high-frequency voltage amplifiers, high performance frequency doublers, radio-frequency subharmonic mixers and multiplier phase detectors have been considered. The agreement between experiment and simulation is quite good in general, although fine adjustment would require further modelling. The intrinsic description of the device given in this thesis serves as a starting point toward a complete GFET model which could incorporate additional non-idealities. Among them, the parasitic effects such as parasitic capacitances, inductances taking into account effects of the probing pads and metal interconnections must be included. A common modelling approach for RF applications is to build subcircuits that include the parasitic elements and connecting them to the intrinsic GFET. These subcircuits should also be linked to process and geometry information to guarantee scalability and prediction capability of the model. For instance, the inclusion of the voltage-dependent contact and access resistances is crucial for getting accurate DC and RF performance predictions. Moreover, it has been realized that an accurate and physical description of mobility is essential for distortion analysis [49]. Further inclusions of many important physical effects such as short-channel and narrow width effects, trapped charge, etc., could be also important. Moreover, the model should correctly predict the HF noise, which is important for the design of, for example, low noise amplifiers. The model should also include *non-quasi-static* (NQS) effects, so it can properly describe the device behaviour at very high-frequencies where the *quasi-static* assumption could break down.

Chapter 4

# Large-signal modelling of bilayer graphene based FETs

The gapless nature of SLG prevents the gate voltage to switch off the transistor, so it is not a suited material for logic applications. However, it is believed that graphene could play a relevant role in analogue high-frequency electronics because of its high carrier mobility and saturation velocity. As mentioned in section 1.1, $f_{Tx}$ up to 427 GHz [56] and $f_{max}$ of 200 GHz [45] have been demonstrated. That maximum oscillation frequency is still low in comparison with other existing technologies because of the absence of a bandgap in graphene prevents proper current saturation, especially at the required short gate lengths. Thus, introducing a bandgap does seem necessary. In this regard, different approaches to open an energy bandgap to graphene have been proposed [17]. An interesting possibility is to get the bandgap through size quantization. That is feasible using graphene nanoribbons (GNRs) [150], [151] for which gaps up to 2.3 eV have been demonstrated [152]. However, the production of GNRs is challenging as advanced lithographic techniques are required to produce narrow ribbons with smooth edges. A second alternative to open an energy bandgap would be to apply strain on the SLG. Raman spectrum studies of strained graphene have shown that a tunable energy bandgap of up to 300 meV can be achieved by applying a 1 % uniaxial strain [153]. A third interesting possibility is offered by BLG, where a gap can be induced either by molecular doping [154], [155] or by applying an external electric field perpendicular to the BLG, which allows to tune the gap with the gate bias [156]–[158].

Among the above-mentioned alternatives, this thesis explores the BLG used as the active part of the transistor and, more specifically, the modelling of the BLG-based FET (BLGFET). In this regard, several models have been developed so far, e.g. Ryzhii *et al.* presented an analytical one based on the Boltzmann kinetic equation and Poisson equation in the weak nonlocality approximation [159], [160]; Cheli *et al.* proposed an analytical model based on the effective mass approximation to calculate the thermionic and interband tunneling components of the current under the ballistic transport assumption [161]; Ghobadi and Abdi investigated the device characteristics by calculating the transmission coefficient through a tight-binding method [162]; and Fiori and Iannaccone carried out a study of the main RF FoMs of a BLGFET through the NanoTCAD ViDES simulator, based on the self-consistent solution of the three-dimensional Poisson and Schrödinger equations by means of the non-equilibrium Green's function formalism [163]. The ballistic assumption in which all these models rely on seems unrealistic for the prototype devices explored so far, which do not fulfil the condition $L << \lambda$, where $L$ refers to the transistor channel length and $\lambda$ is the so-called mean free path. The latter has been estimated as $\lambda \approx 10$ nm at carrier densities of $3 \times 10^{12}$ cm$^{-2}$ for exfoliated BLG deposited on a 300 nm $SiO_2$ substrate and at low temperatures [100]. Hence it is worth reconsidering the carrier transport issue under the light of a DD theory when dealing with the practical situation $L >> \lambda$. So in this chapter, a numerical physics-based large-signal model considering the DD transport approach for the drain current, charge and capacitance of dual-gated BLGFETs is presented [164], pursuing the following goals: (i) understanding of electronic properties of BLG and analysis of the special feature: the tunable bandgap; (ii) evaluating the impact of the bandgap opening on the RF FoMs comparing with the SLG counterpart, (iii) performance assessment and benchmarking against other existing technologies, and (iv) provide guidance for device design. Moreover, few compact models for BLGFETs have been proposed suited to be included even in a standard EDA tool [165]–[167]. However, the reported BLGFET models proposed so far are static models, so they do not provide the circuit dynamics and frequency response, which requires a proper device's charge and capacitance modelling. This is done in section 4.1, where a complete numerical large-signal model of the BLGFET is presented, which gives an appropriate description of the current, charge and capacitances.

The model starts by considering the device electrostatics. For such a purpose the 2D-DOS of BLG has been extracted from an effective two-band Hamiltonian at low energy. Upon application of 1D Gauss's law to the gate stack, the carrier concentration in the bilayer graphene channel can be determined as a function of the applied gate bias. Next, the carrier transport has been considered under the DD approach from which the drain current model can be formulated. Based upon it, the charge associated to each transistor's terminal and a complete capacitive model, guaranteeing charge conservation, have been derived as a final step. Just to be sure that the model captures the experimental evidence, it has been validated against reported experimental results in section 4.2. What is more, main FoMs have been projected to illustrate the feasibility of using BLG in HF electronics. Final conclusions are given in section 4.3.

## 4.1 Numerical modelling of BLGFETs

Taking advantage of the physics behind the bilayer graphene requires a basic understanding of the electrical properties. This section presents a review of the electronic properties of bilayer graphene. To model the drain current, the physical framework considered has been a field-effect approach and DD carrier transport incorporating saturation velocity effects. Using such a physical framework as a basis, the charges and capacitances have been derived guaranteeing charge conservation. The device considered is a four-terminal dual-gate transistor. The bandgap is proportional to the perpendicular electric field, which is directly controlled by the double gate stack. The model is of special interest for analogue and radio-frequency applications.

### 4.1.1 Electronic properties of BLG

Bilayer graphene consists of two coupled monolayers of carbon atoms, each with a honeycomb crystal structure with inequivalent sites $A1$, $B1$ and $A2$, $B2$ on the bottom and top graphene sheets, respectively, arranged according to Bernal AB-stacked (the lower layer $B1$ is directly below an atom, $A2$, from the upper layer) as shown in Figure 4.1. The reciprocal lattice is an hexagonal Bravais lattice, and the first Brillouin zone is an hexagon [107].

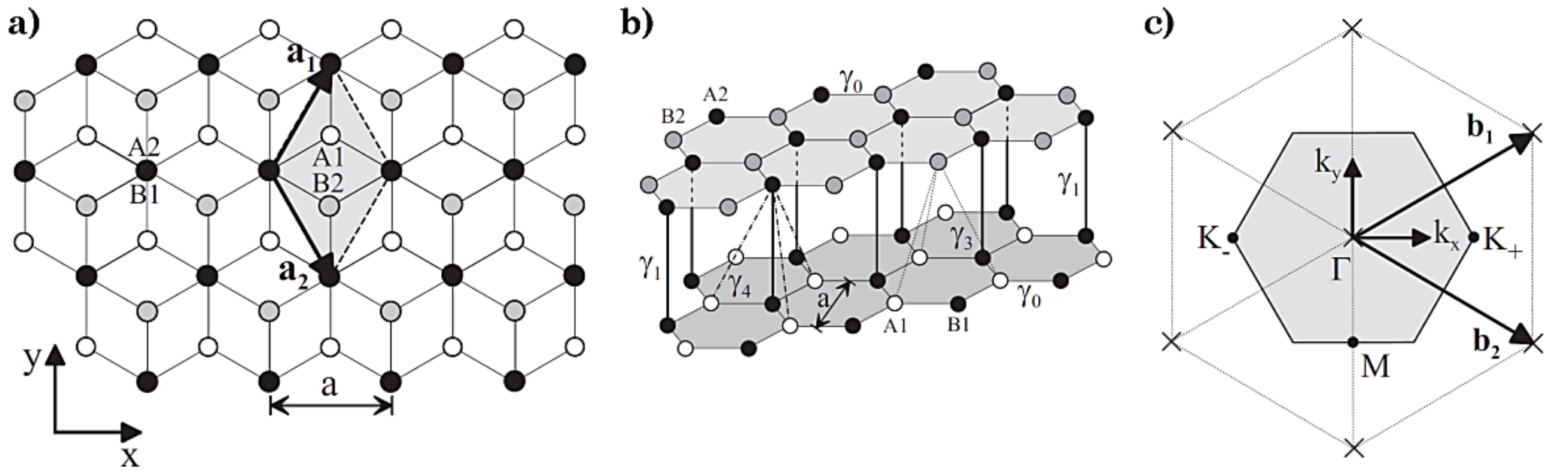

**Figure 4.1** a) Plan and b) side view of the crystal structure of BLG. Atoms $A1$ and $B1$ on the lower layer are shown as white and black circles; $A2$, $B2$ on the upper layer are black and grey, respectively. The shaded rhombus indicates the conventional unit cell; $a_1$ and $a_2$ are primitive lattice vectors. c) Reciprocal lattice of bilayer graphene with lattice points indicated as crosses is shown; $b_1$ and $b_2$ are primitive reciprocal lattice vectors. The shaded hexagon is the first Brillouin zone with Γ indicating the centre, and $K_+$, $K_-$ showing two non-equivalent corners. (Image taken from [107])

In order to compute the electronic band structure of the BLG, the tight-binding model will be described by adapting the Slonczewski-Weiss-McClure parametrization [168] of relevant couplings, taking into account $2p_z$ orbitals on the four atomic sites in the unit cell, labelled as $A1$, $B1$, $A2$, $B2$. In-plane hopping is parameterized by coupling $\gamma_{A1B1} = \gamma_{A2B2} \equiv \gamma_0$ and it leads to the in-plane velocity or Fermi velocity $v_F = (3a\gamma_0/2\hbar)$, where $a$ is the graphene lattice constant. In addition, the strongest interlayer coupling $\gamma_{A2B1} \equiv \gamma_1$ between pairs of orbitals on dimer sites $A2 - B1$ is taken into account, leading to the formation of high energy bands. Parameter $\gamma_{A1B2} \equiv \gamma_3$ that describes interlayer coupling between non-dimer orbitals and parameter $\gamma_{A1A2} = \gamma_{B1B2} \equiv \gamma_4$ that describes interlayer coupling between dimer and non-dimer orbitals (all parameters are shown in Figure 4.1), are not considered in this work because their influence is weak respect to the other couplings. The following Hamiltonian is written near the centres of the valleys [107]:

$$\mathcal{H} = \begin{pmatrix} \epsilon_{A1} & v_F \pi^{\dagger} & 0 & 0 \\ v_F \pi & \epsilon_{B1} & \gamma_1 & 0 \\ 0 & \gamma_1 & \epsilon_{A2} & v_F \pi^{\dagger} \\ 0 & 0 & v_F \pi & \epsilon_{B2} \end{pmatrix} \tag{4.1}$$

where $\pi = \xi p_x + ip_y$, $\pi^{\dagger} = \xi p_x - ip_y$, $p = (p_x, p_y)$ is the momentum measured with respect to the $K$ point, $\xi$ =+1(-1) labels valley $K_+$ ($K_-$). Parameters $\epsilon_{A1}$, $\epsilon_{B1}$, $\epsilon_{A2}$ and $\epsilon_{B2}$ describe the on-site energies on the four atomic sites, that are not equal in the most general case.

At zero magnetic field, Hamiltonian yields four valley-degenerate bands. A simple analytic solution is obtained only considering the interlayer asymmetry between the two layers $U = U_1$-$U_2$ , defined as the difference in the on-site energies of the orbitals on the two layers, where $\epsilon_{A1}$, $\epsilon_{B1}$ = $U_1$ (potential energy of the first layer) and $\epsilon_{A2}$, $\epsilon_{B2}$ = $U_2$ (potential energy of the second layer). The solution could be written as [161]:

$$E(p) = \varepsilon_{\alpha} = \frac{U_1 + U_2}{2} \pm \sqrt{\frac{\gamma_1^2}{2} + \frac{U^2}{4} + v_F^2 p^2 + (-1)^{\alpha} \sqrt{\frac{\gamma_1^4}{4} + v_F^2 p^2 \left(\gamma_1^2 + U^2\right)}} \tag{4.2}$$

with $\alpha$ = 1,2.

In this work, the relevant band for energies near the Fermi level is considered to be the low-energy electronic band structure in the vicinity of the $K$ points at the corners of the first Brillouin zone $E = \varepsilon_1$ [169], by taking into account the assumption based on the intralayer hopping, $\gamma_0$, and the interlayer coupling, $\gamma_1$, are larger than other energies [107]: $\gamma_0, \gamma_1 \gg |E|, v_F p, |U|$; otherwise a four band model of the electronic bands is required in order to obtain the correct physical properties [170], [171]. As a result, the low-energy dispersion relation for BLG reads as:

$$E_{low}(k) = \varepsilon_1 = \frac{U_1 + U_2}{2} \pm \sqrt{\frac{\gamma_1^2}{2} + \frac{U^2}{4} + v_F^2 k^2 \hbar^2 - \sqrt{\frac{\gamma_1^4}{4} + v_F^2 k^2 \hbar^2 \left(\gamma_1^2 + U^2\right)}} \quad (4.3)$$

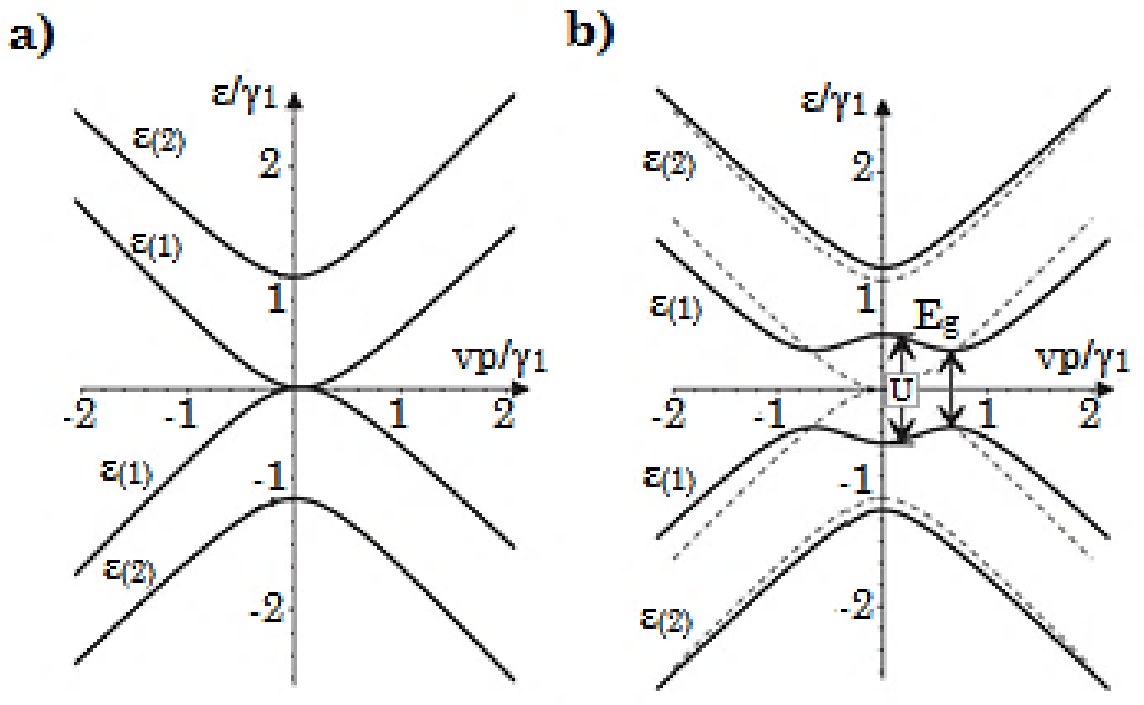

**Figure 4.2** Schematic of the energy dispersion relation near the $K$ point in the presence of a) zero asymmetry $U_1 = U_2 = U = 0$; b) finite layer asymmetry $U$ and $U_1 = -U_2$ resulting in two low bands with "Mexican hat" like shape resulting in a bandgap of $E_g$ (dotted lines show the bands for zero asymmetry). (Image adapted from [172])

In BLG unbiased and undoped $U_1 = U_2 = U = 0$, the Fermi energy is placed on the centre of the band diagram where the conduction and the valence bands touch each other at the $K$ point, as shown in Figure 4.2a. On the other hand, if a gate bias or doping is applied inducing $U_1 \neq U_2$, then $U \neq 0$ and the resulting electronic band diagram is shown in Figure 4.2b, where the asymmetry parameter produce a non-zero bandgap. The induced potentials $U_1 \neq 0$, $U_2 \neq 0$, result in a shifting of the band diagram either upwards or downwards in a quantity $(U_1+U_2)/2$ according to (4.3). This quantity is actually the distance that the band diagram is shifted with respect to the zero point energy, which is the well-known DP. A different approach considered in this work consists of keeping the band diagram centred at the DP and then shifting the Fermi energy as $E_F = -(U_1+U_2)/2$, instead of shifting the band diagram.

Moreover, in case of an unbiased bilayer graphene sheet where both layers have the same molecular doping, it would induce similar energy potentials $U_1 \approx U_2 \neq 0$, resulting in a shifting of the Fermi energy but not inducing a bandgap ($U = 0$). Examples of different configurations are shown in Figure 4.3.

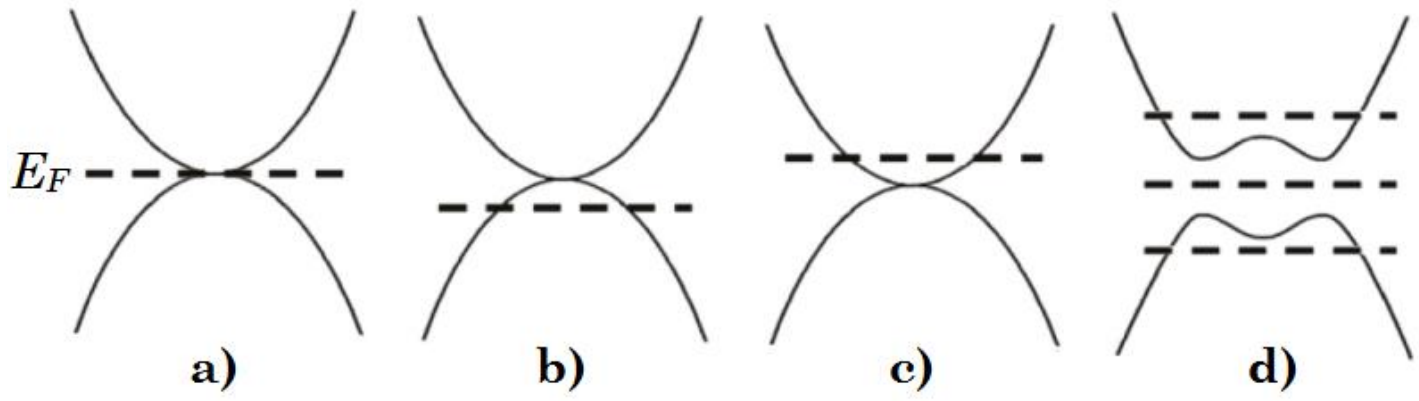

**Figure 4.3** Schematics of energy band diagrams and Fermi energy for a) undoped and unbiased BLG ($U_1 = U_2 = U = 0$; $E_F = 0$), b) unbiased and symmetrical P-doped BLG ($U_1 = U_2 > 0$; $U = 0$; $E_F < 0$), c) unbiased and symmetrical N-doped ($U_1 = U_2 < 0$; $U = 0$; $E_F > 0$) and d) biased and/or doped BLG ($U_1 \neq U_2$; $U \neq 0$).

Turning our attention back to Figure 4.2b, the energy of the low-energy bands exactly at the $K$ point is $E(k = 0) = \pm U/2$. Note that the "Mexican hat" like shape of the low-energy bands means that the true value of the gap, $E_{gap}$, between the conduction and valence bands occurs at finite $k_{min} \neq 0$ away from the $K$ point. From the energy dispersion relation in (4.3), the following expressions can be derived:

$$k_{min} = \frac{|U|}{2v_F}\sqrt{\frac{2\gamma_1^2 + U^2}{\gamma_1^2 + U^2}}; \qquad E_{gap} = 2\left|E\left(\pm k_{min}\right)\right| = \frac{|U|\gamma_1}{\sqrt{\gamma_1^2 + U^2}} \tag{4.4}$$

For huge values of the asymmetry $|U| >> \gamma_1$, the gap saturates at $E_{gap} \approx \gamma_1$, although for modest asymmetry values $|U| << \gamma_1$, the relation is simply $E_{gap} \approx U$.

The 2D-DOS at low energy can also be derived from (4.3), resulting in:

$$DOS_{2D}\left(E\right) = \frac{2|E|}{\pi\hbar^2 v_F^2}\left(1 + \frac{1}{2}\sqrt{\frac{U^2 + \gamma_1^2}{E^2 - E_c^2}}\right) \tag{4.5}$$

where $E_c$ refers to the CB edge. It is worth noticing that if the interlayer asymmetry is zero, then the 2D-DOS of BLG can be expressed as:

$$DOS_{2D}\left(E\right)\Big|_{U=0} = \frac{2|E|}{\pi\hbar^2 v_F^2} + \frac{\gamma_1}{\pi\hbar^2 v_F^2} \tag{4.6}$$

From the derived 2D-DOS in (4.5) both the $n$ and $p$ (-type) carrier concentration can be easily calculated as:

$$n = \frac{\upsilon}{2}\int_{E_c}^{+\infty} DOS_{2D}(E) f(E - E_F)\, dE$$
$$p = \frac{\upsilon}{2}\int_{-\infty}^{E_v} DOS_{2D}(E)\left[1 - f(E - E_F)\right] dE \tag{4.7}$$

where $f$ is the Fermi-Dirac distribution, $\upsilon$ is the band degeneracy; and $E_v$ refers to the VB edge.

### 4.1.2 Electrostatics of BLGFETs

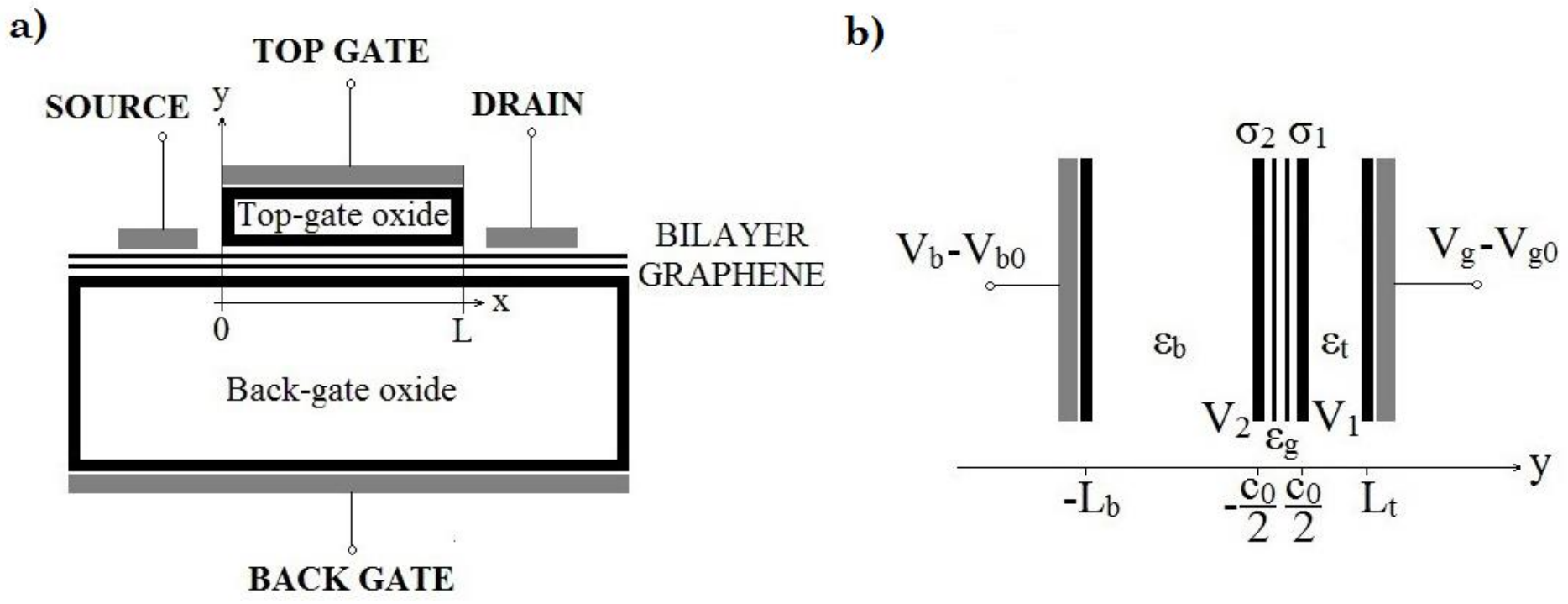

**Figure 4.4** a) Cross-section of a BLGFET. It consists of two graphene sheets playing the role of the active channel. The electrostatic modulation of the carrier concentration in the 2D sheet is achieved via a double-gate stack consisting of top- and back-gate dielectrics and corresponding metal gates. b) Scheme of the BLG-based capacitor showing the relevant physical and electrical parameters, charges and potentials.

The cross-section of a dual-gate BLG-based device is the one depicted in Figure 4.4a. The bilayer graphene sheet plays the role of the active channel between the source and the drain. Just as done in subsection 3.4.2 to get the electrostatic behaviour, the 1D Gauss law's equation is solved along the $y$-axis. Direction $x$ extends from source to drain along the channel length ($L$). Upon application of such an 1D Gauss's law to the double-gate stack shown in Figure 4.4b, the carrier density and potentials on each layer can be gotten from the external gate bias and impurities concentration:

$$C_t\left(V_g - V_{g0} - V_1\right) + C_o\left(V_2 - V_1\right) = -\sigma_1$$
$$C_b\left(V_b - V_{b0} - V_2\right) + C_o\left(V_1 - V_2\right) = -\sigma_2 \tag{4.8}$$

where $C_t = \varepsilon_0\varepsilon_t/(L_t\text{-}c_0/2)$ and $C_b = \varepsilon_0\varepsilon_b/(L_b\text{-}c_0/2)$ are the top and bottom oxide capacitances, respectively; $V_g\text{-}V_{g0}$ and $V_b\text{-}V_{b0}$ are the top- and back-gate voltage overdrive; and $V_{g0}$ and $V_{b0}$ are the flat-band voltages. These quantities comprise work-function differences between the gates and the graphene channel and possible additional charge due to impurities or doping; $V_1$ and $V_2$ are the electrostatic potentials dropped at the first and second graphene layers, respectively; $C_o = \varepsilon_0\varepsilon_g/c_0$ is the graphene parallel plate capacitance, where $c_0$ is the interlayer spacing and $\varepsilon_g$ is an effective dielectric constant for the BLG to characterize charge screening [173]; and $\sigma_1$ and $\sigma_2$ are the charge densities at the first and second graphene layers, respectively.

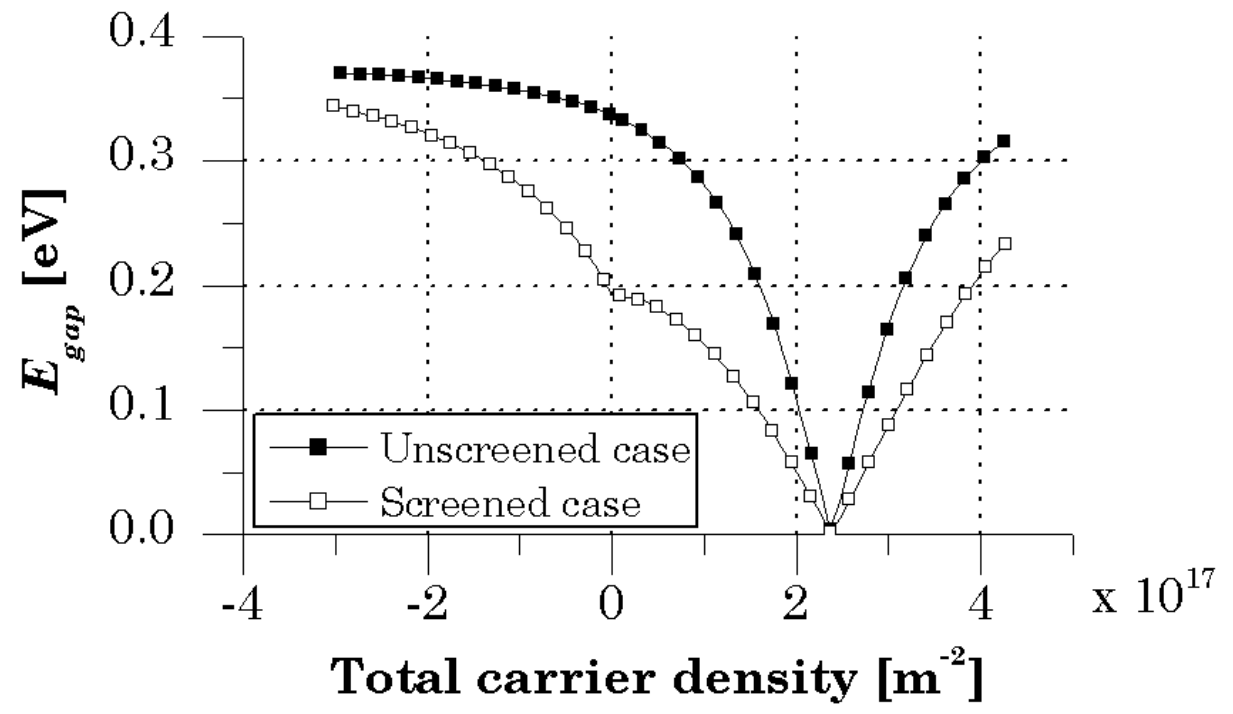

**Figure 4.5** Bandgap vs. carrier density. The solid squares represent the bandgap behaviour for the unscreened case and the open ones for the screened case.

If the perpendicular electric field between two graphene layers is assumed to be unscreened, then the charge density carried by each layer can be written as $\sigma_1 = \sigma_2 = Q_{net}/2$, where $Q_{net}$ is the overall net mobile sheet charge density. That simple assumption is known to overestimate the bandgap [156], [158]. A more accurate way of describing screening effects has been proposed by Edward McCann *et al.* [107] based on a tight-binding model and Hartree's theory. They have found that the individual layer densities are given by:

$$\sigma_{1(2)} = \frac{Q_{net}}{2} \mp \frac{q\gamma_1 U}{4\pi\hbar^2 v_F^{\,2}} \ln\left( \frac{\pi\hbar^2 v_F^{\,2}\left|Q_{net}\right|}{2q\gamma_1^{\,2}} + \frac{1}{2}\sqrt{\left(\frac{\pi\hbar^2 v_F^{\,2} Q_{net}}{q\gamma_1^{\,2}}\right)^2 + \left(\frac{U}{2\gamma_1}\right)^2} \right) \quad (4.9)$$

where $q$ is the elementary charge. Figure 4.5 shows the dependence of the bandgap on the carrier density for both screened and unscreened cases, where it becomes clear that the unscreened hypothesis overestimates the bandgap. From now on, the precise model that considers screening effects is used.

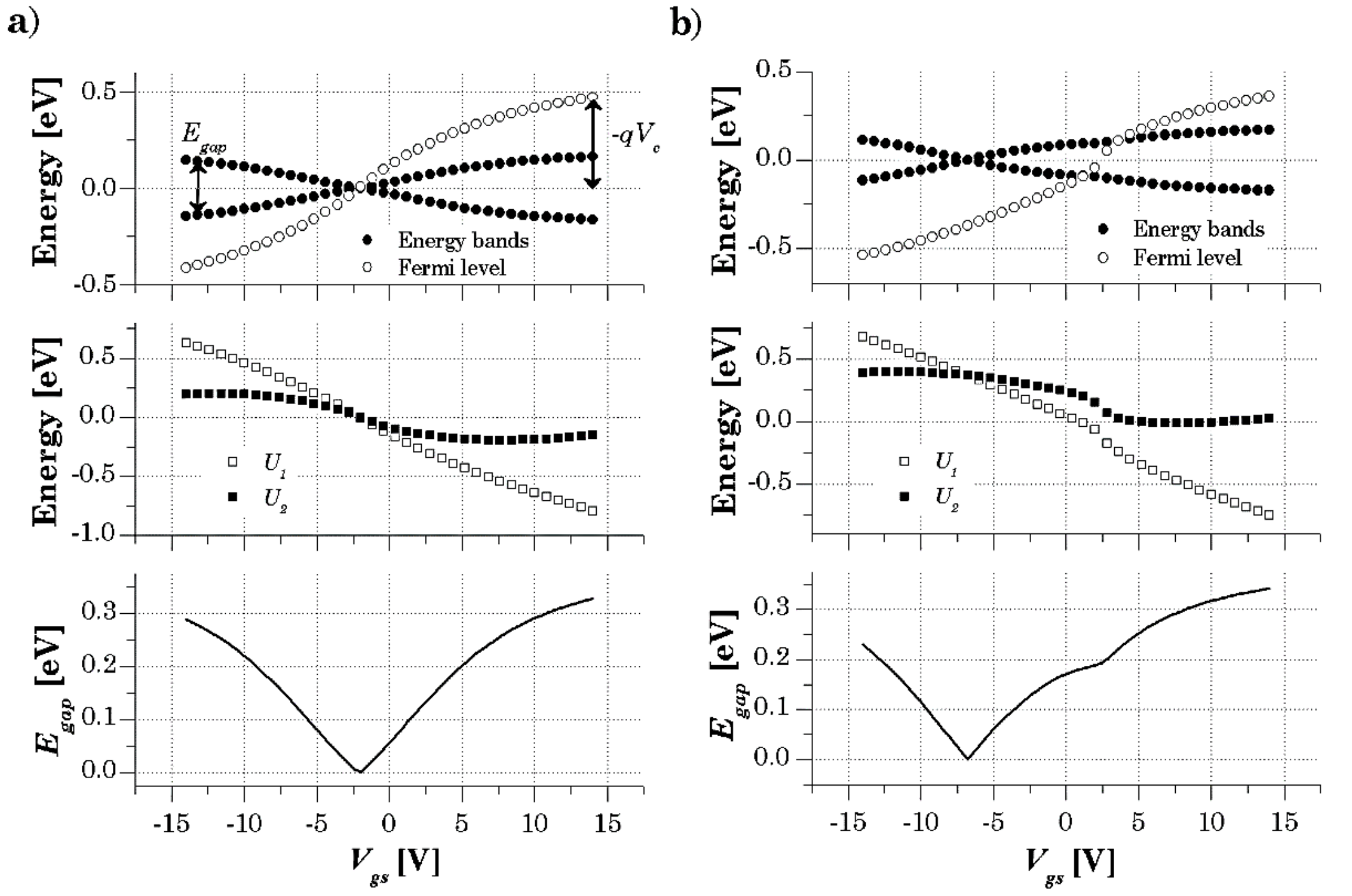

**Figure 4.6** Energy band diagrams; potential energies $U_1$, $U_2$; and bandgap, $E_{gap}$, of a BLGFET as a function of the top-gate bias for two different applied overdrive back-gate biases: a) $V_b$ - $V_{b0}$ = 0 V; b) $V_b$ - $V_{b0}$ = -40 V. In the upper panel, the black circles correspond to the conduction (upper) and valence (lower) band edges and the white circles represents the position of the Fermi level. The voltage drop across the BLG, named as $V_c$, gives the position of the Fermi level respect to the CNP.

External gates are generally used to control the carrier density on a bilayer graphene-based device, just as shown for the monolayer counterpart in Figure 3.4. For the BLG case, both gates also drive the separate layers to different potential energies $U_1$, $U_2$, inducing an interlayer asymmetry, $U$, and shifting of the Fermi energy, $E_F$. This physics can actually be explained in terms of displacement fields. A top and bottom electric displacement field, $D_t$ and $D_b$, respectively, built up upon application of top- and back-gate bias. The average of these quantities, $\Delta D = (D_b+D_t)/2$, breaks the inversion symmetry of the BLG and generates a nonzero bandgap. The difference of both displacement fields, $\partial D = D_b$-$D_t$, shifts $E_F$ and creates a net carrier doping. At the point where $\partial D$ = 0, named as CNP, the Fermi level is located at the middle of the gap, and the corresponding electrical resistance is the highest. Those electric displacement fields ($D$) can be easily calculated as $D_b = \varepsilon_b(V_{bs}-V_{b0})/(L_b\text{-}c_0/2)$ and $D_t = \varepsilon_t(V_{gs}-V_{g0})/(L_t\text{-}c_0/2)$. Figure 4.6 illustrates how the applied gate biases are tuning both the carrier density and the interlayer asymmetry and, ultimately, the bandgap and the Fermi energy. The simulation was done using the parameters from Table 4.1. As explained in subsection 4.1.1, the largest theoretical bandgap that could be reached in BLG, according to (4.4), is limited by the intrinsic interlayer hopping parameter, $\gamma_1$. Experimentally, bandgaps up to 250 meV have been reached [156].

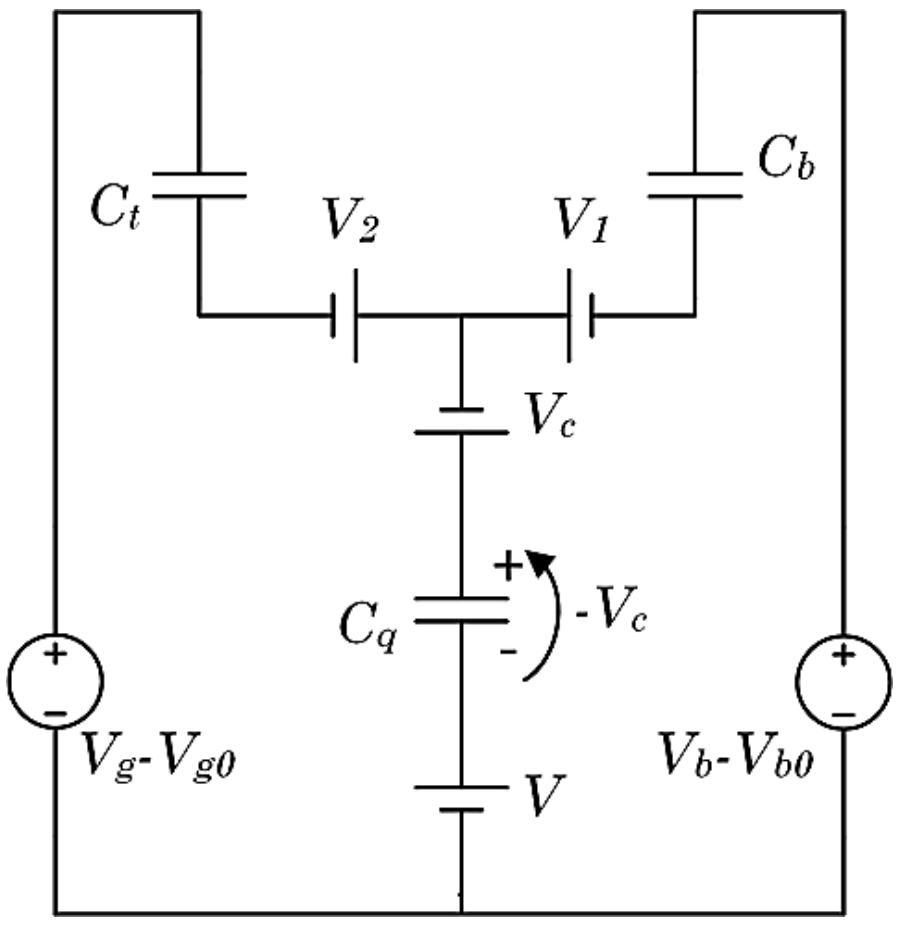

**Figure 4.7** Equivalent capacitive circuit of the BLGFET.

The electrostatics of the BLGFET can be also represented using the equivalent capacitive circuit depicted in Figure 4.7, which has been derived from (4.8) but replacing $V_g$ and $V_b$ by $V_g$ - $V(x)$ and $V_b$ - $V(x)$, respectively, where $V(x)$ is the quasi-Fermi level along the BLG channel. This quantity must fulfil the following boundary conditions: (1) $V(x) = V_s$ at the source end, $x = 0$; (2) $V(x) = V_d$ (drain-source voltage) at the drain end, $x = L$. The potential $-V_c$ in the equivalent circuit represents the SFL respect to the DP or, equivalently, the voltage drop across the quantum capacitance $C_q$, which is the same concept explained for the monolayer counterpart in subsection 4.1.1. Therefore, this quantity is also defined as $C_q = dQ_{net}/dV_c$ and has to do with the 2D-DOS of the BLG. Both quantum capacitance and overall net mobile sheet charge of BLG have been presented in Figure 4.8. Applying circuit laws to the equivalent capacitive circuit, the following straightforward relation is obtained:

$$V(x) = \frac{C_t}{C_t + C_b}\left(V_g - V_{g0} - V_1\right) + \frac{C_b}{C_t + C_b}\left(V_b - V_{b0} - V_2\right) + \frac{Q_{net}(V_c)}{C_t + C_b} \quad (4.10)$$

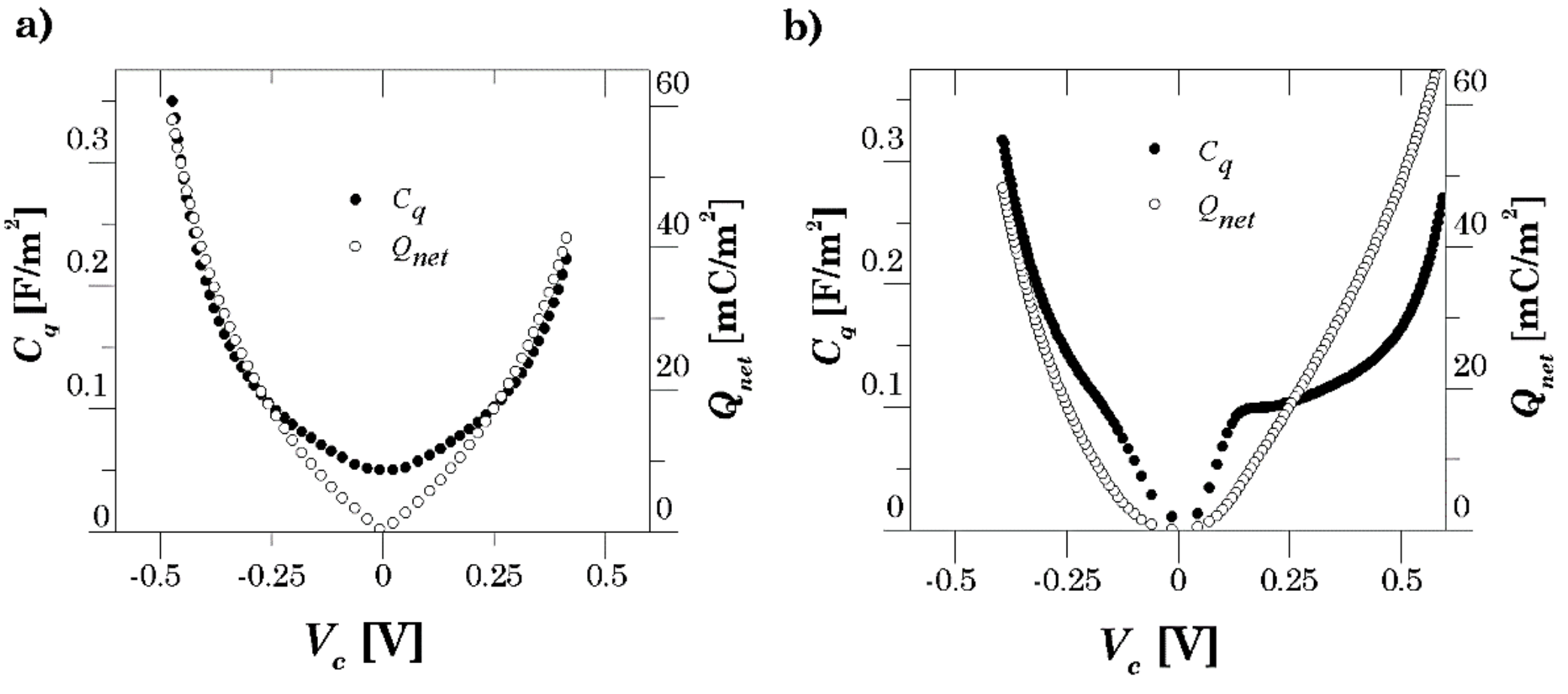

**Figure 4.8** Quantum capacitance and overall net mobile sheet charge density versus the voltage drop across the quantum capacitance for two different applied overdrive back-gate biases: a) $V_b$ - $V_{b0}$ = 0 V; b) $V_b$ – $V_{b0}$ = -40 V. Theoretical results of the BLG quantum capacitance are consistent with calculations in [174], [175].

### 4.1.3 Drift-diffusion transport model of BLGFETs

As current prototype devices present channel lengths greater than the MFP ($L >> \lambda$), which has been estimated as $\lambda \approx$ 10 nm at carrier densities of 3 × $10^{12}$ $cm^{-2}$ for exfoliated BLG deposited on a 300 nm $SiO_2$ substrate at low temperatures [100], to model the drain-to-source current of a BLGFET, a DD transport is assumed under the form of (3.11) and (3.12), where $W$ is the gate width, $Q_{tot}(x) = Q_t(x)+\sigma_{pud}$ is the free carrier sheet density along the bilayer graphene channel at position $x$, $Q_t(x) = q[p(x)+n(x)]$ is the transport sheet charge density, and $\sigma_{pud} = q\Delta^2/\pi\hbar^2 v_F^2$ is the residual charge density due to electron-hole puddles [100], [176]. A soft-saturation model, considering $\beta = 1$ in (3.12), has been also assumed for the drift carrier velocity in BLG adopted consistently with the numerical studies of electronic transport in BLG relying on first-principles analysis and Monte Carlo simulations [177]. Both the effective low-field carrier mobility and saturation velocity, $\mu$ and $v_{sat}$, have been considered constant and independent of the applied electric field, carrier density, or temperature. According to [177], the constant saturation velocity is considered to be $v_{sat} = v_F/\pi$. Then, the drain current can be expressed as:

$$I_{ds} = \mu \frac{W}{L_{eff}} \int_0^{V_{ds}} Q_{tot} dV \tag{4.11}$$

where $L_{eff} = L+\mu\,|V_{ds}|/v_{sat}$ is a correction to the physical channel length to incorporate saturation velocity effects. To get the drain current, it is convenient to solve the above integral using $V_c$ as the integration variable, and consistently express $Q_{tot}$ as a function of $V_c$ in the following way:

$$I_{ds} = \mu \frac{W}{L_{eff}} \int_{V_{cs}}^{V_{cd}} Q_{tot}\left(V_c\right) \frac{dV}{dV_c} dV_c \tag{4.12}$$

where $V_{cs}$ and $V_{cd}$ are obtained from (4.10), with $V_{cs} = V_c|_{V = V_s}$ and $V_{cd} = V_c|_{V = V_d}$. In addition, the quantity $dV/dV_c$ in (4.12) can also be derived from (4.10) and reads as follows:

$$\frac{dV}{dV_c} = -\frac{C_t}{C_t + C_b}\frac{dV_1}{dV_c} - \frac{C_b}{C_t + C_b}\frac{dV_2}{dV_c} + \frac{C_q}{C_t + C_b} \tag{4.13}$$

### 4.1.4 Charge and capacitance models of BLGFETs

An accurate modelling of the intrinsic capacitances of FETs requires an analysis of the charge distribution in the channel versus the terminal bias voltages. In doing so, the terminal charges $Q_g$, $Q_b$, $Q_d$, and $Q_s$ associated with the top-gate, back-gate, drain, and source electrodes of a four-terminal device have been considered. For instance, $Q_g$ can be calculated by integrating $Q_{net_g}(x) = C_t(V_{gs}-V_{g0}-V_1(x)-V(x))$ along the channel and multiplying it by the channel width $W$. This expression for $Q_{net_g}(x)$ has been obtained after applying Gauss's law to the top-gate stack, resulting in (4.14). A similar expression can be found for $Q_b$, so the relation in (3.18) is fulfilled. The Ward-Dutton's linear charge partition scheme is applied in order to guarantee charge conservation and thus the terminal charges can be described as follows:

$$\begin{aligned} Q_g &= WC_t\left[L\left(V_g - V_{g0}\right) - \int_0^L \left[V_1(x) + V(x)\right]dx\right] \\ Q_b &= WC_b\left[L\left(V_b - V_{b0}\right) - \int_0^L \left[V_2(x) + V(x)\right]dx\right] \\ Q_d &= W\int_0^L \frac{x}{L} Q_{net}(x) dx \\ Q_s &= -\left(Q_g + Q_b + Q_d\right) \end{aligned} \tag{4.14}$$

Once the above expressions are conveniently written using $V_c$ as the integration variable according to (3.20), the same capacitance approach developed in subsection 3.4.4 is applied to obtain the 9 independent intrinsic capacitances.

### 4.1.5 Metal – BLG contact resistance model

To reproduce the experimental *I-V* characteristics of a BLGFET, accounting for the voltage drop at the source/drain (S/D) contacts is necessary. State-of-the-art values for the metal-BLG contact resistance are around several hundred of Ω·µm [178]–[181]. To model the metal – BLG contact resistance, the formation of a Schottky barrier between both has been assumed. Whenever an appreciable bandgap exists, the current would be dominated by the thermionic emission of carriers through the Schottky barrier. Hence the current would be proportional to $exp(-q\phi_b/k_BT)$, where $\phi_b$ is the Schottky barrier height, $k_B$ is the Boltzmann constant, and $T$ is the temperature. So, the interfacial contact resistivity ($\rho_c$) between the metal and the BLG can be calculated as [182]:

$$\rho_c(\phi_b) = \frac{k_B}{qTA^*_{metal-BLG}} e^{\frac{q\phi_b}{k_BT}} \tag{4.15}$$

where $A^*_{metal\text{-}BLG}$ is the Richardson constant of the metal–BLG contact, considered here as an empirical fitting parameter. The contact resistance can be expressed as [67]:

$$R_{c-Schottky}(\phi_b) = \frac{\sqrt{\rho_{sh}\rho_c(\phi_b)}}{W} \coth\left(L_c\sqrt{\frac{\rho_{sh}}{\rho_c(\phi_b)}}\right) \tag{4.16}$$

where $L_c$ is the physical contact length and $\rho_{sh} = [q\mu(p+n)]^{-1}$ is the BLG sheet resistivity under the metal. It is possible to define a length $L_T = [\rho_c / \rho_{sh}]^{1/2}$ which physically corresponds to the length of the BLG region underneath the contact where the current mainly flows. Depending on the ratio between $L_c$ and $L_T$ two limit cases might arise: (i) short contact case ($L_c << L_T$), where the resistance is dominated by the interfacial contact resistance and the current flows uniformly across the entire contact; and (ii) long contact case ($L_c >> L_T$), where the resistance is independent of $L_c$, since most of the current flows through the edge of the contact. Figure 4.9a shows the scheme of the physical structure of the metal-BLG contact together with an illustration of the current crowding phenomenon occurring for the long contact case.

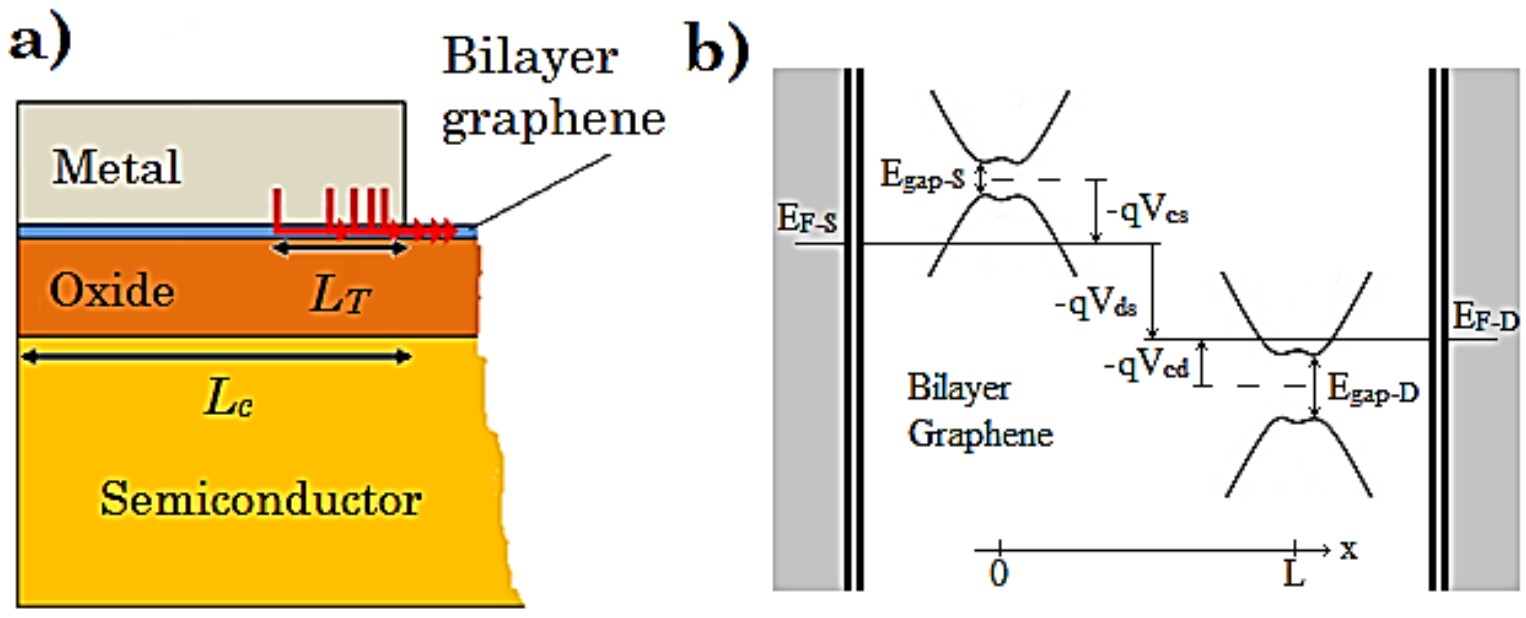

**Figure 4.9** a) Physical structure and scheme of the current crowding effect through the metal-BLG contact, b) Schematics of the band diagram of the metal-BLG contact at the source and drain sides, needed to estimate the Schottky barrier height. The key quantities such as the bandgap size, $E_{gap\text{-}S}$ and $E_{gap\text{-}D}$; the shift of the Fermi level, $V_{cs}$ and $V_{cd}$; and the metal Fermi energy, $E_{F\text{-}S}$ and $E_{F\text{-}D}$, both at the drain and source sides are shown. It is worth noticing that both $E_{F\text{-}S}$ and $E_{F\text{-}D}$ are aligned with the quasi-Fermi level at the source and drain sides, respectively; i.e. $E_{F\text{-}S} = V(0)$ and $E_{F\text{-}D} = V(L)$. The band diagram illustrates a possible mixed p/n-type channel with different bandgap size on each side.

According to the band diagram shown in Figure 4.9b the Schottky barrier height at the source side, which is presented separately for electrons, $\phi_b{}^n$, and for holes, $\phi_b{}^p$, can be calculated as:

$$\phi_{b-S}^{n} = V_{cs} + \frac{E_{gap-S}}{2}; \qquad \phi_{b-S}^{p} = -V_{cs} + \frac{E_{gap-S}}{2} \tag{4.17}$$

An analogous procedure is implemented at the drain side. To quantitatively estimate the effect of the contact resistance, a splitting of the electron and hole contributions to the drain current is necessary. This can be done as follows:

$$\begin{aligned} I_{ds} &= \mu \frac{W}{L_{eff}} \int_{V_{cs}}^{V_{cd}} Q_{tot}\left(V_c\right) \frac{dV}{dV_c} dV_c \\ &= q\mu \frac{W}{L_{eff}} \left[ \int_{V_{cs}}^{V_{cd}} n\left(V_c\right) \frac{dV}{dV_c} dV_c + \int_{V_{cs}}^{V_{cd}} p\left(V_c\right) \frac{dV}{dV_c} dV_c \right] = I_{ds}^n + I_{ds}^p \end{aligned} \tag{4.18}$$

where both $I_{ds}^n$ and $I_{ds}^p$ are the electron and hole contributions, respectively. The intrinsic $V_{gs}$ and $V_{ds}$ are then given by the following equations:

$$\begin{aligned} V_{gs} &= V_{gs,e} - I_{ds}^n\left(V_{gs},V_{ds}\right) R_{c-Schottky}(\phi_{b-S}^n) - I_{ds}^p\left(V_{gs},V_{ds}\right) R_{c-Schottky}(\phi_{b-S}^p) \\ V_{ds} &= V_{ds,e} - I_{ds}^n\left(V_{gs},V_{ds}\right)\left[R_{c-Schottky}(\phi_{b-S}^n) + R_{c-Schottky}(\phi_{b-D}^n)\right] \\ &\quad - I_{ds}^p\left(V_{gs},V_{ds}\right)\left[R_{c-Schottky}(\phi_{b-S}^p) + R_{c-Schottky}(\phi_{b-D}^p)\right] \end{aligned} \tag{4.19}$$

## 4.2 BLGFET large-signal model benchmarking

In this section, the BLGFET drain-current model discussed above is assessed via comparison with the measured electrical behaviour of prototype devices.

The mobility has been considered as an input parameter of the model to fit the experiment. As explained in subsection 4.1.3, it is assumed to be independent of the applied field, carrier density, or temperature, and considered the same for both electrons and holes. It is worth noting that some simulations and experiments have shown that the mobility somehow decreases with the size of the induced bandgap [177], [183], but this refinement has not been included in the model.

The experimental TCs show a non-linear shift of the CNP with the back-gate voltage. This effect is likely to appear because of the presence of charge traps in the gate oxide and/or the BLG interface. So, when a positive $V_b$-$V_{b0}$ is applied to the device, the injection of electrons into the charge traps causes a shift of the CNP towards more positive voltage. On the contrary, applying a negative $V_b$-$V_{b0}$ results in hole injection, so the CNP shifts in the opposite direction. This effect has been reported in [184], [185] for graphene on $SiO_2$ and the strength of it depends upon the swept voltage range, sweep rate, and surrounding conditions. So to capture this CNP shifting effect, a corrective parameter $\beta$ has been introduced in the model to properly modulate the top-gate offset voltage, so $V_{g0}$ is replaced by $V_{g0}+\beta V_b^2$, as proposed in [165], [167].

### 4.2.1 BLG-based device A: drain current model validation

The drain-current model is assessed against the electrical characteristics reported in [157]. The simulations were done using the device's parameters listed in Table 4.1. Figure 4.10 shows both the experimental and predicted TCs and OCs.

**Table 4.1** Input parameters of the BLG-based device A reported in [157].

| Input parameter | Description | Value |
|---|---|---|
| $\gamma_0$ [168] | In-plane hopping parameter | 3.16 eV |
| $\gamma_1$ [168] | Interlayer hopping parameter | 0.381 eV |
| $a$ | Graphene lattice constant | 2.49 Å |
| $c_0$ [186] | Graphene interlayer distance | 3.34 Å |
| $\beta$ | Fitting parameter due to non-linearity in the response of the CNP to the back-gate bias | $3.2 \cdot 10^{-4}$ V$^{-1}$ |
| $T$ | Temperature | 300 K |
| $\mu$ | BLG electron/hole mobility | 114 cm$^2$/Vs |
| $L$ | Gate length | 4 µm |
| $W$ | Gate width | 4 µm |
| $L_t$ | Top-gate oxide thickness | 8 nm |
| $L_b$ | Back-gate oxide thickness | 90 nm |
| $\varepsilon_g$ [173] | Effective BLG relative permittivity | 2.5 |
| $\varepsilon_t$ | Top-gate oxide relative permittivity | 3.9 |
| $\varepsilon_b$ | Back-gate oxide relative permittivity | 3.9 |
| $V_{g0}$ | Top-gate offset voltage | -2.1 V |
| $V_{b0}$ | Back-gate offset voltage | -10 V |
| $\Delta$ | Spatial potential inhomogeneity due to electron-hole puddles | 20 meV |
| $L_c$ | Effective contact length | 4 µm |
| $A_{Ni\text{-}BLG}$ | Richardson constant for Nickel – BLG contact | $5 \cdot 10^4$ A/m$^2$K$^2$ |

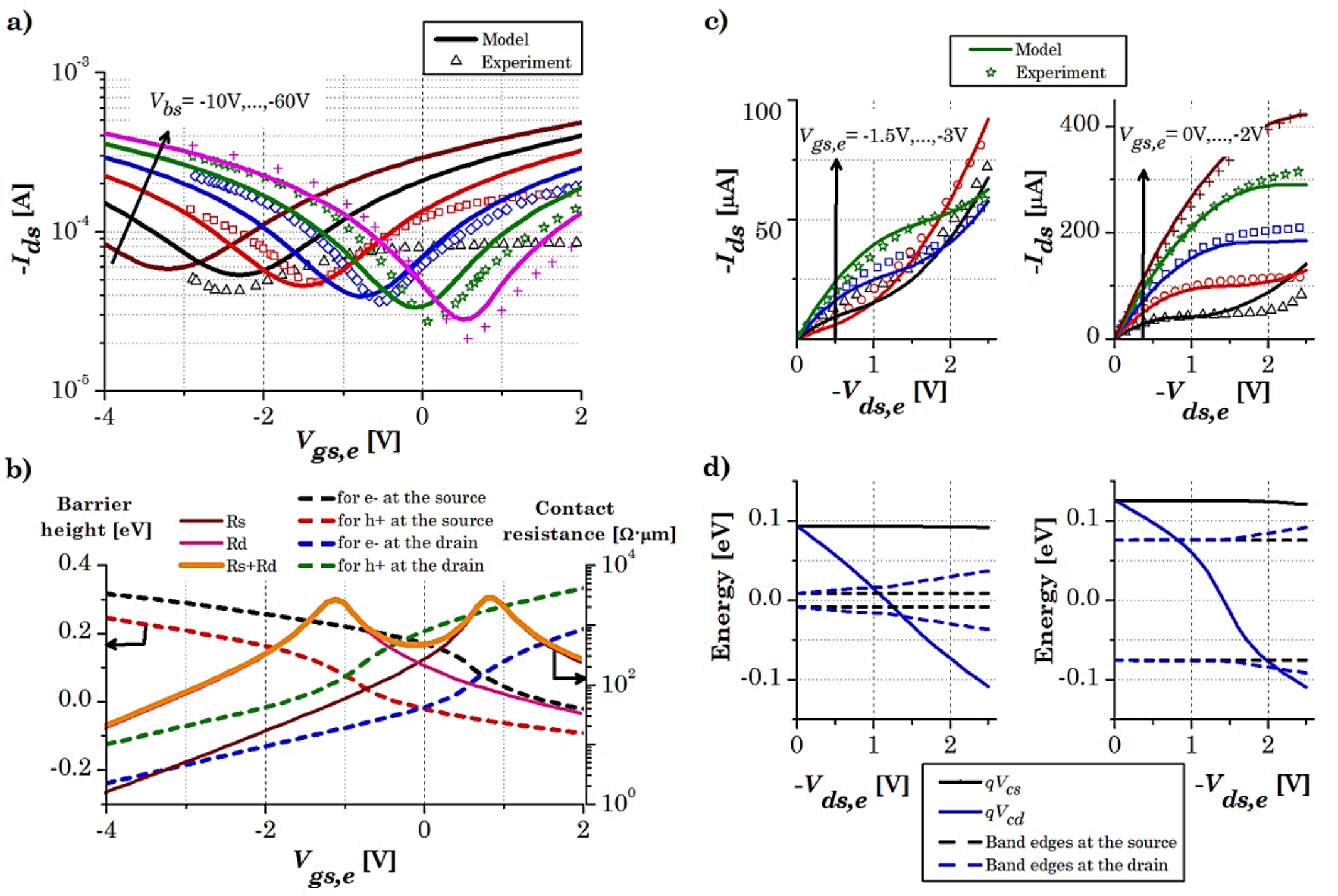

**Figure 4.10** a) Transfer characteristics of the examined device [157] described in Table 4.1 ($V_{ds,e}$ = -2 V). b) Schottky barrier height for both electrons and holes (left axis) and contact resistance (right axis) at the drain and source sides respect to the top-gate bias ($V_{bs}$ = -50 V and $V_{ds,e}$ =-2V). c) Output characteristics for two situations: (left) $V_{bs}$ = -20 V; and (right) $V_{bs}$ = -50 V. d) Evolution of the SFL, conduction and valence band edges at the drain and source sides, versus the drain bias according to the situations described in c): (left) $V_{gs,e}$ = -2.5 V and $V_{bs}$ = -20 V; and (right) $V_{gs,e}$ = -0.5 V and $V_{bs}$ = -50 V.

The electrostatics discussed in subsection 4.1.2 actually corresponds to the DUT. Specifically both Figure 4.6a and Figure 4.8a, depict a situation where the condition $D_b$ = 0 (ZBDC, standing for Zero Bottom electric Displacement field Condition) is fulfilled at $V_{bs} = V_{b0}$ = -10 V, resulting in the brown curve shown in Figure 4.10a. In this case, the charge neutrality condition is reached just at the zero-gap point, where the condition $D_t$ = 0 is fulfilled at $V_{gs} = V_{g0}$, so $\Delta D = \partial D = 0$. Increasing (reducing) the top-gate bias beyond (below) the Dirac voltage results in the Fermi level directly entering into the CB (VB), so there is no any especial advantage of using BLG over SLG. On the other hand, when ZBDC does not apply, a larger current modulation can be obtained. The electrostatics of this situation is illustrated in Figure 4.6b and Figure 4.8b, where now $V_{bs}$-$V_{b0}$ = -40 V, which in turn corresponds to the green curve in Figure 4.10a. In this case, the charge neutrality condition ($\partial D$ = 0) is reached when $D_t = D_b \neq 0$, so $\Delta D \neq 0$ and the CNP happens at some finite energy bandgap. Moving the top-gate bias beyond (below) the CNP results in electron (hole) doping of the BLG together with an induced bandgap that can reach a few hundred of meVs. Importantly, the Fermi level does not directly enter into the CB (VB) beyond (below) the CNP upon application of top-gate bias, but there exists a region where it lies inside the bandgap. So, the combination of these two effects results in larger on-off current ratio than the SLG-based transistor.

The evolution of the Schottky barrier height seen by the electrons and holes at both S/D sides are shown in Figure 4.10b as a function of $V_{gs,e}$. The corresponding $R_c$ is shown as well, broken down into its components $R_s$ and $R_d$, each calculated as the parallel association of the individual contact resistances due to electrons and holes. The back-gate bias is $V_{bs}$ = -50V, far from the ZBDC, and the corresponding electrostatics is plotted in Figure 4.6b. It happens that the highest contact resistance situation is reached at $V_{gs,e}$ = -1.1 V and $V_{gs,e}$ = 0.8 V, corresponding to the pinch-off condition at the drain and source sides, respectively. The Schottky barrier height at these points is just $E_{gap}/2$, as given by (4.17).

Next, the experimental and simulated OCs of the BLGFET near to the ZBDC and far from it are shown in Figure 4.10c. As for the former situation, analysed in Figure 4.10c left, saturation is weak, pretty similar to what is observed in SLG-based transistors. On the contrary, biasing the device far from the ZBDC, results in current saturation over a sizeable range of $V_{sd,e}$ (Figure 4.10c right). Simulations of $V_{cs}$, $V_{cd}$, CB's bottom, VB's top, all of them calculated as a function of $V_{sd,e}$, shown in Figure 4.10d, are helpful to understand why that is happening. As for the near to ZBDC (Figure 4.10d left) the pinch-off condition is reached when $V_{cd}$ = 0. This is happening at $V_{sd,e}$ = 1.2 V. Further increasing (reducing) of $V_{sd,e}$ drives the SFL at the drain side deep into the VB (CB), triggering the current due to holes (electrons). On the other hand, when the transistor is biased far from the ZBDC (Figure 4.10d right), the pinch-off condition now occurs when the SFL at the drain side crosses the middle of a larger gap. This is happening at $V_{sd,e}$ = 1.4 V in the experiment. But now, for a moderate increase (decrease) of $V_{sd,e}$, the SFL at the drain side lies inside the gap, so there won't be appreciable current variation respect to the pinch-off condition, resulting in the observed current saturation. Eventually, if $V_{sd,e}$ is further increased (reduced) beyond (below) the range from 0.8 to 2V, then the SFL at the drain side enters into the VB (CB) and the device gets into the second (first) linear region dominated by holes (electrons). So, the induced gap of the BLG provides a feasible way to virtually extend the pinch-off condition over a larger range of $V_{sd,e}$, which is of upmost technological importance.

### 4.2.2 BLG-based device B: drain current model validation and RF performance outlook

**Table 4.2** Input parameters of the BLG-based device B reported in [187].

| Input parameter | Value | Input parameter | Value |
|---|---|---|---|
| $\gamma_0$ | 3.16 eV | $L_t$ | 19 nm |
| $\gamma_1$ | 0.381 eV | $L_b$ | 300 nm |
| $a$ | 2.49 Å | $\varepsilon_g$ | 2.5 |
| $c_0$ | 3.34 Å | $\varepsilon_t$ | 4.2 |
| $\beta$ | $1.28 \cdot 10^{-4}$ V$^{-1}$ | $\varepsilon_b$ | 3.9 |
| $T$ | 300 K | $V_{g0}$ | 0 V |
| $\mu$ | 1160 cm$^2$/Vs | $V_{b0}$ | 50 V |
| $L$ | 3 μm | $\Delta$ | 30 meV |
| $W$ | 1.6 μm | $A_{Ti\text{-}BLG}$ | $8 \cdot 10^4$ A/m$^2$K$^2$ |
| $L_c$ | 3 μm | | |

The outcome of the drain-current model is again benchmarked against the electrical behaviour of the dual-gated BLGFET reported in [187]. The TCs, shown in Figure 4.11a, were recorded at room temperature by sweeping $V_{gs,e}$ while keeping constant $V_{bs}$. For comparison, the predicted TCs are shown in the same plot. The geometrical and electrical parameters used for the simulations are given in Table 4.2. The model predicts a continuous enhancement of the on-off current ratio expanding from 10 to 100 as $V_{bs}$ goes from 40 V down to -120 V, in correspondence to experimental evidence. According to simulations, the induced gap in the BLG at the CNP goes from 9.7 meV to 195 meV, and the maximum contact resistance goes from 150 Ω·µm to 2.6 kΩ·µm within the explored $V_{bs}$ range.

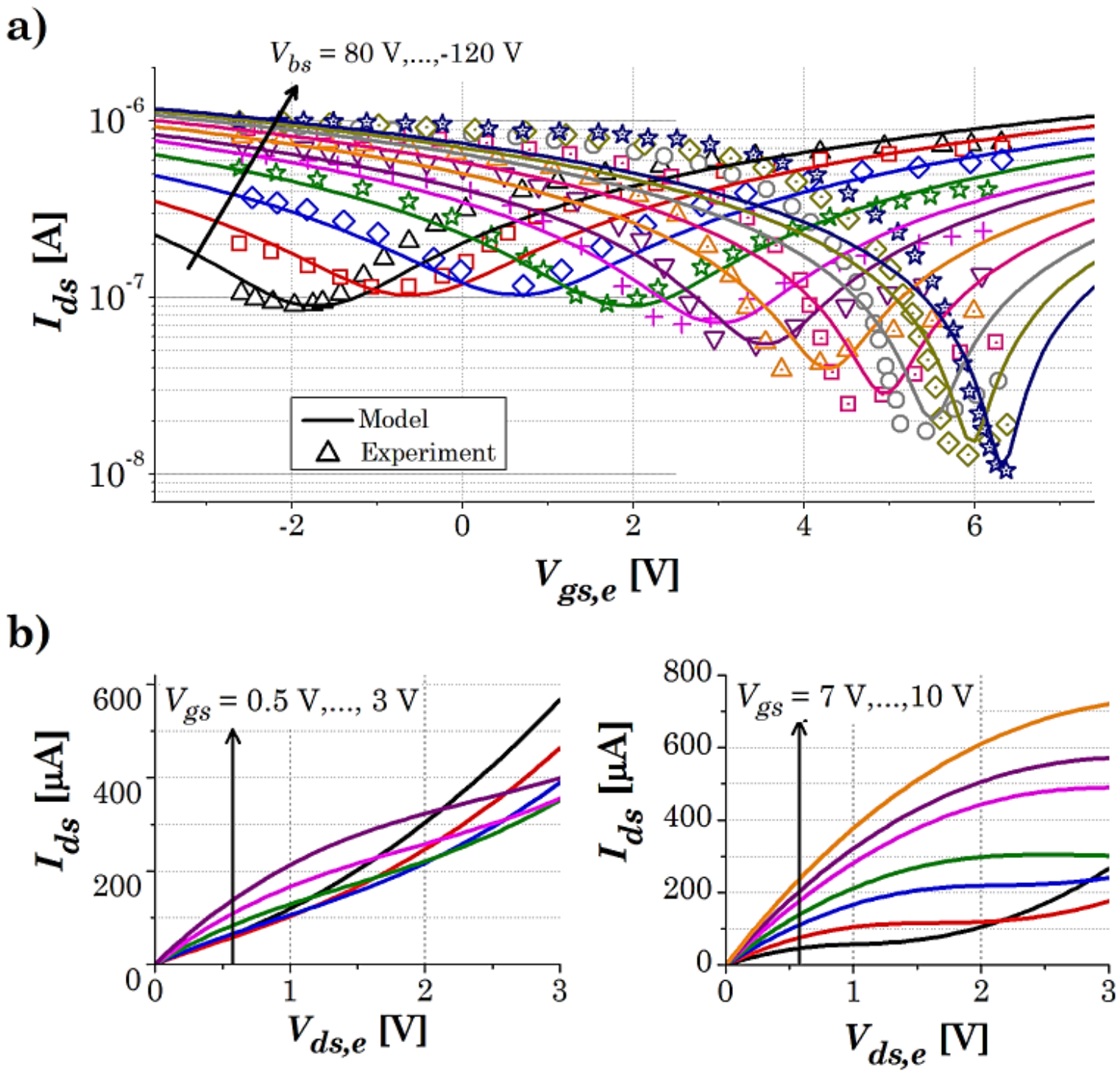

**Figure 4.11** a) Transfer characteristics of the DUT [187] described in Table 4.2 for $V_{ds,e}$ = 1 mV. b) Output characteristics upon application of different $V_{bs}$ resulting in small and large bandgap situations at CNP: (left) $V_{bs}$ = 40 V; and (right) $V_{bs}$ = -100 V.

Regarding the OCs, no experimental data were reported in [187], so only the predicted OCs are shown in Figure 4.11b. The left and right panel correspond to the OCs calculated at $V_{bs}$ = 40 V ≈ $V_{b0}$ and $V_{bs}$ = -100 V << $V_{b0}$, respectively. The induced bandgap is 10 and 175 meV, respectively, so the minimum output conductance is reduced in a factor of 6.7 for the latter case. Finally, predicted $f_{Tx}$ and $f_{max}$ are shown in Figure 4.12 for both $V_{bs}$ under examination. Both RF FoMs have been calculated using (2.7) and (2.9), respectively. Those are tunable with $V_{gs,e}$ showing a peak value of 3.7 and 5.2 GHz, respectively, with a noticeable improvement in a factor of 5 in $f_{max}$ and a factor of 2 in $f_{Tx}$ when the gap goes from 10 to 175 meV. Nevertheless, the device is not yet optimized and there is plenty of room to get higher FoMs. Scaling down of the channel length together with reducing the oxide thickness to keep SCEs under control is necessary.

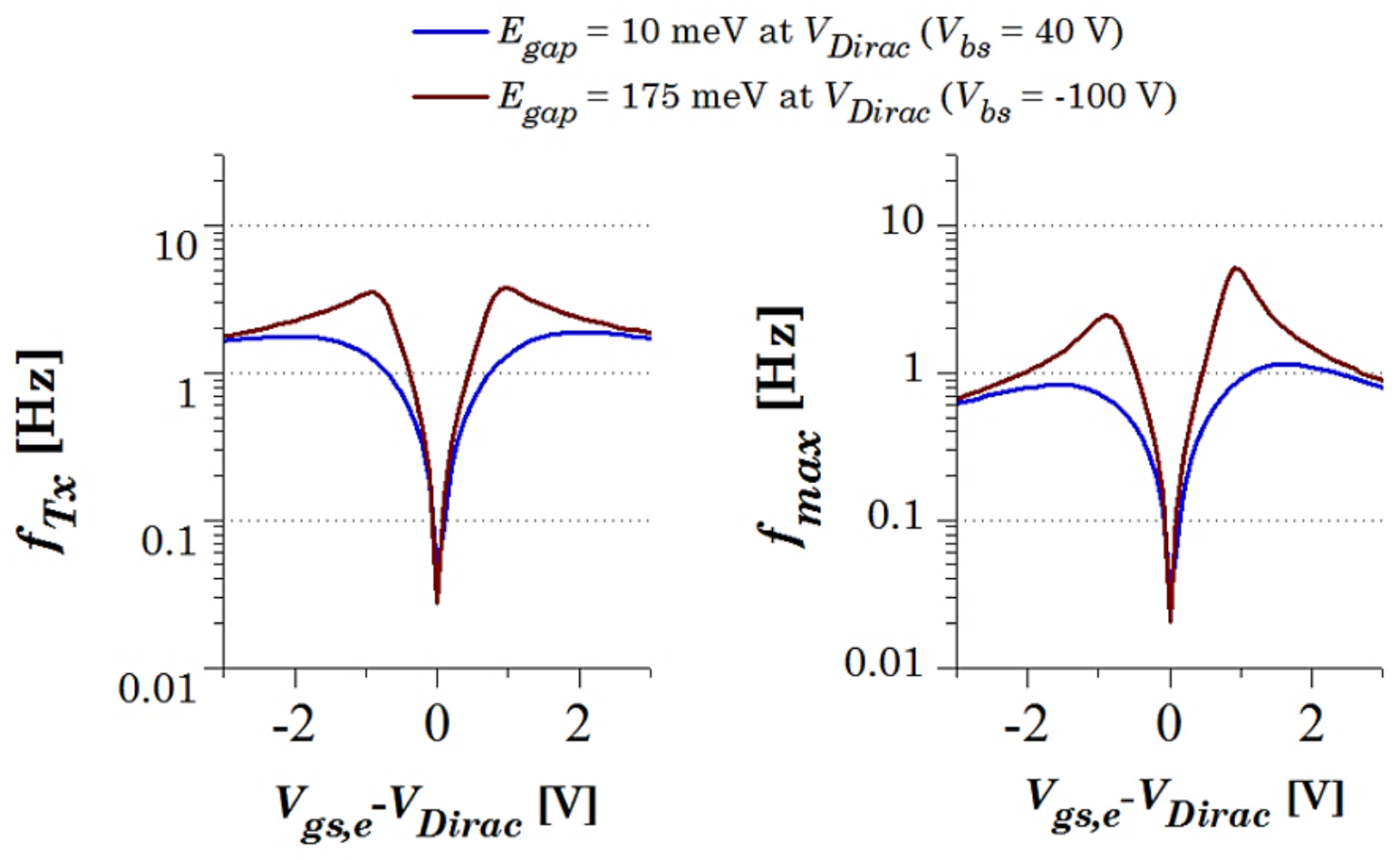

**Figure 4.12** Prediction of a) $f_{Tx}$ and b) $f_{max}$ for the examined device upon application of different $V_{bs}$ resulting in small and large bandgap at the CNP: (blue) $V_{bs}$ = 40 V; and (brown) $V_{bs}$ = -100 V ($V_{ds,e}$ = 1.6 V).

### 4.2.3 BLGFET versus GFET in terms of RF performance

This subsection presents a comparison of the RF performance of two devices: one based on SLG and the other based on BLG. In doing so, the numerical models presented in sections 3.4 and 4.1 are used, respectively. The devices are described via the set of parameters shown in Table 4.3, corresponding to the device reported in [188]. For a fair comparison, the same input parameters have been used for both devices (except those related with the material itself), therefore the same device is considered but replacing the channel material from SLG to BLG.

- *Analysis of the intrinsic capacitances*

First of all, the intrinsic capacitances of the GFET (BLGFET) are analysed as a function of both the gate and drain biases, shown in Figure 4.13a (Figure 4.13b) on the left and on the right, respectively. As for the $C - V_{gs,e}$ characteristics, there are up to three singular points referred as A, B, C in Figure 4.13c (Figure 4.13d) left. Say, for instance the self-capacitance $C_{gg}$, where all three points lie within the simulated $V_{gs,e}$ window. Point A is reached at $V_{gs,e}$ such as $V_{cs} = 0$, so the pinch-off point is just at the source side and the channel is entirely p-type. Further increasing of $V_{gs,e}$ produces the shifting of the pinch-off point towards the middle of the channel where now $V_{cd} = -V_{cs}$, so the half part of the channel close to the source becomes p-type and the other half part close to the drain becomes n-type, resulting in point marked as B. If $V_{gs,e}$ is still further increased, the condition $V_{cd} = 0$ will eventually be reached at the point C. In this case, the pinch-off point has been shifted exactly at the source side and the channel is entirely n-type. Similar discussion could be made for the $C - V_{ds,e}$ characteristics shown in Figure 4.13a (Figure 4.13b) right according to SFLs represented in Figure 4.13c (Figure 4.13d) right. The behaviour discussed so far regarding the intrinsic capacitances of SLG is qualitatively similar to that reported for the BLG case, although quantitative details might differ.

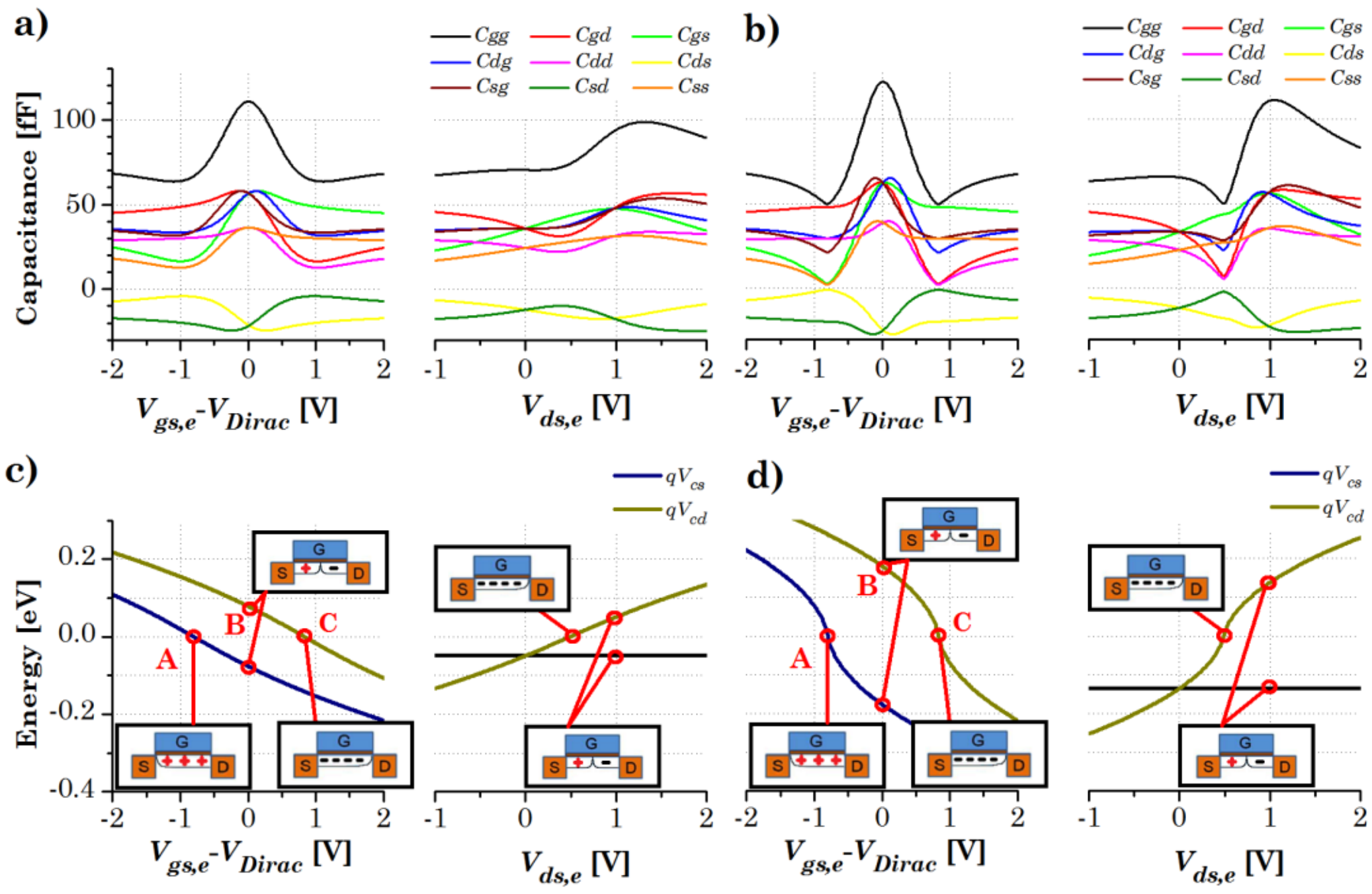

**Figure 4.13** Intrinsic capacitances for the a) SLG- and b) BLG-based device versus the top-gate bias at $V_{ds,e}$ = 1.6V (left) and drain bias at $V_{gs,e}$ =0.3V (right), respectively, for $V_{bs}$ - $V_{b0}$ = 0 V. b) SFL at the drain and source sides for the c) SLG- and d) BLG-based device plotted respect to top-gate bias at $V_{ds,e}$ = 1.6V (left) and drain bias at $V_{gs,e}$ =0.3V (right), respectively.

**Table 4.3** Input parameters of the SLG vs. BLG-based FET benchmarking [188].

| Input parameter | Value | Input parameter | Value |
|---|---|---|---|
| $\gamma_0$ | 3.16 eV | $L_t$ | 12 nm |
| $\gamma_1$ | 0.381 eV | $L_b$ | 300 nm |
| $a$ | 2.49 Å | $\varepsilon_g$ | 2.5 |
| $c_0$ | 3.34 Å | $\varepsilon_t$ | 7.5 |
| $\beta$ | 0 $V^{-1}$ | $\varepsilon_b$ | 3.9 |
| $T$ | 300 K | $V_{g0}$ | -0.8 V |
| $\mu$ | 400 $cm^2$/Vs | $V_{b0}$ | 0 V |
| $L$ | 360 nm | $\Delta$ | 25 meV |
| $W$ | 40 µm | $A_{Ti\text{-}BLG}$ | $8 \cdot 10^4$ $A/m^2K^2$ |
| $L_c$ | 400 nm | | |

- *GFET vs. BLGFET RF performance*

The RF performance of any 2D-FET has to do with the transconductance, output conductance, intrinsic capacitances and extrinsic resistances as given by (2.7) and (2.9). Such extrinsic resistances have been considered constant for both SLG and BLG-based devices in order to make fairer the RF performance comparison, concretely $R_s \cdot W = R_d \cdot W = 500$ Ω·µm and $R_g \cdot L = 4.4$ Ω·µm. Then, a natural question arising is how far the BLG can go respect to its SLG counterpart regarding the RF performance. To answer this question, a BLG channel with a variety of induced bandgaps at the CNP has been considered. This can be done, in practice, by polarizing the device with appropriate $V_{bs}$. To start with, Figure 4.14 shows the calculated $g_m$, $C_{gg}$ and corresponding $f_{Tx}$ as a function of $V_{gs}$, where $C_{gg}$ is the dominant capacitance in defining the RF performance. First observation is that $g_m$ looks like symmetric. This is because the equivalent role played by electrons and holes when positive and negative gate biases, respectively, are applied to the device. In addition, as the gate voltage is varied, $f_{Tx}$ is being modulated by $g_m$. Its value expands over several orders of magnitude depending on the top-gate bias and reaches up to several GHz in this example. The maximum takes place at $V_{gs}$ corresponding to the peak $g_m$, which is around 35 mS, in agreement with the experiment [188]. Moreover, this bias point results in the minimum $C_{gg}$, so $f_{Tx}$ maximizes its value. Around this special point, the advantage of using BLG instead of SLG is clearly observed, so when the induced gap is larger than 220 meV then $f_{Tx}$ scales up in a factor more than 8. Nevertheless, the slightly asymmetry between both peaks of $f_{Tx}$ at negative and positive overdrive gate bias is due to the different output conductance.

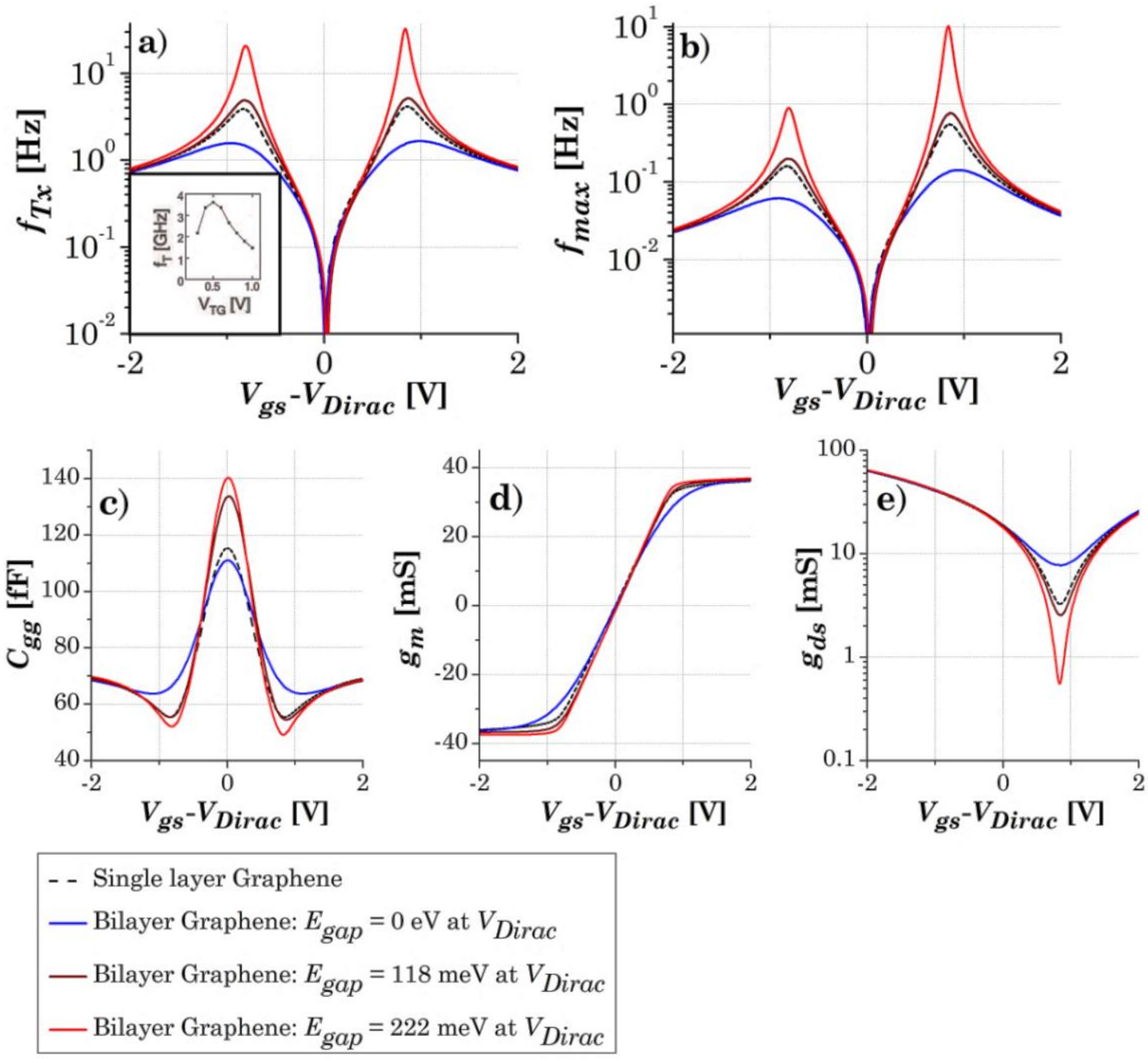

**Figure 4.14** Theoretical calculation of the main RF FoMs such as a) cut-off frequency $f_{Tx}$, and b) maximum oscillation frequency, $f_{max}$. These are shown for a GFET and a BLGFET versus the top-gate overdrive bias at $V_{ds}$ = 1.6 V. Relevant parameters determining the FoMs behavior, such as c) the intrinsic gate capacitance, $C_{gg}$, d) intrinsic transconductance, $g_m$, and e) intrinsic output conductance $g_{ds}$ are also shown. The inset shows the experimental cut-off frequency measured for the GFET in [188].

Next, let us look into the $f_{max}$ behaviour in Figure 4.14b. This FoM critically depends on how good the current saturation is and $g_{ds}$ is serving as key indicator. So, to investigate it, $g_{ds}$ and $f_{max}$ vs $V_{gs}$ for different induced gaps have been also calculated, again by appropriate tuning of $V_{bs}$. As usual, the SLG case has been plotted as a reference. Provided that the gap is larger than one hundred meV, saturation becomes dramatically improved, so $f_{max}$ goes up to the maximum value (several GHz for the examined device). It is interesting to compare this result against the SLG case. The minimum $g_{ds}$ for the GFET is around 3.32 mS. However, for the BLGFET, when the gap size is 222 meV, becomes 6 times smaller and this value ultimately translates into 20 times larger $f_{max}$. This result highlights the importance of current saturation when it comes to optimizing RF FoMs.

## 4.3 Conclusions

The lack of a bandgap in graphene prevents proper current saturation which is linked to the maximum oscillation frequency reached by a GFET. The larger the current saturation is, the less the output conductance is and consequently the larger the maximum oscillation frequency is. In this regard, the possibility of opening a bandgap offered by bilayer graphene is explored. In this chapter, a numerical large-signal model of BLGFET has been presented for the following purposes: (i) understanding of electronic properties of BLG and how to take advantage of its tunable bandgap; (ii) evaluating the impact of the bandgap on the RF FoMs as compared with the SLG counterpart, (iii) performance assessment and benchmarking against other existing technologies, and (iv) provide guidance for device design. For such a purpose, a review of the electronic properties of bilayer graphene has been presented, followed by the electrostatics analysis of a BLGFET-based structure. The model makes full account of the tunable bandgap nature of the BLG and the electric-field screening effects. Then, a drain current model and a charge-based intrinsic capacitance model have been proposed assuming a field-effect model and DD carrier transport. To reproduce experimental *I*-*V* curves, contact resistances have been included considering the Schottky barrier formed between the metal contact and the BLG, which are known to degrade the RF performance.

The large-signal model has been benchmarked against experimental prototype transistors and a comparison between two identical devices based on SLG and BLG has been analysed. The bandgap opening ultimately results into a better switch off together with enhanced drain current saturation as compared with the SLG counterpart. As for the considered bilayer graphene devices, enhancement factors up to 2-20 in either the $f_{Tx}$ or the $f_{max}$ have been found as compared with the equivalent SLG-based devices. What is more important, these enhancement rates are gotten under application of an appropriate back-gate bias $V_b$ producing a bandgap of some hundred meVs at the CNP. However, it is worth noticing that although the drain/source contact and access resistances have been considered to be the same for both SLG and BLG-based devices for the sake of a fair comparison, the mobility has been considered to be the same despite the fact that the mobility somehow decreases with the size of the induced bandgap [177], [183].

It is also worth noticing that the devices considered in this chapter were not optimized to get maximum performance. Optimization requires downscaling of the channel length together with appropriate choice of the insulator thickness and permittivity to keep SCEs under control. The scaling strategy to follow is unknown at the time being. Further investigation of this aspect is needed.

Chapter 5

# General conclusions and outlook

In this thesis, the modelling of 2D material based field-effect transistor has been studied, with a special focus on graphene-based devices. The charge-conserving models proposed comprise a small-signal model for 2D-FETs, a compact large-signal model for GFETs and numerical large-signal models for GFETs and BLGFETs. Taking full advantage of them the following investigation has been performed: (i) analytical calculation of the RF FoMs, (ii) benchmarking among different RF transistor technologies, (iii) DC, AC, transient, *S*-parameters and spectral simulations of GFET-based circuits, (iv) stability analysis of such devices when they are used as a two-port amplifier and (v) thorough investigation of the electronic properties of graphene and bilayer graphene and its impact on the RF performance. As a result, further advance in modelling of 2D material based FETs has been carried out in this thesis. A summary of the main contributions of this thesis is drawn next in section 5.1, together with future prospects, which are given in section 5.2.

## 5.1 Thesis contributions

The main contributions of this thesis, listed by chapter, are summarized:

- Chapter 2, Small-signal model for 2D material based FETs – A small-signal model suited to 2D-FETs that guarantees charge conservation has been proposed. A parameter extraction methodology that includes the metal contact and access resistances has been then proposed. This inclusion is crucial when dealing with low-dimensional FETs. Taking such a small-signal model as a basis, exact analytical expressions for the RF performance of such devices have been provided. Next, a thorough investigation of the scalability and stability of these devices when acting as power amplifiers has been done. This kind of model is of upmost importance when dealing with the first stages of a new technology, helping for fast prototyping and serving as accurate tools to assess the performance of such new 2D-FETs.

- Chapter 3, Large-signal modelling of graphene-based FETs – The key contribution reported in this chapter has been the development of an intrinsic physics-based large-signal compact model of GFETs, ready to be used in conventional EDA tools allowing device-circuit co-design. This compact model has the potential of being a useful tool for designing complex MMICs based on graphene. It is available online in [130] and the source code has been protected under the Benelux Office for Intellectual Property (BOIP) with i-DEPOT number: 083447, keeping the use of the model only for research purposes. Contrary to the small-signal model proposed in Chapter 2, the proposed large-signal compact model is oriented towards more mature 2D-FET technologies of higher TRL, potentially making the circuit design-fabrication cycle more efficient and enabling more complex MMIC designs.

- Chapter 4, Large-signal modelling of bilayer graphene based FETs – A numerical large-signal model of BLGFETs has been proposed to investigate the impact of the BLG tunable gap on the RF performance. The better on-off current ratio, as well as the better current saturation observed in BLG compared to the SLG counterpart, have been qualitatively explained because of the formation of an energy gap at the CNP. A proper biasing of the device is crucial to take advantage of the gap tunability to get the best possible RF performance. The maximum gap that could be opened considering intrinsic BLG is ~ 250 meV. With the same transport properties in both BLG and SLG, a gap of ~220 meV at the CNP could improve the maximum oscillation frequency in a factor of 20 compared to a device based on SLG.

## 5.2 Future outlook

To end the chapter, this section reports several promising research directions to further extend the state-of-the-art of 2D-FET modelling:

- Inclusion of GFET non-idealities. The intrinsic description of GFETs given in this thesis must serve as a starting point toward a complete GFET model which could incorporate additional non-idealities. Among them, (i) an extrinsic description of the device should be carried out. In doing so, a description of the parasitic effects such as parasitic capacitances, inductances taking into account effects of the probing pads, metal interconnections must be included. Likewise, the inclusion of the voltage-dependent contact and access resistances seems to be crucial. (ii) An accurate and physical description of mobility has been realized to be essential for distortion analysis [49]. (iii) An accurate prediction of the HF noise would be very useful for the design of many RF building blocks, as well as, the inclusion of the NQS effect, so the model could properly describe the device behaviour at very high-frequencies where the *quasi-static* assumption breaks down. Finally, (iv) further inclusions of other physical effects such as short-channel and narrow width effects, trapped charge, etc.
- The development of a parameter extraction methodology for the compact large-signal model of GFETs. No matter how accurate a physical model is, it cannot give accurate results unless appropriate values are used for its parameters. Determining these values is not a simple matter because (i) some of these parameters may not be known accurately, (ii) some of them are basically empirical in nature or (iii) even if the value of a physical parameter is known accurately, this value may not be the best one to use in the model to predict a behaviour as close as possible to measurements. Because of all above-mentioned, a suitable parameter extraction methodology should be developed for the compact large-signal model of GFETs.
- Modelling of 2D-FETs. Further experimental validation of 2DMs for high-frequency electronics must be done before choosing the suitable one, but some of them are becoming promising materials for solving the scaling issues and, perhaps even more important, for the future development of flexible electronics. Examples of them are transition metal dichalcogenides, where $MoS_2$ stands out, phosphorene, silicene, etc. The great deal of interest in 2DMs, especially for flexible applications, makes relevant the formulation of appropriate models to help pushing the 2DM technology to the next level. The compact modelling of different technologies would ease the design of complex MMICs as well as integrated RF circuits processed at the back end of line of regular silicon CMOS technology.

# List of publications by topic

## Chapter 2

## Small-signal model for 2D material based FETs

- F. Pasadas, W. Wei, E. Pallecchi, H. Happy and D. Jiménez, "Small-signal model for 2D-material based field-effect transistors targeting radio-frequency applications: the importance of charge conservation," submitted to *IEEE Trans. Electron Devices*, 2017.
- F. Pasadas and D. Jiménez, "Small-signal model for RF graphene transistors," *Graphene2017 International Conference*, Barcelona (Spain), March 2017.
- F. Pasadas and D. Jiménez, "RF performance of graphene field-effect transistors," *11$^{th}$ Spanish Conference on Electron Devices*, Barcelona (Spain), Feb. 2017.

## Chapter 3

## Large-signal modelling of graphene-based FETs

- F. Pasadas and D. Jiménez, "Large-Signal Model of Graphene Field-Effect Transistors – Part I: Compact Modeling of GFET Intrinsic Capacitances," *IEEE Trans. Electron Devices*, vol. 63, no. 7, pp. 2936-2941, Jul. 2016. DOI: 10.1109/TED.2016.2570426

- F. Pasadas and D. Jiménez, "Large-Signal Model of Graphene Field-Effect Transistors – Part II: Circuit Performance Benchmarking," *IEEE Trans. Electron Devices*, vol. 63, no. 7, pp. 2942-2947, Jul. 2016. DOI: 10.1109/TED.2016.2563464
- F. Pasadas and D. Jiménez, "Large-signal model of graphene field-effect transistors," *Graphene2016 Conference*, Genoa (Italy), Apr. 2016.
- F. Pasadas, D. Jiménez, M. Iannazzo and E. Alarcón, "Capacitance compact modelling of four-terminal graphene FETs preserving charge conservation: A circuit-oriented device model benchmark," *Graphene2015 Conference*, Bilbao (Spain), March 2015.

## Chapter 4

# Large-signal modelling of bilayer graphene based FETs

- F. Pasadas and D. Jiménez, "Large-signal model of the bilayer graphene field-effect transistor targeting radio-frequency applications: Theory versus experiment," *J. Appl. Phys.*, vol. 118, no.24, p. 244501, Dec. 2015. DOI: 10.1063/1.4938114
- F. Pasadas and D. Jiménez, "Large-signal model of bilayer graphene field-effect transistors," *Graphene Flagship – Graphene Study*, Kaprun (Austria), March 2015.
- F. Pasadas and D. Jiménez, "Electrostatics and drift-diffusion current model of bilayer graphene field-effect transistors," *10th Spanish Conference on Electron Devices*, Aranjuez (Spain), Feb. 2015.
- F. Pasadas and D. Jiménez, "Electrostatics and drain current model of bilayer graphene field-effect transistors," *Graphene2014 International Conference & Exhibition*, Toulouse (France), May 2014.